\documentclass[preprint,preprintnumbers]{lecnotes}

\usepackage{colortbl,stmaryrd}
\usepackage[T1]{fontenc}   % T1 encoding has pre\UTF{2011}built accented glyphs

\usepackage[utf8]{inputenc}
\usetikzlibrary{positioning, calc, arrows.meta, shadows.blur}
\usepackage{import}
\usepackage{fancyvrb}
\usepackage{array}
\usepackage{graphicx}

\tikzset{
  stage/.style={
    draw,
    thick,
    rounded corners=12pt,
    minimum width=2.8cm,
    minimum height=1.4cm,
    inner sep=2pt,
    font=\sffamily\footnotesize\bfseries,
    align=center%,
  },
  data/.style={stage, fill=blue!15, draw=blue!60!black},
  process/.style={stage, fill=teal!15, draw=teal!60!black},
  model/.style={stage, fill=orange!18, draw=orange!70!black},
  analysis/.style={stage, fill=purple!12, draw=purple!70!black},
  special/.style={stage, fill=green!18, draw=green!60!black},
  connector/.style={ultra thick, -{Latex[length=6pt,width=12pt]}, draw=black}
}
\usepackage[parfill]{parskip}

\DeclarePairedDelimiterX{\klx}[2]{(}{)}{%
  #1\;\delimsize\|\;#2%
}

\newtheorem{lem}{Lemma}

\pdfoutput=1
\usepackage{dsfont}
\usepackage{graphicx}
\usepackage{subcaption}
\usepackage{amsmath}
\usepackage{slashed}
\usepackage{caption}
\usepackage{cleveref}
\usepackage{xcolor,color}
\usepackage{tikz}
\usepackage{tikz-cd}
\usepackage[framemethod=tikz]{mdframed}
\usepackage{hyperref}

\definecolor{DarkSlateBlue}{HTML}{483D8B}
\hypersetup{
    colorlinks,
    linkcolor={DarkSlateBlue},
    citecolor={blue!50!black},
    urlcolor={blue!80!black}
}

\makeatletter
\renewcommand*\env@matrix[1][\arraystretch]{%
  \edef\arraystretch{#1}%
  \hskip -\arraycolsep
  \let\@ifnextchar\new@ifnextchar
  \array{*\c@MaxMatrixCols c}}
\makeatother

\newcommand{\saw}[1]{\left\langle #1 \right\rangle}

\newcommand{\sn}[1]{\S~\ref{sec:#1}}

\newcommand\kinv{\kappa}

\newcolumntype{C}[1]{>{\centering\arraybackslash}m{#1}}

\newcommand{\rankfeat}[2]{%
  \begin{tabular}{@{}c@{}}%
  $#1$\\[-2pt]{\tiny $#2$}%
  \end{tabular}%
}

\newcommand\blfootnote[1]{%
  \begingroup
  \renewcommand\thefootnote{}%
  \footnote{#1}%
  \addtocounter{footnote}{-1}%
  \endgroup
}

\def\shrug{\texttt{\raisebox{0.75em}{\char`\_}\char`\\\char`\_\kern-0.5ex(\kern-0.25ex\raisebox{0.25ex}{\rotatebox{45}{\raisebox{-.75ex}"\kern-1.5ex\rotatebox{-90})}}\kern-0.5ex)\kern-0.5ex\char`\_/\raisebox{0.75em}{\char`\_}}}

\usepackage{xxcolor}
\definecolor{cardinal}{rgb}{0.6,0,0}
\definecolor{darkgreen}{rgb}{0,0.5,0}
\definecolor{golden}{rgb}{0.92, 0.7, 0}
\definecolor{midnight}{rgb}{0, 0, 0.5}
\definecolor{darkblue}{rgb}{0.2, 0, 0.8}

\begin{document} 

\title{
Learning Topological Features of $\widehat Z$-invariants\\
}

\author{Brandon Robinson$^{\dagger,*}_{a}$ \and Shimal Harichurn$^\dagger_{b}$ \and Fabian Ruehle$_{c,d,e}$ \and Sergei Gukov$_{f}$  \and Rak-Kyeong Seong$_{g}$ \and Miranda C. N. Cheng$_{a,h,i}$ }
\date{%
    \bigskip
    \small
    $^a$Institute of Physics, University of Amsterdam, Science Park 904, 1098 XH Amsterdam,
Netherlands\\[7pt]
    $^b$School of Mathematics, Statistics and Computer Science, University of KwaZulu-Natal, South Africa\\[7pt]
    $^c$Department of Physics, Northeastern University, Boston, MA 02115, USA\\[7pt]    
    $^d$Department of Mathematics, Northeastern University, Boston, MA 02115, USA\\[7pt]
    $^e$The NSF AI Institute for Artificial Intelligence and Fundamental Interactions\\[7pt]
    $^f$Richard N. Merkin Center for Pure and Applied Mathematics, California Institute of Technology,
    Pasadena, CA 91125, USA\\[7pt]
    $^g$Department of Mathematical Sciences, and Department of Physics, Ulsan National Institute of Science and Technology, 50 UNIST-gil, Ulsan 44919, South Korea\\[7pt]
    $^h$Institute for Mathematics, Academia Sinica, Taipei 106319, Taiwan\\[7pt]%
    $^i$Korteweg-de Vries Institute for Mathematics, University of Amsterdam, Science Park 105-107, 1098 XG Amsterdam, Netherlands\\[7pt]
    \bigskip
}

\maketitle
\begin{abstract}
Machine learning and data analysis techniques have recently emerged as powerful tools for identifying patterns and formulating conjectures in mathematical research, most notably in the field of low-dimensional topology. In this paper, we initiate a systematic approach to handling mathematical data structured as (truncated) infinite $q$-series, or equivalently, infinite series of integers.
To apply this data analysis pipeline, we construct a comprehensive dataset of $\widehat{Z}$-invariants (homological blocks) for plumbed 3-manifolds. We demonstrate that neural networks can reliably extract essential topological information, such as homology class and underlying graph structure, directly from the $q$-series coefficients.
A central feature of our methodology is a focus on interpretability; by contrasting local gradient sensitivity with global feature relevance, we reveal that the networks learn to bypass complex topological rules in favor of specific spectral and geometric proxies.  Finally, we apply this pipeline to probe homology cobordism, discovering a high-accuracy predictive relationship between the $\widehat{Z}$-invariant exponents and the Heegaard Floer $d$-invariant (correction term). These results suggest that $\widehat{Z}$-invariants capture subtle geometric information regarding cobordism equivalences, warranting a new direction for the study of quantum invariants.

%\blfootnote{$^\star$ Corresponding author: b.j.robinson@uva.nl}
\blfootnote{$^\dagger$ Equal Contribution, ${}^*$ b.j.robinson@uva.nl}

\end{abstract}

\clearpage
\tableofcontents

\section{Introduction}

The interaction between artificial intelligence and pure mathematics has historically focused on automated theorem proving or symbolic computation. In recent years, however, machine learning (ML) has emerged as a distinct and powerful partner in the theoretical research pipeline, expanding the scope of neural network applications from verification to discovery. This paradigm leverages the capacity of neural networks to detect high-dimensional correlations that may elude human intuition, thereby aiding researchers towards finding new, mathematically rigorous conjectures. So far, this strategy has proven particularly fruitful in the field of low-dimensional topology, where vast tabulations of knot invariants provide rich, structured datasets suitable for data-driven analysis. Notable examples include discovering novel relationships between the geometric structure of knots and their signature \cite{Davies2021}, as well as predicting invariants such as the Jones polynomial and hyperbolic volume \cite{Jejjala:2019kio,Craven2021, Craven_2024}. Partially as a result of the availability of knot datasets, the majority of this existing literature focuses on knots and knot complements rather than closed 3-manifolds, and typically targets scalar or polynomial invariants  \cite{Gukov:2020qaj, Hughes2020, Levitt2022,lindsay2025learnability}.  

In this work, we turn to data in the form of infinite $q$-series, or equivalently infinite series of integers, appropriately truncated. This is in part motivated by the so-called $\widehat Z$-invariants of closed 3-manifolds, $\widehat Z(M_3;q)$. 
The $\widehat{Z}$-invariants, often referred to as homological blocks, constitute a family of infinite $q$-series invariants, $\widehat{Z}_b(q) \in q^{\Delta_b} \mathbb{Z} [[q]]$, characterized as power series with integer coefficients that converge within the unit disk $|q| < 1$ \cite{Gukov:2016gkn, Gukov:2017kmk}. These invariants have their origin in M-theory, arising from the physics of M5-branes wrapped on 3-manifolds, where they capture the 3d half-index (the supersymmetric partition function on a cigar background multiplied by the temporal circle) of the 3-dimensional $\mathcal{N}=2$ theory resulting from the compactification. 

Much of the study of the $\widehat{Z}$-invariants, including the present work, focuses on a class of 3-manifolds known as the {\it plumbed 3-manifolds}. 
Given a weighted tree $\Gamma$, one can build a framed link $L(\Gamma) \subseteq S^3$ consisting of copies of unknots chained together whenever vertices are connected by an edge in $\Gamma$ and with framing numbers provided by the weights on the vertices. One can then perform  Dehn surgery on the resulting framed link to obtain a closed 3-manifold $Y(\Gamma)$, called a plumbed manifold. We call $\Gamma$ a \textit{plumbing graph} for the corresponding plumbed manifold. The data of the weighted tree can equivalently be encoded in a matrix, referred to as the {\it plumbing matrix}. 
For a so-called {\it weakly negative plumbed 3-manifolds}, the invariants are defined mathematically through a principal value contour integral of a lattice theta function, appropriately decorated by the contribution from the edges of the plumbing graph. A plumbed manifold is said to be {\it weakly negative} if its plumbing matrix inverse, $M^{-1}$, is negative definite when restricted to the subspace spanned by vertices of degree three or greater. We refer to these  vertices as the high-degree vertices. Concretely, the $\widehat Z$-invariants for these manifolds are given by 
\begin{equation}\label{weak_neg_def}
\widehat Z_b(M_{3};\tau) :=  q^{\Delta'} \oint \prod_{v\in V}\frac{dz_v}{2\pi i z_v}~ \left(z_v-\frac{1}{z_v}\right)^{2-{\rm deg}(v)}~\Theta_b^{M}(\tau;{\bf z})
\end{equation}
for some  $\Delta' \in \mathbb Q$, and some choice of vector $b$ in the lattice, 
where the theta function is given by
\begin{equation}\label{def:theta}
\Theta^M_b (\tau;{\bf z}):=\sum_{\ell \in  2M\mathbb Z^{|V|}\pm b}q^{-\ell^T M^{-1} \ell/4} \, {\bf z}^\ell.
\end{equation}
These invariants are significant across the realms of mathematics and physics, functioning simultaneously as topological invariants and BPS invariants, and they are believed to be closely related to quantum modular forms and (non-rational) vertex operator algebras \cite{Cheng:2018vpl, Cheng:2022rqr, Cheng:2024vou, Bringmann_2019, Bringmann_2020}. They are known to extend the Witten--Reshetikhin--Turaev (WRT) invariants away from the roots of unity, and it is expected that their categorification will yield powerful smooth $4$-manifold invariants.

While various properties of the $\widehat{Z}$-invariants have been intensively studied in recent years, many remain mysterious. For instance, we would like to ask 
\begin{center}
{\it What type of information about the 3-manifold $M_3$ does $\widehat{Z}(M_3)$ encode? And how? }
\end{center}
Given the recent advances in data analysis and machine learning methods, it is natural to probe these questions using these new tools. Such an approach was initiated in the preliminary work \cite{PhysRevD.110.046016}, where the authors performed Principal Component Analysis (PCA) of $\widehat{Z}$, truncated to ten terms, for a small dataset of $20$ plumbed 3-manifolds. 
To greatly scale up such a data-based study of $\widehat{Z}$-invariants, in this paper we first build various datasets of size $10^4$ of plumbed 3-manifolds, each with $\widehat Z$-invariants computed up to the first $10^4$ terms. 
Subsequently, we build an ML-assisted data analysis pipeline to extract information from this type of data.

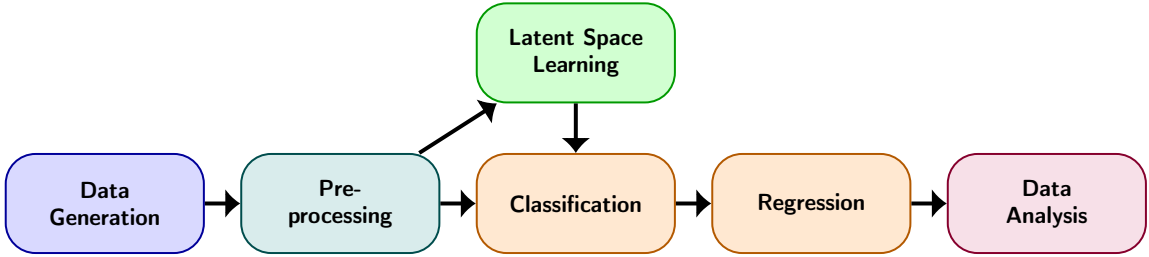
\begin{figure*}[ht!]
    \centering
    \resizebox{\textwidth}{!}{%
\begin{tikzpicture}[node distance=5mm and 5mm]

  % --- Main pipeline boxes ---
  \node[data] (gen) {Data\\Generation};
  \node[process, right=of gen] (pre) {Pre-\\processing};
  \node[model, right=of pre] (cls) {Classification};
  \node[model, right=of cls] (reg) {Regression};
  \node[analysis, right=of reg] (ana) {Data\\Analysis};

  % --- Contrastive learning above classification ---
  \node[special, above=7mm of cls] (con) {Latent Space\\Learning};

  % --- Connectors ---
  \draw[connector] (gen) -- (pre);
  \draw[connector] (pre) -- (cls);
  \draw[connector] (cls) -- (reg);
  \draw[connector] (reg) -- (ana);
  \draw[connector] (pre) -- (con);
  \draw[connector] (con) -- (cls);

\end{tikzpicture}
}
    \caption{Schematic of the ML-assisted data analysis pipeline.}\label{fig:ml-flow-chart}%
\end{figure*}

Beyond standard prediction networks, we place special emphasis on interpreting the mechanisms driving the neural network's decisions. To achieve this, drawing on the  strategy of \cite{halverson2025learning}, we train a separate regression network to map a comprehensive set of mathematically motivated features directly to the logits of the classifier. We then employ data analysis tools, including feature saliency (FS), permutation importance (PI), and Shapley values (SHAP), to decode the information exploited by the classification model. Figure \ref{fig:ml-flow-chart} summarizes the integrated pipeline developed in this work.
 
Recall that the $\widehat{Z}$-invariants in our database are computed using the plumbing graph, or equivalently the plumbing matrix, of a closed 3-manifold $M_3$. 
To test the interpretability methodology, we first choose classification tasks which are straightforward using the plumbing graph/matrix but obscure when just given the resulting infinite $\widehat{Z}$ $q$-series. 
Examples include the identification of integral homology spheres, which correspond to plumbed manifolds with unimodular plumbing matrices, and the determination of the underlying graph topology, such as distinguishing star-shaped graphs from H-shaped configurations. 

To ensure the robustness of the interpretations, we perform many of the experiments in three different data representations --- Full Sequence (including all zero coefficients), Value-Index Pair (excluding zeros), and Latent Space representations trained using contrastive learning with labeled datasets. 
We find that our method can indeed lead to human-understandable insights into how the classification network decodes specific topological information from the infinite $q$-series. 

These insights reveal that the neural network sometimes employs simpler ``surrogate'' features that are merely correlated with the defining mathematical properties, rather than learning the exact theoretical rules. 
For instance, we observe that the network effectively uses the minimum-magnitude eigenvalue of the plumbing matrix as a proxy for the determinant in some of the classification tasks.
This reliance on proxies provides a clear and quantifiable explanation for the classifier's imperfect generalization, as we discuss in detail in \S\ref{sec:topExp} and \S\ref{sec:otherexp}.

Having validated our methodology on basic topological questions, we turn to the exploration of the connection between the $q$-series invariants and the three-dimensional homology cobordism group $\Theta^3_{\mathbb{Z}}$, a subject of prime interest in low-dimensional topology. Indeed, while $\Theta^3_{\mathbb{Z}}$ is related to many important questions about smooth structures on 4-manifolds and other exciting questions, its structure --- for instance, the existence of nontrivial torsion --- remains quite mysterious despite decades of study.

The definition of $\Theta^3_{\mathbb{Z}}$ is deceptively simple: it is the set of oriented integer homology spheres modulo homology cobordism, where two spheres $Y_0$ and $Y_1$ are equivalent if they cobound a smooth 4-manifold $W$, $\partial W = Y_0 \sqcup (-Y_1)$,
such that the inclusion maps $Y_i \hookrightarrow W$ induce isomorphisms on integer homology. Under the connected sum operation, these equivalence classes form an abelian group, with the identity element represented by the standard sphere $S^3$. Some of the basic properties of the homology cobordism group come from Rokhlin's signature theorem, which yields the homomorphism $\mu\colon \Theta^3_{\mathbb{Z}} \to \mathbb{Z}/2$ \cite{Rokhlin1952}, and one has the associated short exact sequence
\[
0 \longrightarrow \ker(\mu) \longrightarrow \Theta^3_{\mathbb{Z}} \xrightarrow{\ \mu\ } \mathbb{Z}/2 \longrightarrow 0.
\]
Galewski--Stern and Matumoto showed that the question of whether this sequence splits is equivalent to the existence of simplicial triangulations for all manifolds of dimension $\ge 5$ \cite{GalewskiStern1980,Matumoto1978}. Manolescu proved that it does \emph{not} split using Pin$(2)$-equivariant Seiberg--Witten Floer theory, thereby resolving the high-dimensional Triangulation Conjecture \cite{Manolescu2016}. This illustrates how the integer homology cobordism group has important applications in low-dimensional topology, albeit through rather advanced tools like gauge theory and Floer homology.

Important developments in understanding the structure of $\Theta^3_{\mathbb{Z}}$ include the construction of explicit infinite-order elements via instanton Floer theory by Fintushel--Stern \cite{FintushelStern1990}, Furuta's proof that $\Theta^3_{\mathbb{Z}}$ contains a free abelian subgroup of infinite rank \cite{Furuta1990}, and Fr{\o}yshov's surjective homomorphism $h\colon \Theta^3_{\mathbb{Z}} \to \mathbb{Z}$ (hence a $\mathbb{Z}$-summand) \cite{Froyshov2002}. 
Heegaard Floer theory provides computable homology cobordism invariants such as the correction term $d$ \cite{OzsvathSzabo2003_dInvariant} and its involutive refinements \cite{HendricksManolescu2017}, and Dai--Hom--Stoffregen--Truong showed that $\Theta^3_{\mathbb{Z}}$ admits an infinite-rank direct summand \cite{DaiHomStoffregenTruong2023}. 
Despite these advances, key questions remain open, so the group continues to serve as a central testing ground for new topological and quantum invariants \cite{Savk2024Survey}, particularly their ability to detect non-trivial cobordism classes that classical invariants fail to distinguish.

After having tested our data analysis methodology on basic topological questions with the $\widehat{Z}$ database we constructed, we turn to the question of probing cobordism information contained in $\widehat{Z}$-invariants. 
A homology cobordism invariant is a group homomorphism $\lambda: \Theta^3_{\mathbb{Z}} \to A$ where $A$ is an abelian group. 
A modern example of such an invariant is the Heegaard Floer correction term, or $d$-invariant, introduced by Ozsv\'ath and Szab\'o \cite{OzsvathSzabo2003_dInvariant}. 
Originating from the grading of the Heegaard Floer homology group $HF^+(Y)$, the $d$-invariant measures the ``spectral shift'' of the infinite rank tower in the homology relative to that of the 3-sphere. For integer homology spheres it takes values in the even integers and provides a powerful obstruction to
slicing knots and embedding manifolds. Partially inspired by previous work \cite{cobordism_invariants_bps, harichurn2024deltaainvariantsnonperturbativecomplex, delta_plumbed_manifolds}, which studied the relationship of $d$ to the leading $q$-power of the $\widehat{Z}$-invariant (the $\Delta$-invariant) and to specializations of the $\widehat{Z}$-invariant, in this work we explore the relation between the $q$-powers of the $\widehat{Z}$-invariant and the $d$ invariant. 

To our surprise, our experiment yielded the following result.  
\begin{center}
     {\it On datasets of Seifert integer homology spheres with $r$ singular fibers (Brieskorn spheres for $r=3$), we found that regression models are able to learn to fit $d$ from the leading exponents of $\widehat{Z}$ and explain $>99.9\%$ of the target variance. By including symmetrized Dedekind sums of the Seifert invariants in the training data, exact integral predictions of $d$ reach nearly $90\%$ accuracy from quadratic polynomials of the important exponents. }
\end{center}
 
Through various diagnostics, we were able to provide a heuristic understanding of how such a  positive result, which might be surprising at first glance, can be plausibly obtained by machine learning. 
Nevertheless, we believe that the result indicates the possibility of some new relation between $\widehat{Z}$-invariants and the $d$-invariant, which in turn suggests an interesting connection between the homology cobordism group and the $\widehat{Z}$-invariants. 

The remainder of this paper is organized as follows. Section~\ref{sec:glossary} summarizes the notation used throughout this work. Section~\ref{sec:data} describes the generation of $\widehat{Z}$ datasets for manifolds plumbed by star- and H-shaped graphs and introduces our input representations. Section~\ref{sec:topExp} details the topological classification experiments, the isolation of driving features, and a control study utilizing the derived features. Section~\ref{sec:cobordism-experiments} investigates the prediction of the Heegaard-Floer $d$-invariant from $q$-series coefficients. Section~\ref{sec:otherexp} presents additional experiments regarding singular fiber counts and $SL_2(\mathbb{Z})$ orbit sizes. Finally, Section~\ref{sec:conclusion} provides a summary of our findings and future directions, followed by technical appendices on data-analysis tools, the multi-seed robustness protocol, spectral and $\min\tau$ bounds, and splice diagrams. 

\section{Glossary}\label{sec:glossary}
\begin{itemize}
    \item[$c_i$] The integer coefficients appearing in the $q$-series expansion.
    \item[$c_{\rm max}$] The maximum coefficient value found in the truncated $q$-series expansion used as input data.
    \item[$\Delta$] The minimal $q$-power of $\widehat{Z}=q^{\Delta} (1+O(q))$.
    \item[$M$] The plumbing matrix  associated with the plumbing graph of the manifold
    \item[$\det$] The determinant of $M$.
    \item[$\varepsilon_{\rm min}$] The eigenvalue of $M$ with the smallest magnitude.
    \item[$\varepsilon_{\rm max}$] The eigenvalue of $M$ with the largest magnitude.
    \item[$\mu_{\varepsilon}$] The mean value of the eigenvalues of $M$.
    \item[$\sigma_{\varepsilon}$] The standard deviation of the eigenvalues of $M$.
    \item[$\gamma_{0}$] The first gap between exponents with nonzero coefficients in the $q$-series.
    \item[$\mu_{\gamma}$] The mean of the gaps between exponents of nonzero coefficients in the $q$-series.
    \item[$\sigma_{\gamma}$] The standard deviation of the gaps between exponents of nonzero coefficients in the $q$-series.
    \item[$d$] The Heegaard-Floer homology cobordism invariant, the ``correction term".
    \item[$\alpha$] The size of the $SL_2(\mathbb{Z})$ orbit associated with the invariant.
    \item[BAS] Balanced Accuracy Score: The average recall across all classes, used to evaluate classification performance on imbalanced datasets.
    \item[PCC] Pearson Correlation Coefficient: A metric measuring the linear correlation between predicted and ground truth values, ranging from -1 to 1.
    \item[$R^2$] Coefficient of Determination: A regression metric quantifying the proportion of variance in the observed data explained by the model.
    \item[$\rho$] Spearman Rank Correlation: A metric assessing the monotonic relationship between ground truth and predictions by calculating the correlation of their rank vectors.
    \item[$\vartheta$] Sign Accuracy: A metric evaluating whether a regression output falls on the correct side of a decision boundary (typically zero), effectively treating it as a binary classifier.
    \item[$S_{\mathrm{MAE}}$] Complement Normalized MAE Score: A metric derived from the Mean Absolute Error, normalized by the range and inverted so that higher values indicate better performance.
    \item[$Q_{\mathrm{composite}}$] A weighted composite score combining $R^2$, $\rho$, $\vartheta$, and $S_{\mathrm{MAE}}$ to assess how well a regressor characterizes a decision boundary.
    \item[FS] Feature Saliency: A gradient-based interpretability method that weights input features by their gradient contribution to the network's output.
    \item[PI] Permutation Importance: A model-agnostic method that quantifies feature importance by measuring the drop in model performance when a feature's values are randomly shuffled.
    \item[SHAP] Shapley Additive Explanations: A game-theoretic approach to interpretability that distributes the model's prediction deviation from the average among features based on their marginal contributions.
    \item[SHTL] Semi-Hard Triplet Loss: A loss function used in contrastive learning that focuses on "semi-hard" negatives (negatives farther from the anchor than the positive but within a margin) to improve latent space clustering.
\end{itemize}

\section{Data}
\label{sec:data}

This section details the data generation process, the resulting datasets, and the input representations used in our experiments. We name the datasets based on the topology of their underlying plumbing graphs and homological property. Each sample consists of the $q$-series expansion of the $\widehat{Z}_b(M_3)$ invariant, truncated (for the topology datasets of  Sec.~\ref{sec:topExp}) to the first $10^4$ coefficients --- the cobordism and orbit-size datasets of later sections use shorter windows described there, and the orbit-size analysis additionally assumes square-free $b_1b_2b_3$ ---, for a plumbed 3-manifold $M_3$ equipped with a fixed Spin$^c$-structure $b$\footnote{For all samples, we select the Spin$^c$ structure $b$ defined by the degree vector, satisfying $b_v = \deg(v)$ for every vertex $v$.}. In total, our aggregate dataset comprises 63,603 such truncated series, derived from plumbing graphs containing one or two high-degree vertices, which we refer to as the star-shaped and H-shaped graphs, respectively. To ensure the quality of the graph topology labels in the dataset, all of the generated graphs have been passed through an automated algorithm to check that they are in minimal form.

\subsection{Datasets}
\label{subsec:datasets}

We begin by outlining the datasets constructed for our analysis. A key constraint is the computational cost of generating data for graphs with multiple high-degree nodes; consequently, our dataset is predominantly composed of star-shaped graphs (roughly 71\%), namely weighted trees with a single high-degree vertex. 
Nevertheless, with sample sizes of $O(10^4)$ for each category, the data remains robust enough to support the machine learning tasks described in this paper.

\begin{table}[ht]
    \centering
    \begin{tabular}{c|c|c}\toprule
          & \textbf{Homology Sphere} & \textbf{Non-Sphere} \\\midrule
         \textbf{Star}& 13228 & 31706 \\\hline
         \textbf{H-shaped} & 7952 & 10717 \\\bottomrule
    \end{tabular}
    \caption{Summary of datasets split between homology and graph topology labels used for binary classification tasks.}
    \label{tab:dataset-binary}
\end{table}

\begin{table}[ht]
    \centering
    \begin{tabular}{c|c|c|c|c|c}\toprule
          $\#$ singular fibers&  3 & 4 & 5 & 6 & 7\\\midrule
         $\#$ samples  & 6000 & 5998& 5988& 5984& 5978 \\\bottomrule

    \end{tabular}
    \caption{Data for multi-class experiments learning the number of singular fibers of Seifert manifolds from $\widehat{Z}$.}
    \label{tab:dataset-SEIF}
\end{table}

\subsubsection{Star-shaped (Seifert) Datasets}

Star-shaped plumbing graphs correspond to Seifert fibered 3-manifolds. 
In this paper, we restrict  our attention to those satisfying the weakly negative condition allowing for straightforward computation of the $\widehat{Z}$-invariants. 

\begin{figure}[ht]
    \centering
    \begin{tikzpicture}
    \begin{scope}[every node/.style={circle, draw, fill=black, inner sep = 0pt, minimum size=5pt}]
        \node (1) at (-2, 0) {};
        \node (2) at (0, 2) {};
        \node (3) at (0, 1) {};
        \node (4) at (0, -2) {};
        \node (5) at (2, 2) {};
        \node (6) at (2, 1) {};
        \node (7) at (2, -2) {};
        \node (8) at (4, 2) {};
        \node (9) at (4, 1) {};
        \node (10) at (4, -2) {};
        \node (11) at (6, 2) {};
        \node (12) at (6, 1) {};
        \node (13) at (6, -2) {};
        \node (14) at (0, -0.5) {};
        \node (15) at (2, -0.5) {};
        \node (16) at (4, -0.5) {};
        \node (17) at (6, -0.5) {};
    \end{scope}
    \draw (1)--(2);
    \draw (1)--(3);
    \draw (1)--(4);
    \draw (2)--(5);
    \draw (3)--(6);
    \draw (4)--(7);
    \draw[dashed] (5)--(8);
    \draw[dashed] (6)--(9);
    \draw[dashed] (7)--(10);
    \draw (8)--(11);
    \draw (9)--(12);
    \draw (10)--(13);
    \draw[dashed] (1)--(14);
    \draw[dashed] (14)--(15);
    \draw[dashed] (15)--(16);
    \draw[dashed] (16)--(17);
% vertex labels
% weight labels
\node (1B) at (-2, 0.5) [text=black]{$b$};
\node (2B) at (0, 1.25) [text=black]{$-w_{{2}, 1}$};
\node (3B) at (0, 2.5) [text=black]{$-w_{1, 1}$};
\node (4B) at (0, -1.5) [text=black]{$-w_{{r}, 1}$};
\node (2B) at (2, 1.25) [text=black]{$-w_{{2}, 2}$};
\node (3B) at (2, 2.5) [text=black]{$-w_{1, 2}$};
\node (4B) at (2, -1.5) [text=black]{$-w_{{r}, 2}$};
\node (2B) at (4, 1.25) [text=black]{$-w_{{2}, l_{2}-1}$};
\node (3B) at (4, 2.5) [text=black]{$-w_{{1}, l_{1}-1}$};
\node (4B) at (4, -1.5) [text=black]{$-w_{{r}, l_{r}-1}$};
\end{tikzpicture}
    \caption{A star-shaped graph representing a Seifert manifold. Vertex labels have been omitted.}
    \label{fig:seifert_star_graph}
\end{figure}
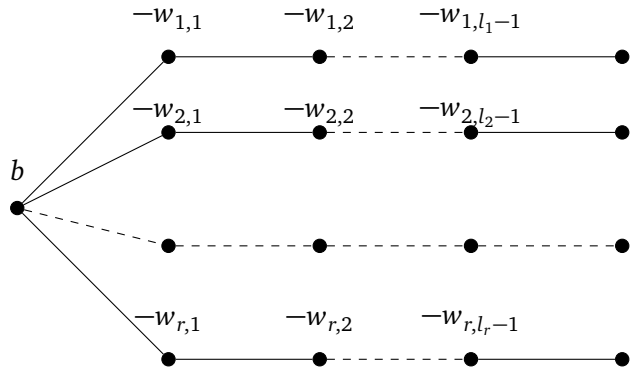

Specifically, for a star-shaped graph as shown in Figure \ref{fig:seifert_star_graph}, we require that on each of the $r$ legs (chains of degree-2 vertices terminating in a degree-1 vertex), the weights satisfy the continued fraction decomposition:\[w_{i, 1} - \cfrac{1}{w_{i, 2} - \cfrac{1}{\cdots - \cfrac{1}{w_{i, l_i}}}} = \frac{b_i}{a_i}\]for coprime integers $a_i, b_i$ satisfying $0 < a_i < b_i$. Additionally, we impose the condition that the weight of the central vertex (of degree $r$) satisfies $b < 0$. Under these constraints, the resulting plumbed manifold corresponds to the Seifert fibered space $M(b; \frac{a_1}{b_1}, \dots, \frac{a_r}{b_r})$. We refer to the pairs $(a_i, b_i)$ as the \textit{Seifert invariants} of the manifold. Note that the orbifold Euler number $e \in \mathbb{Q}$ is given by:\begin{equation}\label{eq:seifert_euler_b}e = b + \sum_{i=1}^r \frac{a_i}{b_i}.\end{equation} Crucially, the condition $b < 0$ typically ensures that the plumbing matrix is weakly negative, a requirement for the convergence of the $\widehat{Z}$-invariant.

The normalization condition $0 < a_i < b_i$ serves to fix the ambiguity arising from 3D Neumann moves. By imposing these constraints, we ensure that each Seifert manifold in our dataset is represented by a unique 'canonical' plumbing graph, eliminating redundancy in the input data.

\textbf{Star-Non-Sphere Dataset}

In the case of star-shaped graphs with three legs, we perform a brute-force search over the $b_i$ in the range $1 < b_i < 60$ for $i = 1, 2, 3$ and find the  corresponding $a_i$s such that the pairs $(b_i, a_i)$ are Seifert invariants for some Seifert manifold.

When generating star-shaped graphs with more than three legs, we adopt a streamlined sampling strategy. We arbitrarily select a central weight $b \in [-10, -1]$ and base weights $b_i$ in the range $1 < b_i < 50$, while fixing all numerators $a_i = 1$. Since $1 < b_i$, the pairs $(b_i, 1)$ automatically satisfy the condition for valid Seifert invariants ($0 < a_i < b_i$). Geometrically, setting $a_i=1$ corresponds to legs of length one; consequently, the resulting plumbing graphs contain no degree-2 vertices. Exploiting this strategy, we populate the \textit{Star-Non-Sphere} dataset. In cases where the determinant of the resulting plumbing matrix is $\pm 1$, this strategy produces a homology sphere, and we instead add it to the \textit{Star-Homology Sphere Dataset}. 

\textbf{Star-Homology Sphere Dataset}

Recall that the manifold corresponding to the plumbing graph in Figure \ref{fig:seifert_star_graph} is an integral homology sphere if and only if the determinant of the corresponding plumbing matrix is $\pm 1$. In the Seifert case, this condition implies that the orbifold Euler number $e$ and the base weights $b_i$ must satisfy:\begin{equation}\label{eq:generalized_brieskorn}e \prod_{i=1}^r b_i = \pm 1.\end{equation}While the general topological definition allows for either sign, the weakly negative condition required for the $\widehat{Z}$-invariants restricts us to the case where the orbifold Euler number is negative. Consequently, we enforce the condition $e \prod b_i = -1$. Crucially, with the sign fixed to $-1$, for any set of pairwise coprime integers $\{b_1, \dots, b_r\}$, the numerators $a_i$ are uniquely determined by this equation combined with the normalization conditions $b \in \mathbb{Z}$ and $0 < a_i < b_i$. We generate datasets for central vertex degree $r = 4, 5, 6$ by performing a brute-force search over possible tuples $(b_1, \dots, b_r)$ that admit such solutions. To ensure uniqueness and reduce the search space, we impose the ordering $2 \le b_1 < b_2 < \dots < b_r$. Specifically, for $r = 4$, we search for configurations where the largest weight satisfies $b_r \le 29$. For $r = 5$ and $r = 6$, we restrict the search to $b_r \le 23$. For each valid tuple found, we compute the unique corresponding $a_i$ values to reconstruct the full plumbing graph.

The case of three singular fibers admits an alternative description: \[\Sigma(b_1, b_2, b_3) = \{(x_1, x_2, x_3) \in \mathbb{C}^3 \mid x_1^{b_1}+x_2^{b_2}+x_3^{b_3} = 0 \} \cap S^5\]
and in this case the $\widehat{Z}$-invariant admits a particularly simple expression, allowing for faster computations. 

Define $m := b_1b_2b_3$ and 
\begin{align*}
\alpha_{\epsilon_1,\epsilon_2,\epsilon_3} = b_1b_2b_3 \left( 1-\sum_{i=1}^{3} (-1)^{\epsilon_i} {\frac{1}{b_i}} \right) 
, ~~ \epsilon_i\in \{0,1\}. 
\end{align*}
We then define \textit{false-theta functions} \begin{equation}\label{theta-func-1}
    \widetilde{\vartheta}_{m,a}(q) := \sum_{n=0}^\infty \psi^{(a)}_{2m}(n)\,q^{\frac{n^2}{4m}}\,, \qquad\widetilde{\vartheta}_{m,a}(q)\in q^{\frac{a^2}{4m}}\mathbb{Z}[[q]]\,,
\end{equation} where 
\begin{equation}\label{theta-func-2}
\psi^{(a)}_{2m}(n) = \begin{cases}
\  \pm1  \   \text{ if } \  n=  \pm a ~{\rm mod}~2m \\
\ \ \  0 \ \text{ otherwise}.
\end{cases}    
\end{equation}

Then for $M_3=\Sigma(b_1, b_2, b_3)$  we have
\begin{equation}\label{z-hat-brieskorn}
    \widehat{Z}(q) =  q^{\tilde\Delta}\cdot \left[q^{-\frac{\alpha_{0, 0, 0}^2}{4m}} \sum_{\substack{\epsilon_i \in \{0,1\}\\ \sum_{i=1}^3 \epsilon_i \equiv 0 (2) }}\widetilde{\vartheta}_{m,\alpha_{\epsilon_1,\epsilon_2,\epsilon_3} }(q) \right] +p(q) 
\end{equation} 
where $p(q)=0$ for $(b_1,b_2,b_3)\neq (2,3,5)$ and is a simple $q$-monomial when $(b_1,b_2,b_3)= (2,3,5)$. 
The factor $\tilde\Delta$ can be computed  
\begin{equation}\label{eq:delta-brieskorn}
  \tilde  \Delta= \frac{1}{4}\left(\sum_{i=1}^3 h_i  - 3\nu - \operatorname{Tr}(M) - \frac{b_2b_3}{b_1} - \frac{b_1b_3}{b_2} - \frac{b_1b_2}{b_3} \right) + \frac{\alpha_{0, 0, 0}^2}{4m}
\end{equation}
where $M$ is the plumbing matrix for $\Sigma(b_1, b_2, b_3)$, and 
 $h_i$ is the absolute value of the determinant of the plumbing matrix of the graph obtained by deleting the terminal vertex (a vertex with degree 1) on the $i$-th leg from the plumbing graph $\Gamma$ of $\Sigma(b_1, b_2, b_3)$, and $\nu$ is the number of vertices in $\Gamma$.  

Using this approach, we compute $\widehat{Z}$ for a subset of the Brieskorn spheres $\Sigma(b_1, b_2, b_3)$ with pairwise coprime $2 \leq b_1 < b_2 < b_3$ satisfying $b_1 \leq 41$, $b_2 \leq 99$ and $b_3 \leq 115$. In particular a subset of manifolds $\Sigma(b_1,b_2, b_3)$, will be used in the $SL_2(\mathbb{Z})$ orbit learning process discussed in \S\ref{subsec:LearningOrbit} and further specifics will be discussed therein. 

Using the methods described above to find $\Sigma(b_1, \dots, b_r)$ for $3 \leq r \leq 6$, we build the \textit{Star-Homology-Sphere} dataset. 

\subsubsection{H-shaped Datasets}

Next we turn to the datasets of H-shaped graphs, which are, as the name suggests, weighted graphs with two degree three vertices. 

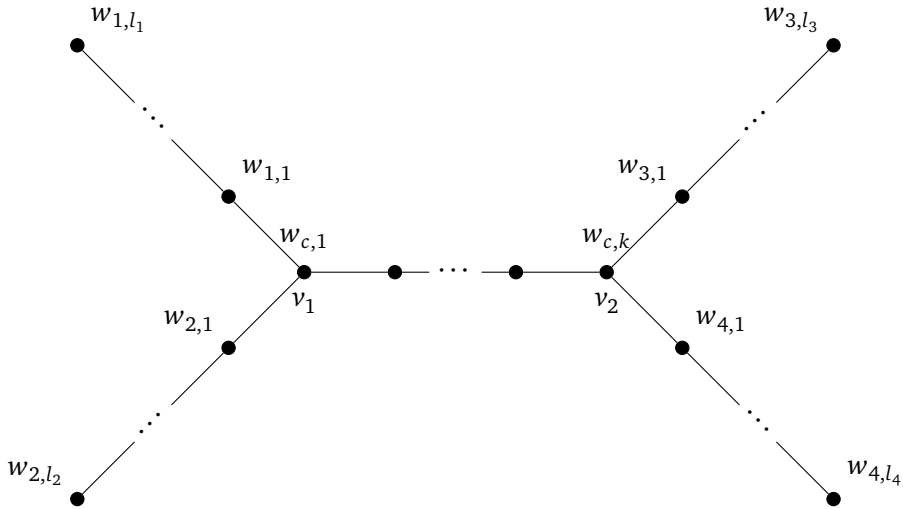
\begin{figure}[ht]
    \centering
    \begin{tikzpicture}
    % Vertices styling
    \begin{scope}[every node/.style={circle, draw, fill=black, inner sep=0pt, minimum size=5pt}]
        % Central nodes
        \node (v1) at (-2, 0) {};
        \node (v2) at (2, 0) {};
        
        % Central path nodes
        \node (u1) at (-0.8, 0) {};
        \node (uk) at (0.8, 0) {};
        
        % Arm 1 nodes (Top Left)
        \node (a11) at (-3, 1) {};
        \node (a1l) at (-5, 3) {};
        
        % Arm 2 nodes (Bottom Left)
        \node (a21) at (-3, -1) {};
        \node (a2l) at (-5, -3) {};
        
        % Arm 3 nodes (Top Right)
        \node (a31) at (3, 1) {};
        \node (a3l) at (5, 3) {};
        
        % Arm 4 nodes (Bottom Right)
        \node (a41) at (3, -1) {};
        \node (a4l) at (5, -3) {};
    \end{scope}

    % Ellipsis representing hidden nodes in the chains
    \node[inner sep=3pt] (dotsC) at (0, 0) {$\cdots$};
    \node[rotate=-45, inner sep=3pt] (dots1) at (-4, 2) {$\cdots$};
    \node[rotate=45, inner sep=3pt] (dots2) at (-4, -2) {$\cdots$};
    \node[rotate=45, inner sep=3pt] (dots3) at (4, 2) {$\cdots$};
    \node[rotate=-45, inner sep=3pt] (dots4) at (4, -2) {$\cdots$};

    % Edges mapping the chains
    \draw (v1) -- (u1); \draw (u1) -- (dotsC); \draw (dotsC) -- (uk); \draw (uk) -- (v2);
    \draw (v1) -- (a11); \draw (a11) -- (dots1); \draw (dots1) -- (a1l);
    \draw (v1) -- (a21); \draw (a21) -- (dots2); \draw (dots2) -- (a2l);
    \draw (v2) -- (a31); \draw (a31) -- (dots3); \draw (dots3) -- (a3l);
    \draw (v2) -- (a41); \draw (a41) -- (dots4); \draw (dots4) -- (a4l);

    % Vertex labels (Nodes)
    \node[below=4pt] at (v1) {$v_1$};
    \node[below=4pt] at (v2) {$v_2$};

    % Weight labels (Euler Numbers) - Minus signs removed
    \node[above=4pt] at (v1) {$w_{c,1}$};
    \node[above=4pt] at (v2) {$w_{c,k}$};
    
    \node[above=2pt] at (u1) {};
    \node[above=2pt] at (uk) {};
    
    % Shifted labels to be perpendicular to the arm lines
    \node[above right=2pt] at (a11) {$w_{1,1}$};
    \node[above right=2pt] at (a1l) {$w_{1,l_1}$};
    
    \node[above left=2pt] at (a21) {$w_{2,1}$};
    \node[above left=2pt] at (a2l) {$w_{2,l_2}$};
    
    \node[above left=2pt] at (a31) {$w_{3,1}$};
    \node[above left=2pt] at (a3l) {$w_{3,l_3}$};
    
    \node[above right=2pt] at (a41) {$w_{4,1}$};
    \node[above right=2pt] at (a4l) {$w_{4,l_4}$};
    \end{tikzpicture}
    \caption{A generic H-shaped plumbing graph.}
    \label{fig:h_shaped_expanded}
\end{figure}

\textbf{H-shaped non-sphere dataset.}
The generic H-shaped dataset is generated from integer tuples $(p_1,p_2,p_3,p_4,\mathfrak{a},\mathfrak{b},w_{c,1},w_{c,J})$ satisfying the following constraints.  We take
\[
2\leq p_1,p_3\leq 9,\quad
p_1+1\leq p_2\leq 11,\quad
p_3+1\leq p_4\leq 11,
\]
with $\gcd(p_1,p_2)=\gcd(p_3,p_4)=1$.  The parameters $\mathfrak{a}$ and $\mathfrak{b}$ are chosen in the range $1\leq \mathfrak{a},\mathfrak{b}\leq 49$  subject to $\gcd(\mathfrak{a},p_1p_2)=1$ and $\gcd(\mathfrak{b},p_3p_4)=1$, and the two central weights satisfy $-5\leq w_{c,1}, w_{c,J} \leq -2$. Finally, we impose the positivity condition $\mathfrak{a}\mathfrak{b}-p_1p_2p_3p_4>0$.

Given such a tuple, we construct an H-shaped plumbing tree of the form shown in Figure~\ref{fig:h_shaped_expanded}.  The four arms are determined by the residues
\[
\beta_1 \equiv p_2 \mathfrak{a} \pmod {p_1},\quad
\beta_2 \equiv p_1 \mathfrak{a} \pmod {p_2},\quad
\beta_3 \equiv p_4 \mathfrak{b} \pmod {p_3},\quad
\beta_4 \equiv p_3 \mathfrak{b} \pmod {p_4}.
\]
For each $i=1,\ldots,4$, the weights along the $i^{\rm th}$ arm are obtained from the negative continued-fraction expansion of $p_i/\beta_i$.  Thus, if
\[
\frac{p_i}{\beta_i}=[-w_{i,1},\ldots,-w_{i,l_i}],
\]
then the corresponding arm carries vertex weights $w_{i,1},\ldots,w_{i,l_i}$.

The two high-valence vertices carry weights $w_{c,1}$ and $w_{c,J}$.  Between them we insert a linear chain of $k-2$ vertices, each of weight $-2$, where
\[
J=2+\max\left\{1,\left\lfloor \log\bigl(\mathfrak{a}\mathfrak{b}-p_1p_2p_3p_4+1\bigr)\right\rfloor\right\}.
\]
This produces an H-shaped plumbing tree with two Seifert-type ends joined by a variable-length central chain. We only retain plumbing trees which are in minimal form (under Kirby--Neumann moves) and whose plumbing matrices are strictly negative-definite and satisfy absolute determinant $<1000$.

\textbf{H-shaped Homology Sphere Dataset from Splice Diagrams}

As mentioned before, prior work has performed PCA analysis on a small dataset (20 instances)  of $\widehat Z$-invariants of homology spheres with H-shaped plumbing graphs \cite{PhysRevD.110.046016}. 
This limitation arose from the difficulty of generating random integer matrices with determinant $\pm 1$, a necessary condition for the plumbed manifold to be an integral homology sphere. In this work, we overcome this bottleneck by leveraging an algorithm based on splice diagrams. This approach allows us to efficiently generate a large, diverse dataset of H-shaped homology spheres for ML analysis.

A \textit{splice diagram} is a finite tree graph with finitely many vertices. 
It contains only vertices of degree 1 (referred to as \textit{leaves}) or degree at least 3 (referred to as \textit{nodes}).
Furthermore, to each edge departing a node we associate an integer weight. In what follows, we use $(v,u)$ to denote the edge from vertex $v$ to vertex $u$, and $w_{vu}$ the corresponding weight  in case $v$ is a node. 

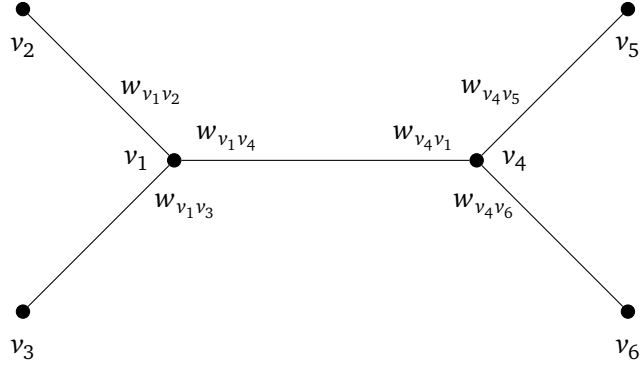
\begin{figure}[ht]
    \centering
    \begin{tikzpicture}
    \begin{scope}[every node/.style={circle, draw, fill=black, inner sep = 0pt, minimum size=5pt}]
        \node (1) at (0, 0) {};
        \node (2) at (4, 0) {};
        \node (3) at (-2, 2) {};
        \node (4) at (-2, -2) {};
        \node (5) at (6, 2) {};
        \node (6) at (6, -2) {};
    \end{scope}
    \draw (1)--(2);
    \draw (1)--(3);
    \draw (1)--(4);
    \draw (2)--(5);
    \draw (2)--(6);
% vertex labels
\node (1A) at (-0.5, 0) {$v_1$};
\node (2A) at (4.5, 0) {$v_4$};
\node (3A) at (-2, 1.5) {$v_2$};
\node (4A) at (-2, -2.5) {$v_3$};
\node (5A) at (6, 1.5) {$v_5$};
\node (6A) at (6, -2.5) {$v_6$};
% weight labels
\node (1B) at (0.7, 0.3) [text=black]{$w_{v_1v_4}$};
\node (2B) at (-0.3, 0.9) [text=black]{$w_{v_1v_2}$};
\node (3B) at (0.15, -0.6) [text=black]{$w_{v_1v_3}$};
\node (4B) at (3.3, 0.3) [text=black]{$w_{v_4v_1}$};
\node (5B) at (4.2, 0.9) [text=black]{$w_{v_4v_5}$};
\node (6B) at (4.1, -0.6) [text=black]{$w_{v_4v_6}$};
\end{tikzpicture}
    \caption{A splice diagram with two nodes and four leaves.}    \label{fig:dual_node_splicing_diagram}
\end{figure}

For any edge connecting two nodes, we define the edge determinant as the product of the two weights on that edge minus the product of the weights on all other edges incident to those nodes. For the edge $(v_1, v_4)$ in Figure \ref{fig:dual_node_splicing_diagram}, the edge determinant is given by:\[{\rm Det}_{(v_1, v_4)} = w_{v_1v_4}w_{v_4v_1} - \left( \prod_{u \in N(v_1) \setminus \{v_4\}} w_{v_1u} \right) \left( \prod_{z \in N(v_4) \setminus \{v_1\}} w_{v_4z} \right)=w_{v_1 v_4}w_{v_4 v_1} - w_{v_4 v_5}w_{v_4 v_6}w_{v_1 v_2}w_{v_1 v_3} ,\]where $N(v)$ denotes the set of neighbors of $v$.

The utility of splice diagrams lies in the following result: if a splice diagram satisfies (a) the weights around any node are positive and pairwise coprime, (b) the weight on any edge ending on a leaf is $>1$, and (c) all edge determinants are positive, then there exists an explicit algorithm to convert the splice diagram into a plumbing graph for a negative-definite integral homology sphere. This algorithm, detailed in \cite[Section 9]{neumann2005complex}, is reviewed in Appendix \ref{appendix:splicing_diagrams}.

%\textcolor{red}{MC: suggested version\\
The H-shaped homology spheres in Tab.~\ref{tab:dataset-binary} were generated as follows. Let $\mathcal{P} = \{P_\alpha\}_{\alpha=1}^\infty $ denote the sequence of prime numbers $\{2, 3, 5, \dots \}$. We initialize an H-shaped splice diagram as in Figure~\ref{fig:dual_node_splicing_diagram} and assign weights in two steps. First, for the central edge connecting the two nodes, we set $w_{v_1v_4} = P_\alpha$ and $w_{v_4v_1} = P_\beta$, where the prime indices $\alpha$ and $\beta$ are sampled uniformly from integers in $[11, 51]$ and $[11, 31]$ respectively, subject to the constraint $\alpha \neq \beta$. Next, for the external edges connecting a node $v$ to a leaf $u$, we sample the weights $w_{vu}$ uniformly from the set $\{2, 3, 5, 7, 11, 13\}$. To satisfy the conditions for the resulting manifold to be a homology sphere, we compute the edge determinant for each generated configuration and retain only those instances where it remains strictly positive. Finally, we apply the conversion algorithm to transform these valid splice diagrams into standard H-shaped plumbing graphs.
%}

\subsection{Data Representations}
To ensure the robustness of our interpretation analysis, in our experiments we examine the impact of varying input formats on the learned features. In this subsection we introduce the different data representations used in this paper. 

In this paper, we fix a cutoff, given by $N=10,000$, in the $q$-series expansion: we write an infinite $q$-series as
\[ q^\Delta\left( \sum_{n=0}^{N-1} c_n q^n + O(q^{10,000})\right). \]
The data is normalized by a multiplication of $\mathbb C$-number such that the first coefficient is $c_0=1$.  

This fixed cutoff imposes a potentially significant limitation. For sufficiently sparse $q$-series, particularly those corresponding to star-shaped graphs, the first gap $\gamma_0$ may exceed the cutoff $N$. In extreme cases, the truncation discards all non-trivial coefficients, resulting in a degenerate input vector $[1, 0, \dots, 0]$. Without auxiliary features such as $\Delta$ to resolve these ambiguities, the resulting data collisions may confound the network, most notably in the binary classification of homology spheres. We discuss this and other potential issues posed by this data preprocessing in Section~\ref{sec:topExp}.

\subsubsection{Full Sequence Representation}

In the full sequence representation, we encode the input data as arrays of shape $(N,)$, containing the coefficients $(c_0, \dots, c_{N-1})$. The primary advantage of this representation is its simplicity: a single array of fixed shape fully captures the truncated data, where the array indices map directly to the $q$-series exponents.

\subsubsection{Value-Index Pair Representation}
\label{subsubsec:value-index-pair}

In this representation, we compress the data by storing only the nonzero coefficients. Specifically, the input is defined as a sequence of tuples $(c_i, n_i)$, where each tuple represents a non-vanishing term $c_i q^{n_i}$ (with $c_i \neq 0$). To ensure a uniform input dimension across the dataset, we identify the maximum sequence length and pad all shorter sequences to the right with zeros, resulting in a fixed dataspace dimension $N=10{,}474$. This approach allows us to process inputs of varying sparsity.

\subsubsection{Latent Space Representation}
\label{subsubsec:Latent Space Representation}

As an alternative to the direct sequence inputs described above, we investigate the impact of data compression through a learned, structured latent space. Specifically, we employ a network trained using contrastive learning to map the high-dimensional input space onto a low-dimensional (typically two-dimensional) manifold. This learned embedding allows us to examine whether the topological features driving the classification tasks can be condensed into a compact geometry without loss of interpretability. See Appendix \ref{appendix:contrastive_learning} for the details of the contrastive learning techniques used.

\section{Topology Experiments}\label{sec:topExp}

As illustrated in the data analysis pipeline schematic (Figure~\ref{fig:ml-flow-chart}), the goal of our experiments is to extract interpretable topological features from the $\widehat{Z}$-invariant $q$-series through a sequence of modular stages. Following the generation of raw data in \S 3, the \emph{preprocessing} stage converts these series into one of three input representations: full sequence, value-index pair, or latent space representations. 

Depending on the choice of representation, the pipeline proceeds along two distinct paths:
\begin{itemize}
    \item \textbf{Direct Classification (\S\ref{sec:sparseRep} \& \S\ref{sec:denseRep}):} The preprocessed full sequence or value-index pair vectors are fed directly into the classifier network.
    \item \textbf{Latent Space Classification (\S\ref{sec:latentRep}):} The data first passes through a Latent Space Learning module to generate low-dimensional embeddings, which then serve as inputs for the classifier.
\end{itemize}

The pipeline concludes with a dedicated \emph{interpretability} phase. Here, we train a regression network to approximate the classifier's decision values (logits). Crucially, this regressor does not see the raw $q$-series; instead, its inputs are restricted to a set of candidate features we selected based on their mathematical plausibility (see Glossary). By determining which of these human-understandable candidates are necessary to reconstruct the decision boundary, we isolate the specific topological information driving the classifier's performance.

To validate the insights gained from this analysis, we include a set of control experiments in \S\ref{subsec:classification-using-features}. In one experiment, we bypass the $q$-series inputs entirely and train classifiers directly on the derived topological features, allowing us to verify whether the explicitly learned decision boundaries align with those reconstructed by the regression pipeline. In another experiment, described in \S\ref{subsubsec:discFullSeq}, we utilize a teacher--student (margin-reconstruction) protocol in which the teacher is trained on $q$-series data and student classifiers, trained
only on the derived topological features, are asked to reproduce its decision boundary.

At the end of the pipeline, the \emph{data analysis} stage assesses feature importance using three complementary metrics: Feature Saliency (FS), Permutation Importance (PI), and Shapley Additive Explanations (SHAP). The patterns of consensus and discrepancies observed among these metrics constitute the primary findings of this section.

\subsection{Task Description}\label{sec:classTask}

In this section, we define the supervised learning tasks probing  the global topological information encoded in the $\widehat{Z}$-invariants: distinguishing homology spheres from non-spheres (HOM), and identifying the underlying plumbing graph topology (TOP). 
We selected these two classification tasks because they represent topological questions that are both theoretically significant and well-understood. 

\paragraph{Homology Classification}

The \textbf{Homology Classification (HOM)} task evaluates the network's capacity to distinguish integral homology spheres from non-spherical manifolds based on their truncated $q$-series. The ground truth labels separate manifolds satisfying $|\det M|=1$ (homology spheres) from those with $|\det M| \neq 1$, where $M$ denotes the plumbing matrix of the plumbed manifold. This distinction is significant because it probes the network's sensitivity to a global property that is often considered obfuscated in the coefficients of the $q$-series invariant. To disentangle the influence of the plumbing graph structure from the homological properties, we first restrict the experiment to datasets with fixed graph topologies: \textbf{HOM-Star} (star-shaped graphs) and \textbf{HOM-H} (H-shaped graphs). We then assess the robustness of the learned features by performing the classification on the full, heterogeneous dataset (\textbf{HOM-All}).

\paragraph{Graph Topology Classification}
The \textbf{Graph Topology Classification (TOP)} task investigates whether the truncated $\widehat{Z}$-invariant retains sufficient information to recover the underlying plumbing graph structure. Specifically, we train the network to distinguish between Star-shaped and H-shaped graphs, testing its sensitivity to the complexity of the plumbing graph. To determine if the network's learning strategy depends on the homological properties of the manifold, we partition the experiments into three distinct cases. The first two experiments restrict the dataset based on the homology class label: \textbf{TOP-Sph} (restricted to homology spheres) and \textbf{TOP-Non} (restricted to non-spheres). Finally, we perform the same classification experiment on the full, heterogeneous dataset (\textbf{TOP-All}). 

Each experiment is conducted across all three input representations: full sequence, value-index pair, and latent space.

\paragraph{Multi-class Classification}
To probe the non-linear geometry of the latent space, we combine the graph topology and homology labels into a single four-class classification problem (HOM-Star, HOM-H, TOP-Sph, TOP-H). Unlike the binary tasks, which admitted linear solutions, distinguishing these four clusters requires a non-linear classifier. We therefore employ a RBF-Kernel Support Vector Machine (SVM) trained on the latent embeddings derived from the value-index pair representation, which yielded the highest performance.

To interpret these non-linear boundaries, we train a Random Forest (RF) regressor to reconstruct the SVM decision values using our standard set of topological features. We assess the fidelity of this reconstruction using the composite score defined in \sn{performance-metrics}.

\paragraph{Experimental protocol and reporting.}
All results in this section are reported using the multi-seed robustness protocol described in Appendix~\ref{app:interpretability_robustness}.  For each experiment we aggregate over classifier seeds, regressor seeds, and interpretability seeds, and report mean values with standard deviations.  Classifier performance is measured by balanced accuracy score (BAS).  The feature-to-logit reconstruction estimator is evaluated by both $R^2$ and the Pearson correlation coefficient (PCC) on the train and validation splits: $R^2$ measures calibrated reconstruction of the scaled logits, while the PCC measures alignment with the learned logit geometry up to affine rescaling. Feature saliency, permutation importance, and SHAP are summarized by average feature rank across the interpretability ensemble.

The supervised classifier is trained on one of three truncated $\widehat{Z}$ representations: the full coefficient sequence, the serialized value-index representation, or a low-dimensional contrastive latent representation.  In the full-sequence and value-index cases, the classifier is a fully connected \texttt{ReLU} network with one hidden layer of width $64$, trained with cross-entropy loss using Adam with learning rate $10^{-3}$, batch size $128$, and early stopping on balanced accuracy.  In the contrastive-learning case, the high-dimensional input is first mapped to a five-dimensional latent space by a \texttt{ReLU} encoder trained with semi-hard triplet loss; the downstream classifier is then trained on this latent representation using the same validation criterion, with hidden layer width $32$.

For interpretability, we extract the pre-softmax logits of the trained classifier and use them as a dataset-level coordinate system for the learned decision geometry.  A second \texttt{ReLU} network maps ten abstract features derived from the plumbing matrix and the $q$-series to the scaled classifier logits.  This reconstruction estimator has architecture $(d,w)=(4,128)$, where $d$ is the number of affine layers and $w$ is the hidden-layer width.  It is trained with mean-squared error using Adam with learning rate $10^{-4}$, batch size $64$, and an epoch cap of $3000$.  We use the validation-selected checkpoint for reporting and interpretation.  The train metrics measure the signal strength of the ten-feature reconstruction of the learned logit geometry, while the validation metrics measure the stability of this reconstruction under the regressor split.

\begin{table}[ht!]
\centering
\small
\begin{tabular}{c|c|c|c|c}
\toprule
\textbf{Stage} &
\textbf{Input} &
\textbf{Architecture} &
\textbf{Batch} &
\textbf{Learning rate} \\
\midrule
Classifier &
Full Sequence &
$(2,64)$ &
$128$ &
$10^{-3}$ \\
\hline
Classifier &
Value-Index &
$(2,64)$ &
$128$ &
$10^{-3}$ \\
\hline
Contrastive encoder &
Full Sequence &
$(4,128)$ &
$128$ &
$10^{-3}$ \\
\hline
CL classifier &
Latent Space &
$(2,32)$ &
$128$ &
$10^{-3}$ \\
\hline
Regressor &
Abstract features &
$(4,128)$ &
$64$ &
$10^{-4}$ \\
\bottomrule
\end{tabular}
\caption{
Summary of the network architectures and hyperparameters used in the seed-robust topology and homology experiments.  The architecture column records $(d,w)$, where $d$ is the number of affine layers and $w$ is the hidden-layer width.  Classifiers are trained with cross-entropy loss and selected by balanced accuracy.  The contrastive encoder is trained with semi-hard triplet loss, and the regressor is trained with mean-squared error to reconstruct the scaled classifier logits from the ten abstract features.
}
\label{tab:summary-1}
\end{table}

\subsection{Full Sequence Representation}\label{sec:sparseRep}

In this subsection, we evaluate the classification tasks defined in \S\ref{sec:classTask} using the full sequence representation. We present results for both the stratified experiments (e.g., \textbf{HOM-Star}, \textbf{TOP-Sph}) and the combined datasets (\textbf{HOM-All}, \textbf{TOP-All}).

%%%%%%%%%%%%%%%%%%%%%%%%%%%%%%%%%%%%%%%%%%%%%%%%%%%%%%%%%%%%%%%%%%%%%%%%%%%%%%%%
% Multi-seed table for full-sequence
%%%%%%%%%%%%%%%%%%%%%%%%%%%%%%%%%%%%%%%%%%%%%%%%%%%%%%%%%%%%%%%%%%%%%%%%%%%%%%%%
\begin{table}[ht!]
\centering
\small
\setlength{\tabcolsep}{3pt}
\resizebox{\textwidth}{!}{%
\begin{tabular}{c|c|c|c|ccc}
\toprule
\textbf{Experiment} &
\textbf{BAS} &
$\mathbf{R^2}$ &
\textbf{PCC} &
\textbf{FS} &
\textbf{PI} &
\textbf{SHAP} \\
\midrule
HOM-Star & $0.928\,\pm\,0.003$ & \shortstack{$\mathrm{tr}: 0.430\,\pm\,0.253$\\$\mathrm{val}: 0.193\,\pm\,0.290$} & \shortstack{$\mathrm{tr}: 0.885\,\pm\,0.123$\\$\mathrm{val}: 0.565\,\pm\,0.476$} & \rankfeat{\varepsilon_{\rm min}}{1.8}, \rankfeat{\sigma_{\gamma}}{4.1}, \rankfeat{\Delta}{4.1} & \rankfeat{\varepsilon_{\rm min}}{2.3}, \rankfeat{\sigma_{\varepsilon}}{3.2}, \rankfeat{\varepsilon_{\rm max}}{4.3} & \rankfeat{\varepsilon_{\rm min}}{1.4}, \rankfeat{\sigma_{\varepsilon}}{3.1}, \rankfeat{\varepsilon_{\rm max}}{4.4} \\
\hline
HOM-H & $0.923\,\pm\,0.009$ & \shortstack{$\mathrm{tr}: 0.769\,\pm\,0.023$\\$\mathrm{val}: 0.396\,\pm\,0.039$} & \shortstack{$\mathrm{tr}: 0.879\,\pm\,0.014$\\$\mathrm{val}: 0.742\,\pm\,0.013$} & \rankfeat{\mu_{\gamma}}{1.3}, \rankfeat{\sigma_{\gamma}}{1.7}, \rankfeat{\varepsilon_{\rm min}}{3.5} & \rankfeat{\sigma_{\gamma}}{1.7}, \rankfeat{\mu_{\gamma}}{3.4}, \rankfeat{\varepsilon_{\rm min}}{3.8} & \rankfeat{\sigma_{\gamma}}{2.0}, \rankfeat{\varepsilon_{\rm min}}{2.2}, \rankfeat{\mu_{\gamma}}{2.5} \\
\hline
HOM-All & $0.952\,\pm\,0.003$ & \shortstack{$\mathrm{tr}: 0.698\,\pm\,0.035$\\$\mathrm{val}: 0.572\,\pm\,0.029$} & \shortstack{$\mathrm{tr}: 0.852\,\pm\,0.025$\\$\mathrm{val}: 0.798\,\pm\,0.026$} & \rankfeat{\det}{1.5}, \rankfeat{\varepsilon_{\rm min}}{2.1}, \rankfeat{c_{\rm max}}{3.6} & \rankfeat{\sigma_{\varepsilon}}{2.5}, \rankfeat{\varepsilon_{\rm max}}{2.8}, \rankfeat{\varepsilon_{\rm min}}{3.8} & \rankfeat{\varepsilon_{\rm min}}{1.4}, \rankfeat{\det}{2.0}, \rankfeat{\sigma_{\gamma}}{3.1} \\
\hline
TOP-Sph & $0.964\,\pm\,0.004$ & \shortstack{$\mathrm{tr}: 0.837\,\pm\,0.012$\\$\mathrm{val}: 0.725\,\pm\,0.017$} & \shortstack{$\mathrm{tr}: 0.916\,\pm\,0.007$\\$\mathrm{val}: 0.863\,\pm\,0.010$} & \rankfeat{\mu_{\gamma}}{1.0}, \rankfeat{c_{\rm max}}{2.5}, \rankfeat{\sigma_{\gamma}}{3.6} & \rankfeat{\mu_{\gamma}}{1.0}, \rankfeat{\sigma_{\gamma}}{2.0}, \rankfeat{\Delta}{3.4} & \rankfeat{\mu_{\gamma}}{1.0}, \rankfeat{\sigma_{\gamma}}{2.0}, \rankfeat{\det}{3.6} \\
\hline
TOP-Non & $0.971\,\pm\,0.009$ & \shortstack{$\mathrm{tr}: 0.759\,\pm\,0.019$\\$\mathrm{val}: 0.600\,\pm\,0.029$} & \shortstack{$\mathrm{tr}: 0.876\,\pm\,0.010$\\$\mathrm{val}: 0.819\,\pm\,0.012$} & \rankfeat{\det}{1.0}, \rankfeat{\varepsilon_{\rm min}}{2.9}, \rankfeat{\sigma_{\gamma}}{3.5} & \rankfeat{\varepsilon_{\rm min}}{2.1}, \rankfeat{\det}{2.8}, \rankfeat{\mu_{\varepsilon}}{3.7} & \rankfeat{\det}{1.0}, \rankfeat{\varepsilon_{\rm min}}{2.4}, \rankfeat{\sigma_{\gamma}}{3.8} \\
\hline
TOP-All & $0.978\,\pm\,0.004$ & \shortstack{$\mathrm{tr}: 0.761\,\pm\,0.018$\\$\mathrm{val}: 0.616\,\pm\,0.013$} & \shortstack{$\mathrm{tr}: 0.876\,\pm\,0.010$\\$\mathrm{val}: 0.816\,\pm\,0.010$} & \rankfeat{\det}{1.0}, \rankfeat{\varepsilon_{\rm min}}{2.2}, \rankfeat{\sigma_{\gamma}}{3.0} & \rankfeat{\varepsilon_{\rm min}}{2.6}, \rankfeat{\sigma_{\gamma}}{3.0}, \rankfeat{\sigma_{\varepsilon}}{3.3} & \rankfeat{\sigma_{\gamma}}{1.6}, \rankfeat{\varepsilon_{\rm min}}{2.1}, \rankfeat{\det}{2.3} \\
\bottomrule
\end{tabular}%
}
\caption{
Multi-seed full-sequence results for the topology and homology experiments.  BAS is reported as mean $\pm$ standard deviation over classifier seeds.  The $R^2$ and Pearson correlation (PCC) columns report train and validation reconstruction quality of the feature-to-logit regressor, stacked within each cell and averaged over the classifier--regressor seed ensemble.  The final columns report the three lowest mean feature ranks for feature saliency (FS), permutation importance (PI), and SHAP, where the subscript gives the average rank over the normalized interpretability ensemble.
}
\label{tab:summary-2}
\end{table}

\subsubsection{Homology Classification}
\label{subsubsec:FS-homology-classification}

{\bf HOM-Star}

The \textbf{HOM-Star} dataset comprises 44,934 samples, consisting of 13,228 homology spheres and 31,706 non-spheres. Given the approximate 2.4:1 class imbalance, to remove any population effects we compared models trained with a resampled class-balanced dataset against those trained on the raw, imbalanced distribution. We observed no significant performance deviation; consequently, we report results derived from the full, imbalanced dataset.

Naively, this classification task appears non-trivial because the $q$-series for homology spheres and non-spheres often share identical asymptotic growth. This is particularly apparent for weakly negative plumbed manifolds with three singular fibers in our dataset, as all of them have $q$-series invariants with coefficients $c_0, c_1, \dots$ that are either $-1$, $+1$, or $0$. While one might expect the network to rely on explicit features like coefficient gaps to resolve this ambiguity, our subsequent analysis reveals that it exploits more abstract derived features.

We trained the classifier using the architecture specified in the first row of Tab.~\ref{tab:summary-1}, with a $90/10$ train-test split. The model achieved a mean Balanced Accuracy Score (BAS) of $0.928\,\pm\,0.003$, with the  confusion matrix shown in Figure~\ref{fig:FS_confusion}. Figure~\ref{fig:FS_logits} visualizes the classifier's output logits for a single representative seed; these values serve as the target variables for the regression-based interpretability analysis discussed below.

\paragraph{Discussion of Features}
Mathematically, integral homology spheres are defined by the condition $|\det M|=1$. However, the non-vanishing variance observed  along the principal axis of the classifier logits suggests that the network does not rely on this discrete invariant alone; if it did, the homology sphere samples would collapse to a single point rather than spanning a segment. To isolate the actual decision drivers, we trained a regression network on the full set of 10 interpretable features, achieving a mean fit of 
\[(R^2_{\rm tr},R^2_{\rm val},{\rm PCC}_{\rm tr}, {\rm PCC}_{\rm val})==(0.430\pm0.253,\,0.193\pm0.290,\,0.885\pm0.123,\,0.565\pm0.476).
\]
These performance metrics suggest that the classifier is robust, but the ten-feature logit reconstruction is weak and high variance. The feature rankings obtained from the regressor network alone are thus only suggestive, but will be corroborated by further experiments.

The interpretability analysis of this regressor reveals a divergence between metrics. Beyond the minimum magnitude eigenvalue $\varepsilon_{\rm min}$, Feature Saliency (FS) prioritizes spectral information about the gaps between non-trivial terms in the $q$-series, i.e. their standard deviation $\sigma_\gamma$, and the unseen leading power $\Delta$, while the statistical metrics (PI and SHAP) consistently identify spectral properties---specifically  $\varepsilon_{\rm min}$, the standard deviation of the eigenvalues $\sigma_\varepsilon$, and maximum magnitude eigenvalue $\varepsilon_{\rm max}$---as the dominant predictors.

The prominence of $\varepsilon_{\rm min}$ is notable
and allows for a clear explanation: for the plumbing matrices in our dataset, we observe that the condition $|\det M|=1$ is typically achieved by a spectrum where one eigenvalue is significantly smaller than the others.  This empirical observation can be placed on rigorous footing. By combining Cramer's rule for the diagonal entries of $M^{-1}$ with the Rayleigh quotient bound on the spectral radius, one obtains a strict upper bound on $|\varepsilon_{\rm min}|$ in terms of the Seifert data (see Appendix~\ref{app:spectral_bound} for the full derivation):
\begin{lem}\label{prop:spectral_bound}
Let $M$ be the plumbing matrix of a three-legged star-shaped graph associated to a Seifert manifold with singular fiber orders $b_1, b_2, b_3$. Then
\begin{equation}\label{eq:spectral_bound_main}
    |\varepsilon_{\rm min}| \le \frac{|\det M|}{b_1 b_2 b_3}.
\end{equation}
\end{lem}
\noindent For Brieskorn spheres, $|\det M| = 1$ and the $b_i$ are pairwise coprime integers $\ge 2$, so $b_1 b_2 b_3 \ge 30$, forcing $|\varepsilon_{\rm min}| \le 1/30$---typically much smaller for generic Seifert homology spheres. The bound thus provides a structural explanation for the network's reliance on $\varepsilon_{\rm min}$: rather than learning the determinant directly, the network exploits a spectral consequence of the homology sphere condition that is guaranteed by the underlying geometry. 

The caveat is that the relatively poor average performance of the regressor networks places any definitive claim that the full-sequence networks robustly learn $\varepsilon_{\rm min}$ as a meaningful surrogate discriminating feature on weak footing. We will see in further experiments with much stronger regressor signals that $\varepsilon_{\rm min}$ is frequently identified as a leading regressor surrogate to explain the decision boundary separating homology spheres from non-spheres.

The secondary feature, $\sigma_\varepsilon$, captures the spread of the spectrum. In particular, neither $\varepsilon_{\rm min}$ nor $\sigma_\varepsilon$ directly captures the determinant. This indicates that the network solves the classification task by augmenting the primary proxy ($\varepsilon_{\rm min}$) with independent spectral information ($\sigma_\varepsilon$), rather than using the determinant or relying solely on a direct proxy for the determinant.

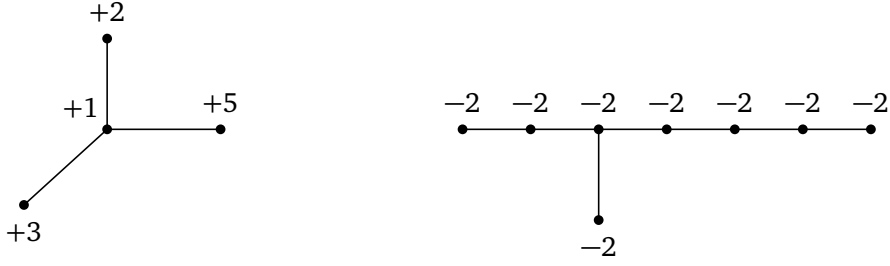
\begin{figure}[ht]
    \centering
 \begin{tikzpicture}[
  vertex/.style={circle,fill,inner sep=1.3pt},
  edge/.style={line width=0.6pt}
]

% --- Left: Σ(2,3,5) plumbing with a trivalent central node (weight +1; positive-weight presentation)
\begin{scope}[xshift=-3.8cm]
  \coordinate (L0) at (0,0);          % central (valence 3)
  \coordinate (Lr) at (1.5,0);        % right arm leaf
  \coordinate (Lu) at (0,1.2);        % top arm leaf
  \coordinate (Ll) at (-1.1,-1.0);    % bottom-left arm leaf

  \draw[edge] (L0)--(Lr) (L0)--(Lu) (L0)--(Ll);

  \node[vertex] at (L0) {};
  \node[vertex] at (Lr) {};
  \node[vertex] at (Lu) {};
  \node[vertex] at (Ll) {};

  % weights
  \node at ($(L0)-(0.35,-0.28)$) {$+1$};
  \node at ($(Lr)+(0,0.35)$) {$+5$};
  \node at ($(Lu)+(0,0.35)$) {$+2$};
  \node at ($(Ll)+(0,-0.35)$) {$+3$};
\end{scope}

% --- Right: E8 plumbing (all weights -2) with a single trivalent node
\begin{scope}
  % horizontal chain of 7 nodes
  \foreach \i in {1,...,7} {
    \coordinate (R\i) at (0.9*\i,0);
  }
  % vertical branch from the 3rd node (the trivalent vertex)
  \coordinate (Rv) at (0.9*3,-1.2);

  \draw[edge] (R1)--(R2)--(R3)--(R4)--(R5)--(R6)--(R7) (R3)--(Rv);

  % nodes and weights
  \foreach \i in {1,...,7} {
    \node[vertex] at (R\i) {};
    \node at ($(R\i)+(0,0.35)$) {$-2$};
  }
  \node[vertex] at (Rv) {};
  \node at ($(Rv)+(0,-0.35)$) {$-2$};
\end{scope}

\end{tikzpicture}
\caption{Two plumbings of $\Sigma(2,3,5)$ equivalent under Kirby--Neumann moves.}
    \label{fig:plumbing-graphs-235}
\end{figure}

This reliance on eigenvalue statistics highlights a tension between machine learning efficacy and topological invariance. Unlike the determinant, individual eigenvalues are not invariant under Kirby--Neumann moves; as exemplified in Figure~\ref{fig:plumbing-graphs-235}, equivalent plumbing graphs can yield distinct spectra, while still satisfying the above lemma. The network effectively ignores the invariant determinant in favor of spectral data, identifying a statistical shortcut that is valid for most, but not all, of the dataset. This suggests the model is exploiting distribution-specific correlations rather than learning the exact topological definition, resulting in high but imperfect classification performance and failure to learn the crucial invariance property. 

In particular, it is important to note that the spectral bound of Lemma~\ref{prop:spectral_bound} is one-directional: it guarantees that Brieskorn spheres \emph{must} have a small $|\varepsilon_{\rm min}|$, but does not exclude the possibility that certain non-spheres also possess a small minimum-magnitude eigenvalue. This asymmetry---where the network mistakes a necessary condition for being a sphere for a sufficient one---provides a structural explanation for the observed generalization error of the classifier: the network's reliance on $\varepsilon_{\rm min}$ as a proxy is mathematically well-motivated but inherently incomplete.

\textbf{HOM-H.}

The dataset for the HOM-H experiment consists of plumbed manifolds defined by \(H\)-shaped plumbing graphs.  It contains \(18{,}669\) samples, comprising \(7{,}952\) homology spheres and \(10{,}717\) non-spheres.

We trained the classifier detailed in Tab.~\ref{tab:summary-1} using a \(90/10\) train-test split, achieving a mean BAS of \(0.923\pm0.009\) on the test set.  The resulting confusion matrix and representative single-seed logit distribution are visualized in Figs.~\ref{fig:FS_confusion} and \ref{fig:FS_logits}, respectively.  To interpret these decisions, we trained a regression network to reconstruct the classifier logits from the full set of ten derived features.  The regressor achieves a substantially more stable fit than in HOM-Star, with
\[
(R^2_{\rm tr},R^2_{\rm val},{\rm PCC}_{\rm tr}, {\rm PCC}_{\rm val})=(0.769\pm0.023,0.396\pm0.039,0.879\pm0.014,0.742\pm0.013).
\]
Thus, while the abstract features do not fully reconstruct the calibrated two-logit geometry, they preserve a stable monotone relation with the classifier's decision function.

Although the classification objective is the same as in HOM-Star---distinguishing homology spheres from non-spheres---the \(H\)-graph sector changes the feature geometry of the problem.  In the star sector, the leading recurring feature is \(\varepsilon_{\min}\), suggesting a low-end spectral proxy for the determinant condition.  In the \(H\)-graph sector, the attribution pattern shifts toward gap statistics.  The full-sequence interpretability metrics in Tab.~\ref{tab:summary-2} consistently rank \(\mu_\gamma\), \(\sigma_\gamma\), and \(\varepsilon_{\min}\) among the leading features.  Feature saliency ranks \(\mu_\gamma\) and \(\sigma_\gamma\) first and second, with average ranks \(1.3\) and \(1.7\), followed by \(\varepsilon_{\min}\) at rank \(3.5\).  Permutation importance places \(\sigma_\gamma\) first and retains \(\mu_\gamma\) and \(\varepsilon_{\min}\) among the top three.  SHAP similarly keeps the same trio at the top, with \(\sigma_\gamma\), \(\varepsilon_{\min}\), and \(\mu_\gamma\) all closely ranked.

This indicates that the HOM-H classifier does not rely on a single low-end eigenvalue proxy.  Instead, it recruits gap statistics to resolve examples whose coarse spectral behavior is ambiguous.  The \(H\)-graph dataset contains non-spherical examples whose low-end spectrum can partially resemble that of homology spheres, so \(\varepsilon_{\min}\) alone is not a sufficiently stable discriminator.  The gap statistics provide complementary information about the sparsity profile of the \(q\)-series, allowing the classifier to separate examples whose eigenvalue summaries are similar but whose coefficient support differs.

Geometrically, this suggests that in the \(H\)-graph sector the homology-sphere decision is mediated not only by the determinant constraint, but also by how the quadratic form \(\ell^T M^{-1}\ell\) organizes the support of the \(q\)-series.  Changes in the spectrum of \(M\), especially near small eigenvalues, affect the directions along which this quadratic form grows slowly.  The resulting changes in coefficient spacing and support density are summarized by the gap features \(\mu_\gamma\) and \(\sigma_\gamma\).  The interpretability results therefore point to a mixed spectral--sparsity mechanism: \(\varepsilon_{\min}\) remains relevant as a low-end spectral feature, but the more stable HOM-H signal is carried by the gap profile of the \(q\)-series.

\textbf{HOM-All.}

Finally, we generalize the homology-sphere task by training on the combined heterogeneous dataset, asking the network to identify homology spheres without first conditioning on the plumbing graph topology.  The classifier achieves a mean BAS of \(0.952\pm0.003\), improving over both HOM-Star and HOM-H individually.  A representative single-seed logit distribution for the combined dataset is shown in Fig.~\ref{fig:FS_logits}.  This performance is notable because the combined dataset contains graph-topology heterogeneity: the classifier must learn a homology-sphere decision rule that is stable across both star-shaped and \(H\)-shaped plumbing graphs.

The logit-regression analysis indicates that this heterogeneous task is substantially better captured by the abstract feature dictionary than either HOM-Star or HOM-H alone. The regressor achieves 
\[
(R^2_{\rm tr},R^2_{\rm val},{\rm PCC}_{\rm tr}, {\rm PCC}_{\rm val})=(0.698\pm0.035,0.572\pm0.029, 0.852\pm0.025,0.798\pm0.026).
\]
Thus, in the pooled homology task, the derived spectral and support features recover a meaningful fraction of the classifier's two-logit geometry, not merely its hard labels.

The feature rankings in Tab.~\ref{tab:summary-2} show that the combined task does not reduce to a single spectral proxy.  Feature saliency ranks \(\det\) first, followed by \(\varepsilon_{\min}\) and \(c_{\rm max}\).  SHAP similarly ranks \(\varepsilon_{\min}\) and \(\det\) among the leading features, with \(\sigma_\gamma\) also appearing near the top.  Permutation importance emphasizes eigenvalue dispersion and scale, ranking \(\sigma_\varepsilon\), \(\varepsilon_{\max}\), and \(\varepsilon_{\min}\) highest.  Taken together, the attribution methods suggest that the classifier combines the explicit determinant signal with low-end spectral information and distributional eigenvalue statistics.

This differs from the individual HOM-Star and HOM-H sectors.  HOM-Star is dominated by low-end eigenvalue information but is poorly reconstructed by the ten-feature logit regressor, indicating that the present abstract dictionary captures only part of the star-sector decision geometry.  HOM-H has a more stable logit reconstruction, but its leading attributions shift toward gap statistics \((\mu_\gamma,\sigma_\gamma)\) together with \(\varepsilon_{\min}\).  In HOM-All, the pooled task stabilizes around features that are robust across both graph families: \(\det\) supplies the direct homological invariant, while \(\varepsilon_{\min}\), \(\sigma_\varepsilon\), and \(\varepsilon_{\max}\) supply spectral proxies that remain informative when the graph topology varies.  The appearance of \(\sigma_\gamma\) in SHAP further indicates that support-gap information still contributes to the combined decision, but it is no longer the sole organizing feature.

We therefore interpret HOM-All as evidence that the full-sequence classifier learns a hybrid determinant--spectral rule.  The determinant provides the coarse homology-sphere signal, while low-end and dispersion statistics of the spectrum help resolve topology-dependent confounders across the star and \(H\)-graph sectors.  The improved validation \(R^2\) and PCC indicate that this hybrid rule is more faithfully represented in the ten-feature dictionary than the corresponding sector-specific HOM-Star rule.

\subsubsection{Graph Topology Classification}

We now turn from homology-sphere detection to a different classification objective: identifying the underlying plumbing graph topology from the \(q\)-series.  In these experiments the labels distinguish star-shaped and \(H\)-shaped plumbing graphs.  The results show that the classifier again achieves high balanced accuracy, but the features used to reconstruct its logits differ from those in the HOM experiments.  In particular, the topology classifiers use a sector-dependent mixture of gap statistics, determinant information, and low-end spectral data.

\paragraph{TOP-Sph.}
We first isolate the homology-sphere subset, containing \(N=21{,}180\) samples, and ask whether the network can distinguish the underlying plumbing graph topology---star-shaped versus \(H\)-shaped---solely from the \(q\)-series.  The classifier achieves a high and stable mean BAS of \(0.964\pm0.004\).  The corresponding confusion matrix in Fig.~\ref{fig:FS_confusion} shows a mild asymmetry, with a slight tendency to classify some \(H\)-graphs as star-shaped graphs.

The logit-regression stage gives one of the clearest interpretable reconstructions in the full-sequence experiments:
\[
(R^2_{\rm tr},R^2_{\rm val},{\rm PCC}_{\rm tr}, {\rm PCC}_{\rm val})=(0.837\pm0.012,0.725\pm0.017,0.916\pm0.007,0.863\pm0.010).
\]
Thus, for homology spheres, the ten-dimensional abstract feature dictionary reconstructs a substantial fraction of the teacher's two-logit geometry.

The attribution results show a decisive shift away from the determinant/eigenvalue proxies emphasized in the homology classification tasks.  In TOP-Sph, the leading feature is the mean gap \(\mu_\gamma\), which is ranked first by FS, PI, and SHAP.  The next stable signal is \(\sigma_\gamma\), which PI and SHAP rank second, while FS also includes \(c_{\rm max}\) and \(\sigma_\gamma\) among the top-ranked features.  This indicates that, once \(|\det M|=1\) is fixed, the graph-topology decision is primarily encoded in the sparsity and spacing structure of the \(q\)-series rather than in determinant variation.

We interpret this reliance on gap statistics through the analytic structure of the \(\widehat Z\) expansion.  The exponents are governed by values of the quadratic form \(\ell^T M^{-1}\ell\) for lattice vectors \(\ell\), as in \eqref{weak_neg_def}.  Since all samples in TOP-Sph are homology spheres, determinant information cannot distinguish the two graph topologies.  Instead, differences in the spectrum and anisotropy of \(M^{-1}\) can change the rate at which \(\ell^T M^{-1}\ell\) grows along different lattice directions.  This changes the spacing and density of accessible exponents in the \(q\)-series.  The strong ranking of \(\mu_\gamma\) and \(\sigma_\gamma\) suggests that the full-sequence classifier detects topology through this induced gap profile.

\paragraph{TOP-Non.}
We next restrict to non-homology spheres and again distinguish star-shaped from \(H\)-shaped plumbing graphs.  The classifier achieves a high, stable mean BAS of \(0.971\pm0.009\).  The logit-regression reconstruction remains strong, although weaker than in TOP-Sph: \[(R^2_{\rm tr},R^2_{\rm val},{\rm PCC}_{\rm tr}, {\rm PCC}_{\rm val})=(0.759\pm0.019, 0.600\pm0.029, 0.876\pm0.010,0.819\pm0.012).\]

The attribution pattern changes substantially from TOP-Sph.  In the homology-sphere sector, the fixed determinant forces the classifier to rely on gap statistics.  In the non-spherical sector, determinant variation becomes available again, and the classifier uses a hybrid spectral signature.  Feature saliency ranks \(\det\) first, followed by \(\varepsilon_{\min}\) and \(\sigma_\gamma\).  Permutation importance instead places \(\varepsilon_{\min}\) first, followed by \(\det\) and \(\mu_\varepsilon\).  SHAP again ranks \(\det\) first and \(\varepsilon_{\min}\) second, with \(\sigma_\gamma\) also appearing among the leading features.  Across methods, the stable signal is therefore not a pure gap statistic: it is the combination of determinant information, low-end spectral information, and a secondary gap contribution.

This should be interpreted as a sector-dependent shortcut rather than a failure mode.  In TOP-Sph, topology must be inferred from the \(q\)-series spacing structure because the determinant is fixed.  In TOP-Non, the graph topology is correlated with broader spectral properties of the plumbing matrix, including \(\det\) and \(\varepsilon_{\min}\).  The classifier exploits these available invariants, while still retaining some sensitivity to gap structure through \(\sigma_\gamma\).  Thus the non-spherical topology decision combines coarse determinant information with low-end spectral data and residual sparsity information.

\paragraph{TOP-All.}
Finally, we train on the combined topology dataset, asking the classifier to distinguish the plumbing graph topology without filtering by homology class.  The classifier achieves the strongest topology performance, with mean BAS \(0.978\pm0.004\).  The confusion matrix in Fig.~\ref{fig:FS_confusion} and the representative logit distribution in Fig.~\ref{fig:FS_logits} show performance consistent with the high accuracy observed in the sector-specific topology experiments.

The logit-regression stage reflects the mixture of signals seen in TOP-Sph and TOP-Non: 
\[(R^2_{\rm tr},R^2_{\rm val},{\rm PCC}_{\rm tr}, {\rm PCC}_{\rm val})=(0.761\pm0.018, 0.616\pm0.013, 0.876\pm0.010,0.816\pm0.010).
\]
The feature rankings confirm that the pooled topology classifier does not rely on a single mechanism.  Feature saliency ranks \(\det\) first, followed by \(\varepsilon_{\min}\) and \(\sigma_\gamma\).  Permutation importance ranks \(\varepsilon_{\min}\), \(\sigma_\gamma\), and \(\sigma_\varepsilon\) among the leading features.  SHAP ranks \(\sigma_\gamma\), \(\varepsilon_{\min}\), and \(\det\) highest.  Thus the combined topology task retains both components: determinant and low-end eigenvalue information inherited from the non-spherical sector, together with gap statistics inherited from the homology-sphere topology task.

We therefore interpret TOP-All as a hybrid topology classifier.  In the homology-sphere sector, topology is most cleanly expressed through the gap profile of the \(q\)-series, especially \(\mu_\gamma\) and \(\sigma_\gamma\).  In the non-spherical sector, determinant variation and low-end spectral information become available and provide strong graph-topology proxies.  When both sectors are pooled, the full-sequence classifier combines these mechanisms: \(\det\) and \(\varepsilon_{\min}\) capture coarse spectral differences between graph families, while \(\sigma_\gamma\) preserves the sparsity/gap signal that remains informative even after the homology class is no longer fixed.

\subsubsection{Discussion of Full Sequence Results}
\label{subsubsec:discFullSeq}

Synthesizing the results across the full sequence representation experiments, distinct patterns emerge regarding the network's learning strategy and the interpretability of its decisions.

% --- Updated Tab.: feature moments ---
\begin{table}[ht!]
\centering
\small
\begin{tabular}{c|c|c}
\toprule
\textbf{Feature} & \textbf{Skewness} & \textbf{Kurtosis}\\
\midrule
$\det$ & 0.00 & 17.73 \\
\hline
$\varepsilon_{\rm min}$ & 1.65 & 9.98 \\
\hline
$\varepsilon_{\rm max}$ & 5.15 & 45.54 \\
\hline
$\mu_{\varepsilon}$ & 2.01 & 10.48 \\
\hline
$\sigma_{\varepsilon}$ & 5.40 & 50.47 \\
\hline
$\Delta$ & 10.93 & 148.67 \\
\hline
$\gamma_0$ & 7.58 & 77.67 \\
\hline
$\mu_{\gamma}$ & 3.90 & 27.89 \\
\hline
$\sigma_{\gamma}$ & 2.64 & 13.35 \\
\hline
$c_{\rm max}$ & 49.44 & 3052.40 \\
\bottomrule
\end{tabular}
\caption{
Higher moments of the ten-feature distribution. Skewness and kurtosis are computed on the raw feature matrix used in the topology experiments, before the interpretability scores are ranked. Large values indicate heavy-tailed or strongly non-normal feature marginals.
}
\label{tab:feature-moments}
\end{table}

\textbf{Sensitivity vs. Relevance.}
A consistent contrast emerges between interpretability metrics measuring \emph{local sensitivity} and those  measuring \emph{global relevance}.  FS measures the local gradient of the model output with respect to the standardized input features, whereas PI and SHAP probe the effect of feature perturbations or feature coalitions on predictive performance.  Consequently, FS should be interpreted as a local sensitivity diagnostic rather than a direct measure of global explanatory power.

In our experiments, this distinction is most visible in the behavior of $c_{\rm max}$.  FS assigns relatively high importance to $c_{\rm max}$ in cases where PI and SHAP do not give it comparable scores and where direct ablation or single-feature predictive tests do not identify it as a comparably strong explanatory feature.  We therefore do not interpret the FS ranking alone as evidence, in those cases, that $c_{\rm max}$ controls the decision of the classifier network.  Rather, the discrepancy indicates that FS is more prone than PI or SHAP to select features that induce local output sensitivity without necessarily carrying global predictive relevance.

However, this should not be read as a claim that FS \emph{always} selects high-skewness or high-kurtosis features\footnote{All scalar features are standardized before being passed to the regression network.  Thus the saliency comparison is not driven by raw scale differences between input coordinates, although higher moments and outlier structure can still affect local gradients after standardization.}.  The feature-moment analysis in Tab.~\ref{tab:feature-moments} shows that the situation is more nuanced.  For example, $\mu_\gamma$ is not distinguished by unusually large moments compared with other inputs.  The robust empirical statement is instead that, among the three interpretability metrics, FS is the most local and hence the most susceptible to selection effects.  PI and SHAP provide more stable characterization of global relevance, and in the main experiments, they point away from raw coefficient scale features toward spectral invariants of the plumbing matrix or gap data.

We quantified this effect by computing a bias score,
\begin{equation}\label{eq:bias_score}
\mathrm{bias} = \frac{1}{2}\left(\rho_{\mathrm{skew}} + \rho_{\mathrm{kurt}}\right),
\end{equation}
obtained by averaging the Spearman rank correlations ($\rho_{\mathrm{skew}}$, $\rho_{\mathrm{kurt}}$) between
a metric's feature-importance magnitudes and the feature $|\mathrm{skewness}|$ and kurtosis values, so that
positive scores indicate a preference for high-moment features. This analysis reveals that all three metrics
are on average \emph{anti}-biased (TOP-Sph excepted), preferring features with stable marginals (Tab.~\ref{tab:metric-biases}). The contrast is therefore relative: FS is consistently the \emph{least} anti-biased,
attaining the highest bias score in five of the six experiments, while SHAP is the most strongly anti-biased.

% the above replaces the following
\begin{comment}
    \begin{equation}\label{eq:bias_score}
    \mathrm{bias} = -\frac{1}{2}(\rho_{\mathrm{skew}} + \rho_{\mathrm{kurt}}) , 
\end{equation}
obtained by averaging the Spearman rank correlation ($\rho$) between the feature importance ranking and the rankings for feature skewness ($\rho_{\mathrm{skew}}$) and kurtosis ($\rho_{\mathrm{kurt}}$), respectively. This analysis reveals that FS is generally biased towards features with prominent higher moments (Tab.~\ref{tab:metric-biases}). Conversely, PI and SHAP, which measure global predictive contribution, consistently exhibit an anti-bias, indicating a preference for stable features.
\end{comment}

\begin{table}[ht!]
\centering
\small
\begin{tabular}{c|c|c|c}
\toprule
\textbf{Experiment} & \textbf{FS} & \textbf{PI} & \textbf{SHAP}\\
\midrule
HOM-Star & $-0.200\,\pm\,0.416$ & $-0.315\,\pm\,0.184$ & $-0.425\,\pm\,0.190$ \\
\hline
HOM-H & $-0.481\,\pm\,0.100$ & $-0.456\,\pm\,0.137$ & $-0.577\,\pm\,0.117$ \\
\hline
HOM-All & $-0.194\,\pm\,0.157$ & $-0.505\,\pm\,0.142$ & $-0.709\,\pm\,0.115$ \\
\hline
TOP-Sph & $0.304\,\pm\,0.200$ & $0.201\,\pm\,0.122$ & $-0.017\,\pm\,0.176$ \\
\hline
TOP-Non & $-0.391\,\pm\,0.253$ & $-0.769\,\pm\,0.111$ & $-0.761\,\pm\,0.058$ \\
\hline
TOP-All & $-0.362\,\pm\,0.064$ & $-0.666\,\pm\,0.174$ & $-0.769\,\pm\,0.028$ \\
\bottomrule
\end{tabular}
\caption{
Multi-seed higher-moment bias scores for the full-sequence interpretability metrics.  Entries are reported as mean $\pm$ standard deviation over the classifier--regressor--interpretability seed ensemble.  Positive values indicate that an attribution method preferentially ranks features with large skewness or kurtosis, while negative values indicate preference for features with more stable marginals.  Scores are computed via \eqref{eq:bias_score} from Spearman correlations between each metric's feature-importance vector and the feature-moment rankings of Tab.~\ref{tab:feature-moments}.
}
\label{tab:metric-biases}
\end{table}

\paragraph{Margin Reconstruction of the full-sequence teachers.}
The logit-regression experiments above test whether the continuous two-logit geometry of the full-sequence classifiers can be reconstructed from the abstract feature dictionary.  This is a stringent requirement: a low-dimensional feature set may fail to reproduce calibrated logits while still capturing the teacher's hard decision boundary.  We therefore introduce a normalized \emph{Margin Reconstruction} (MR) protocol.  In this protocol the source teachers are not retrained.  Instead, the teachers, train/validation splits, and teacher BAS values are inherited directly from the multi-seed full-sequence experiment. Student classifiers are then trained only on prescribed abstract feature subsets and are asked to reproduce the source teacher's hard decisions on the same held-out split.

\begin{table}[ht!]
\centering
\small
\setlength{\tabcolsep}{3.2pt}
\renewcommand{\arraystretch}{1.12}
\begin{tabular}{c|c|c|c|c|c|c|c}
\toprule
\textbf{Task} &
\textbf{Student feats.} &
\textbf{$B_{10}$ feats.} &
\textbf{Full} &
\textbf{$B_5$} &
\textbf{$B_{10}$} &
\textbf{$B_{20}$} &
\textbf{$E_{50}$} \\
\midrule
HOM-Star
& $\varepsilon$, log.
& $\varepsilon_{\min},\,\sigma_{\varepsilon}$
& $.726\pm.258$
& $0.584\pm0.075$
& $.604\pm.127$
& $.623\pm.177$
& $.762\pm.237$ \\
\hline
HOM-H
& $\det$, RF
& $\det$
& $.920\pm.007$
& $0.576 \pm 0.037$
& $.633\pm.015$
& $.716\pm.032$
& $.973\pm.011$ \\
\hline
HOM-All
& $\det$, RF
& $\det$
& $.953\pm.005$
& $0.737 \pm 0.021$
& $.809\pm.014$
& $.869\pm.017$
& $.970\pm.005$ \\
\hline
TOP-Sph
& $\gamma$, log.
& $\mu_{\gamma},\,\gamma_0$
& $.969\pm.005$
& $ 0.682\pm 0.040$
& $.762\pm.030$
& $.850\pm.026$
& $.990\pm.003$ \\
\hline
TOP-Non
& $\det+\gamma$, MLP
& $\sigma_{\gamma},\,\mu_{\gamma}$
& $.986\pm.003$
& $0.861 \pm 0.035$
& $.909\pm.029$
& $.932\pm.028$
& $.985\pm.005$ \\
\hline
TOP-All
& $\gamma$, log.
& $\sigma_{\gamma},\,\mu_{\gamma}$
& $.977\pm.002$
& $0.784 \pm 0.013$
& $.856\pm.011$
& $.905\pm.011$
& $.985\pm.003$ \\
\bottomrule
\end{tabular}
\caption{
Normalized margin reconstruction for the full-sequence teachers, with feature attribution on the
\(B_{10}\) teacher-boundary slice.  The source teachers and train/validation splits are inherited
directly from the full-sequence experiments.
Students are trained to reproduce the source teacher's hard decisions using only the indicated
abstract feature set.  ``Full'' reports balanced agreement with the source teacher on the held-out
split.  \(B_5\), \(B_{10}\) and \(B_{20}\) restrict agreement to the 5\%, 10\%, and 20\% smallest-margin held-out
samples, where the margin is \(|\ell_1-\ell_0|\).  \(E_{50}\) reports agreement on the 50\%
largest-margin held-out samples.  ``\(B_{10}\) feats.'' lists the leading features under
the average of normalized saliency, SHAP, and permutation-drop attribution scores.  Here
\(\varepsilon\)-stats denotes
\((\varepsilon_{\min},\varepsilon_{\max},\mu_{\varepsilon},\sigma_{\varepsilon})\), and
\(\gamma\)-stats denotes \((\gamma_0,\mu_{\gamma},\sigma_{\gamma})\).
}
\label{tab:boundaryfirst_zseed_margin_profile_attribution}
\end{table}

Table~\ref{tab:boundaryfirst_zseed_margin_profile_attribution} reports the resulting margin profile.  The column ``Full'' gives balanced agreement with the source teacher on the full held-out split.  The columns \(B_5\), \(B_{10}\), and \(B_{20}\) restrict this agreement to the 5\%, 10\%, and 20\% smallest-margin validation samples, where the source-teacher margin is \(|\ell_1-\ell_0|\).  The column \(E_{50}\) reports agreement on the 50\% largest-margin samples.  Thus the table separates easy-region agreement from genuine boundary-local reconstruction.  A surrogate that is accurate only on \(E_{50}\) captures coarse global structure; a surrogate that remains accurate on \(B_5\) and \(B_{10}\) captures the teacher's local decision geometry.

The homology tasks show two different behaviors.  For HOM-H and HOM-All, determinant-based students recover the full-sequence teachers with high global agreement.  HOM-H has full teacher agreement \(0.920\pm0.007\) and true-label BAS \(0.997\pm0.001\), but its agreement drops to \(0.576\pm0.037\) on \(B_5\).  This indicates that \(\det\) captures the dominant global rule but not all boundary-local refinements of the teacher.  HOM-All is stronger: the determinant-only student retains substantial agreement throughout the boundary profile, with \(0.737\pm0.021\) on \(B_5\), \(0.809\pm0.014\) on \(B_{10}\), and \(0.869\pm0.017\) on \(B_{20}\).

The topology tasks are more consistently explained by gap statistics.  TOP-Sph and TOP-All select the \(\gamma\)-statistics feature set \((\gamma_0,\mu_\gamma,\sigma_\gamma)\), while TOP-Non selects \(\det+\gamma\)-statistics.  Their margin profiles show monotone improvement from the hardest boundary slice to the easy region, with strong full-split agreement in all cases.  TOP-Non is the strongest boundary reconstruction result: the selected \(\det+\gamma\)-statistics MLP achieves \(0.861\pm0.035\) on \(B_5\), \(0.909\pm0.029\) on \(B_{10}\), \(0.932\pm0.028\) on \(B_{20}\), and \(0.986\pm0.003\) on the full held-out split.  TOP-All is also robust, with \(\sigma_\gamma\) and \(\mu_\gamma\) dominating the \(B_{10}\)-slice attribution and agreement rising from \(0.784\pm0.013\) on \(B_5\) to \(0.905\pm0.011\) on \(B_{20}\).

The attribution-augmented column resolves the selected feature groups into active components.  For HOM-Star, the selected feature set is the eigenvalue-statistic tuple \((\varepsilon_{\min},\varepsilon_{\max},\mu_\varepsilon,\sigma_\varepsilon)\), with \(\varepsilon_{\min}\) and \(\sigma_\varepsilon\) dominating the \(B_{10}\)-slice attribution.  However, the corresponding surrogate remains weak and unstable: full teacher agreement is only \(0.726\pm0.258\), with \(0.584\pm0.075\) on \(B_5\) and \(0.604\pm0.127\) on \(B_{10}\).  We therefore treat HOM-Star as the hard case rather than as a successful low-dimensional reconstruction.  This agrees with the poor multi-feature logit-regression result and suggests that the HOM-Star full-sequence teacher uses structure not captured by the current determinant/eigenvalue/gap/\(c_{\max}\) abstract dictionary.

Overall, MR gives a more refined interpretation than either classifier accuracy or logit regression alone.  Determinant information explains much of the HOM-H and HOM-All decision structure, gap statistics explain most of the topology decision structure, and HOM-Star remains the principal unresolved feature-discovery case.

\textbf{Spectral Reconstruction.}
The signals identified by the interpretability metrics suggest that the network learns to recover spectral information about the plumbing matrix \(M\) from the \(q\)-series, but that the relevant spectral summary depends on the classification objective and on the sector of the dataset.

In the HOM experiments, the classifier is asked to distinguish homology spheres, where \(|\det M|=1\), from non-spheres.  The full-sequence interpretability results show that the network does not rely only on the determinant itself.  Instead, it repeatedly uses low-end and dispersion information in the spectrum, especially \(\varepsilon_{\min}\), \(\varepsilon_{\max}\), and \(\sigma_\varepsilon\), together with determinant information in the pooled HOM-All task.  These features act as spectral proxies for the determinant constraint: they help distinguish genuine unit-determinant examples from plumbing matrices whose \(q\)-series statistics partially mimic the homology-sphere sector.  The normalized MR analysis is consistent with this interpretation.  For HOM-H and HOM-All, determinant-based students reconstruct the global teacher decisions well, while HOM-Star remains only weakly reconstructed by the current eigenvalue-statistic dictionary, indicating that the star-sector homology decision contains additional structure not captured by these coarse abstract features alone.

In the TOP experiments, the mechanism is more sector-dependent.  For homology spheres, where the determinant is fixed, the dominant signal shifts toward gap statistics of the \(q\)-series.  In TOP-Sph, \(\mu_\gamma\) is ranked first by FS, PI, and SHAP, with \(\sigma_\gamma\) appearing as the next stable feature.  This supports the geometric picture that differences between \(H\)-graphs and star graphs affect the anisotropy of the quadratic form \(\ell^T M^{-1}\ell\) in \eqref{weak_neg_def}.  Greater anisotropy can produce slower growth along selected directions and hence a different sparsity/gap profile in the resulting \(q\)-series.  The classifier detects this topology-dependent anisotropy through the gap statistics, rather than through determinant variation.

For non-spheres, however, the determinant is no longer fixed, and the topology decision is not a pure gap test.  In TOP-Non, the full-sequence attribution table repeatedly identifies \(\det\) and \(\varepsilon_{\min}\) as leading features, with \(\sigma_\gamma\) appearing as a secondary gap statistic.  Thus the non-spherical topology classifier combines determinant information, low-end spectral information, and residual gap structure.  The pooled TOP-All task reflects the same hybrid mechanism: \(\det\), \(\varepsilon_{\min}\), and \(\sigma_\gamma\) all appear across the attribution methods.  In short, gap statistics dominate the fixed-determinant homology-sphere topology task, while determinant and eigenvalue summaries become essential once the non-spherical sector is included.

\subsection{Value-Index Pair Representation}\label{sec:denseRep}

In order to test the effect of data representation, we repeat the six classification tasks using the value-index pair representation described in \S\ref{subsubsec:value-index-pair}.  In this representation, the network receives both coefficient values and their exponent indices, rather than a fixed full-sequence array in which the exponent is encoded by position.  This provides a useful robustness check: if the same abstract features remain predictive across the full-sequence and value-index representations, then the learned signal is unlikely to be an artifact of the dense array indexing.

For these experiments we use a fully connected MLP classifier with \((\text{depth},\text{width})=(2,64)\), and the same feature-to-logit interpretability pipeline with a regression network of size \((4,128)\).\footnote{Classifiers were trained with early stopping after 20 epochs of no improvement in validation BAS.  Regressors used early stopping after 100 epochs of no improvement in validation \(R^2\).}  The results are summarized in Tab.~\ref{tab:summary-3}.  Across all six tasks, the value-index representation improves the classifier performance and substantially improves the feature-to-logit reconstruction quality.  In particular, the validation \(R^2\) values are uniformly high, ranging from \(0.851\pm0.030\) in HOM-All to \(0.987\pm0.004\) in TOP-Sph, and the validation PCC values are at least \(0.927\) in every experiment.  Thus, in the value-index representation, the ten derived features recover not only the hard decisions but also a large fraction of the learned logit geometry.

%%%%%%%%%%%%%%%%%%%%%%%%%%%%%%%%%%%%%%%%%%%%%%%%%%%%%%%%%%%%%%%%%%%%%%%%%%%%%%%%
% Multi-seed table for value-index
%%%%%%%%%%%%%%%%%%%%%%%%%%%%%%%%%%%%%%%%%%%%%%%%%%%%%%%%%%%%%%%%%%%%%%%%%%%%%%%%
\begin{table}[ht!]
\centering
\small
\setlength{\tabcolsep}{3pt}
\resizebox{\textwidth}{!}{%
\begin{tabular}{c|c|c|c|ccc}
\toprule
\textbf{Experiment} &
\textbf{BAS} &
$\mathbf{R^2}$ &
\textbf{PCC} &
\textbf{FS} &
\textbf{PI} &
\textbf{SHAP} \\
\midrule
HOM-Star & $0.999\,\pm\,0.001$ & \shortstack{$\mathrm{tr}: 0.940\,\pm\,0.013$\\$\mathrm{val}: 0.920\,\pm\,0.018$} & \shortstack{$\mathrm{tr}: 0.969\,\pm\,0.007$\\$\mathrm{val}: 0.958\,\pm\,0.011$} & \rankfeat{\Delta}{1.3}, \rankfeat{\mu_{\gamma}}{1.7}, \rankfeat{\det}{3.7} & \rankfeat{\mu_{\gamma}}{1.2}, \rankfeat{\varepsilon_{\rm min}}{2.0}, \rankfeat{\det}{3.5} & \rankfeat{\varepsilon_{\rm min}}{1.4}, \rankfeat{\mu_{\gamma}}{1.6}, \rankfeat{\det}{3.0} \\
\hline
HOM-H & $0.967\,\pm\,0.004$ & \shortstack{$\mathrm{tr}: 0.971\,\pm\,0.023$\\$\mathrm{val}: 0.951\,\pm\,0.031$} & \shortstack{$\mathrm{tr}: 0.986\,\pm\,0.011$\\$\mathrm{val}: 0.976\,\pm\,0.015$} & \rankfeat{\mu_{\gamma}}{1.0}, \rankfeat{\sigma_{\gamma}}{2.0}, \rankfeat{\Delta}{3.0} & \rankfeat{\mu_{\gamma}}{1.7}, \rankfeat{\Delta}{2.5}, \rankfeat{\sigma_{\gamma}}{2.8} & \rankfeat{\mu_{\gamma}}{1.0}, \rankfeat{\sigma_{\gamma}}{2.0}, \rankfeat{\det}{3.3} \\
\hline
HOM-All & $0.985\,\pm\,0.002$ & \shortstack{$\mathrm{tr}: 0.892\,\pm\,0.029$\\$\mathrm{val}: 0.851\,\pm\,0.030$} & \shortstack{$\mathrm{tr}: 0.948\,\pm\,0.014$\\$\mathrm{val}: 0.927\,\pm\,0.013$} & \rankfeat{\det}{1.0}, \rankfeat{\varepsilon_{\rm min}}{2.0}, \rankfeat{\sigma_{\gamma}}{3.5} & \rankfeat{\sigma_{\varepsilon}}{2.2}, \rankfeat{\varepsilon_{\rm min}}{2.4}, \rankfeat{\mu_{\gamma}}{3.3} & \rankfeat{\varepsilon_{\rm min}}{1.5}, \rankfeat{\det}{2.3}, \rankfeat{\mu_{\gamma}}{2.6} \\
\hline
TOP-Sph & $1.000\,\pm\,0.000$ & \shortstack{$\mathrm{tr}: 0.990\,\pm\,0.002$\\$\mathrm{val}: 0.987\,\pm\,0.004$} & \shortstack{$\mathrm{tr}: 0.995\,\pm\,0.001$\\$\mathrm{val}: 0.994\,\pm\,0.002$} & \rankfeat{\mu_{\gamma}}{1.0}, \rankfeat{\sigma_{\gamma}}{2.2}, \rankfeat{\gamma_0}{3.0} & \rankfeat{\mu_{\gamma}}{1.1}, \rankfeat{\sigma_{\gamma}}{2.0}, \rankfeat{\gamma_0}{3.1} & \rankfeat{\mu_{\gamma}}{1.0}, \rankfeat{\sigma_{\gamma}}{2.1}, \rankfeat{\Delta}{3.0} \\
\hline
TOP-Non & $1.000\,\pm\,0.000$ & \shortstack{$\mathrm{tr}: 0.979\,\pm\,0.003$\\$\mathrm{val}: 0.975\,\pm\,0.004$} & \shortstack{$\mathrm{tr}: 0.990\,\pm\,0.002$\\$\mathrm{val}: 0.988\,\pm\,0.002$} & \rankfeat{\mu_{\gamma}}{1.4}, \rankfeat{\Delta}{3.2}, \rankfeat{\det}{3.6} & \rankfeat{\mu_{\gamma}}{1.4}, \rankfeat{\sigma_{\gamma}}{2.3}, \rankfeat{\varepsilon_{\rm min}}{3.7} & \rankfeat{\mu_{\gamma}}{1.0}, \rankfeat{\varepsilon_{\rm min}}{2.2}, \rankfeat{\sigma_{\gamma}}{3.3} \\
\hline
TOP-All & $1.000\,\pm\,0.000$ & \shortstack{$\mathrm{tr}: 0.960\,\pm\,0.003$\\$\mathrm{val}: 0.952\,\pm\,0.005$} & \shortstack{$\mathrm{tr}: 0.980\,\pm\,0.002$\\$\mathrm{val}: 0.977\,\pm\,0.002$} & \rankfeat{\Delta}{1.8}, \rankfeat{\sigma_{\gamma}}{2.5}, \rankfeat{\varepsilon_{\rm min}}{3.8} & \rankfeat{\varepsilon_{\rm min}}{2.3}, \rankfeat{\mu_{\gamma}}{2.6}, \rankfeat{\sigma_{\gamma}}{3.1} & \rankfeat{\sigma_{\gamma}}{1.0}, \rankfeat{\mu_{\gamma}}{2.1}, \rankfeat{\varepsilon_{\rm min}}{3.0} \\
\bottomrule
\end{tabular}%
}
\caption{
Multi-seed value-index results for the topology and homology experiments.  BAS is reported as mean $\pm$ standard deviation over classifier seeds.  The $R^2$ and Pearson correlation (PCC) columns report train and validation reconstruction quality of the feature-to-logit regressor, stacked within each cell and averaged over the classifier--regressor seed ensemble.  The final columns report the three lowest mean feature ranks for feature saliency (FS), permutation importance (PI), and SHAP, where the subscript gives the average rank over the normalized interpretability ensemble.
}
\label{tab:summary-3}
\end{table}

\subsubsection{Homology Classification}

\noindent\textbf{HOM-Star.}
We begin by training the value-index pair classifier to distinguish homology spheres from non-spheres within the star-shaped graph dataset.  The classifier is trained on \(10{,}474\)-dimensional value-index inputs and achieves nearly perfect classification, with mean BAS \(0.999\pm0.001\).  This is a substantial improvement over the corresponding full-sequence HOM-Star experiment, where the classifier remained accurate but the logit geometry was poorly reconstructed by the ten-feature regressor.

In the value-index representation, the feature-to-logit regressor achieves
\[
(R^2_{\rm tr},R^2_{\rm val},{\rm PCC}_{\rm tr}, {\rm PCC}_{\rm val})=(0.940\pm0.013,
0.920\pm0.018,
0.969\pm0.007,
0.958\pm0.011).
\]
Thus the ten derived features now reconstruct the value-index classifier logits extremely well.  This does not mean that the raw network uses no information beyond the ten features; rather, it means that the part of the learned decision function relevant to the classifier logits is accurately summarized by these spectral and support statistics.

The attribution metrics indicate that the value-index HOM-Star classifier uses a mixed determinant--gap--spectral signature.  Feature saliency ranks \(\Delta\) and \(\mu_\gamma\) highest, followed by \(\det\).  Permutation importance ranks \(\mu_\gamma\) first, then \(\varepsilon_{\min}\), then \(\det\).  SHAP ranks \(\varepsilon_{\min}\) first, with \(\mu_\gamma\) and \(\det\) close behind.  Thus, unlike the full-sequence HOM-Star case, the value-index representation does not isolate a single low-end eigenvalue proxy.  Instead, it makes the HOM-Star decision highly linearly reconstructible from a combination of first-gap information, mean gap information, determinant information, and \(\varepsilon_{\min}\).  The robust appearance of \(\mu_\gamma\) across all three methods suggests that exposing the exponent indices directly allows the classifier to use the support spacing of the \(q\)-series more efficiently than in the full-sequence representation.

\noindent\textbf{HOM-H.}
Next, we evaluate the value-index pair classifier on the \(H\)-graph homology task.  The model achieves BAS \(0.967\pm0.004\), improving over the full-sequence HOM-H classifier.  The regressor also becomes highly accurate:
\[
(R^2_{\rm tr},R^2_{\rm val},{\rm PCC}_{\rm tr}, {\rm PCC}_{\rm val})=(0.971\pm0.023,\,0.951\pm0.031,0.986\pm0.011,\,0.976\pm0.015).
\]
Thus, for \(H\)-graphs, the value-index classifier's logits are almost completely reconstructed by the ten abstract features.

The feature rankings are also more coherent than in the full-sequence case.  FS, PI, and SHAP all identify \(\mu_\gamma\) as a leading feature, with average ranks \(1.0\), \(1.7\), and \(1.0\), respectively.  The gap dispersion \(\sigma_\gamma\) is also consistently important, appearing second under FS and SHAP and third under PI.  The unseen leading power \(\Delta\) is emphasized by FS and PI, while SHAP includes \(\det\) as the third-ranked feature.  This pattern strengthens the interpretation developed in the full-sequence HOM-H analysis: in the \(H\)-graph sector, homology-sphere detection is not governed solely by a low-end eigenvalue proxy.  Instead, the classifier uses the support-spacing structure of the \(q\)-series, especially \(\mu_\gamma\) and \(\sigma_\gamma\), together with residual determinant information.

\noindent\textbf{HOM-All.}
We next train on the combined homology dataset, asking the value-index classifier to distinguish homology spheres from non-spheres without conditioning on the graph topology.  The classifier achieves BAS \(0.985\pm0.002\), again improving over the full-sequence representation.  The feature-to-logit regressor gives
\[
(R^2_{\rm tr},R^2_{\rm val},{\rm PCC}_{\rm tr}, {\rm PCC}_{\rm val})=(0.892\pm0.029,0.851\pm0.030,0.948\pm0.014,0.927\pm0.013).
\]
The ten-feature dictionary therefore remains highly predictive even in the pooled, heterogeneous homology task.

The attribution metrics show a hybrid rule.  Feature saliency ranks \(\det\) first and \(\varepsilon_{\min}\) second, with \(\sigma_\gamma\) third.  SHAP similarly ranks \(\varepsilon_{\min}\), \(\det\), and \(\mu_\gamma\) as the leading features.  Permutation importance emphasizes eigenvalue dispersion, ranking \(\sigma_\varepsilon\), \(\varepsilon_{\min}\), and \(\mu_\gamma\) highest.  Hence the value-index HOM-All classifier combines the explicit determinant signal with low-end spectral information and gap information.  This is consistent with the pooled full-sequence HOM-All experiment, but the value-index representation makes the relationship between the classifier logits and the abstract features much sharper.

\subsubsection{Graph Topology Classification}

We next ask the value-index classifier to distinguish star-shaped from \(H\)-shaped plumbing graphs.  This is a particularly useful representation check.  In the full-sequence representation, exponent information is encoded implicitly by the array position.  In the value-index representation, the coefficients and exponent indices are supplied as paired data.  If graph topology were detected only through artifacts of the dense array coordinate system, one would expect a collapse in performance or a substantial change in the reconstructed features.  Instead, the topology tasks become almost perfectly classified, and the feature-to-logit regressors achieve very high validation \(R^2\) and PCC.

\noindent\textbf{TOP-Sph.}
For the homology-sphere subset, the value-index topology classifier achieves perfect separation, with BAS \(1.000\pm0.000\).  The logit regressor also becomes nearly exact:
\[
(R^2_{\rm tr},R^2_{\rm val},{\rm PCC}_{\rm tr}, {\rm PCC}_{\rm val})=(0.990\pm0.002, 0.987\pm0.004, 0.995\pm0.001,0.994\pm0.002).
\]
The attribution metrics are essentially unanimous.  FS, PI, and SHAP all rank \(\mu_\gamma\) first, and all identify \(\sigma_\gamma\) as the next stable feature.  The third feature varies between \(\gamma_0\) and \(\Delta\), but the dominant signal is clearly the gap profile.

This reproduces and strengthens the full-sequence TOP-Sph interpretation.  Once the determinant is fixed by restricting to homology spheres, topology is encoded primarily in the spacing structure of the \(q\)-series.  The value-index representation makes this signal even cleaner: the classifier can access exponent locations explicitly, and its logit geometry is almost completely reconstructed by the gap statistics \(\mu_\gamma\), \(\sigma_\gamma\), and related first-gap information.

\noindent\textbf{TOP-Non.}
For non-homology spheres, the value-index topology classifier again achieves perfect separation, with BAS \(1.000\pm0.000\).  The regressor remains highly accurate:
\[
(R^2_{\rm tr},R^2_{\rm val},{\rm PCC}_{\rm tr}, {\rm PCC}_{\rm val})=(0.979\pm0.003,0.975\pm0.004,0.990\pm0.002,0.988\pm0.002).
\]
Unlike the full-sequence TOP-Non experiment, where determinant and \(\varepsilon_{\min}\) were the most stable recurring features, the value-index representation shifts the dominant reconstructed signal toward gap statistics.  All three attribution methods rank \(\mu_\gamma\) first or near first: FS and PI give it average rank \(1.4\), while SHAP ranks it first.  PI and SHAP also retain spectral information through \(\varepsilon_{\min}\), and FS includes \(\det\), but these appear as secondary rather than primary signals.

Thus TOP-Non and TOP-Sph are no longer aligned with the full-sequence distinction between a gap-driven sphere sector and a determinant/eigenvalue-driven non-sphere sector.  In the value-index representation, the non-spherical topology classifier can still use determinant and low-end spectral information, but its logit geometry is reconstructed most cleanly by the \(q\)-series gap profile.  This suggests that explicitly providing exponent indices makes topology-dependent support spacing accessible even when determinant variation is present.

\noindent\textbf{TOP-All.}
Finally, in the pooled topology task, the value-index classifier again achieves BAS \(1.000\pm0.000\).  The regressor gives
\[
(R^2_{\rm tr},R^2_{\rm val},{\rm PCC}_{\rm tr}, {\rm PCC}_{\rm val})=(0.960\pm0.003,0.952\pm0.005,0.980\pm0.002,0.977\pm0.002).
\]
The pooled value-index topology task therefore remains almost perfectly reconstructible from the abstract features.

The feature rankings show a mixed but gap-centered rule.  FS ranks \(\Delta\), \(\sigma_\gamma\), and \(\varepsilon_{\min}\) highest.  PI ranks \(\varepsilon_{\min}\), \(\mu_\gamma\), and \(\sigma_\gamma\) highest.  SHAP ranks \(\sigma_\gamma\) first, followed by \(\mu_\gamma\) and \(\varepsilon_{\min}\).  Thus the stable features across methods are \(\sigma_\gamma\), \(\mu_\gamma\), and \(\varepsilon_{\min}\), with the unseen leading power \(\Delta\) prominent in FS.

We interpret this as a representation-dependent strengthening of the gap signal.  In the full-sequence TOP-All experiment, the reconstructed topology signal was hybrid, involving \(\det\), \(\varepsilon_{\min}\), and \(\sigma_\gamma\).  In the value-index representation, determinant information becomes less central in the pooled topology attribution, while gap statistics become the dominant reconstructed features.  This is consistent with the purpose of the value-index input: by supplying exponent positions explicitly, it makes the support-spacing structure of the \(q\)-series easier for the classifier to exploit.  The topology signal is therefore not lost when the dense array coordinate system is removed; instead, it becomes more cleanly associated with the gap profile of the \(q\)-series.

\subsection{Latent Space Representation}\label{sec:latentRep}

\subsubsection{Binary Classification}\label{subsubsec:binary_latent}

We first evaluate classifiers trained on contrastive latent embeddings.  In these experiments, the encoder is trained from the \(10{,}000\)-dimensional full-sequence representation using semi-hard triplet loss and outputs a \(5\)-dimensional embedding.  The classifier is then trained on the standardized \(5\)-dimensional latent coordinates.  This creates a severe information bottleneck, compressing the original full-sequence input by a factor of \(2000\).  The purpose of this experiment is to test whether the contrastive objective distills the \(q\)-series into a lower-dimensional representation whose classifier logits remain reconstructible from the ten derived spectral and support features.

\textbf{Homology Classification (Star, H, and Combined).}
For the homology classification tasks, the latent-space classifiers achieve strong performance, although they generally do not match the nearly perfect value-index classifiers.  The BAS values are \(0.940\pm0.005\) for HOM-Star, \(0.960\pm0.008\) for HOM-H, and \(0.957\pm0.004\) for HOM-All.  Thus the contrastive bottleneck preserves most of the discriminative information needed to identify homology spheres, despite reducing the input dimension from \(10{,}000\) to \(5\).

The main advantage of the latent representation is not raw classifier accuracy, but the stability and interpretability of the learned logit geometry.  In HOM-H, for example, the latent classifier slightly underperforms the value-index classifier, but its logits are highly reconstructible from the ten-feature dictionary:
\[
(R^2_{\rm tr},R^2_{\rm val},{\rm PCC}_{\rm tr}, {\rm PCC}_{\rm val})=(0.991\pm0.002,0.899\pm0.021,0.996\pm0.001,0.952\pm0.009).
\]
The same pattern holds across the homology tasks.  HOM-Star improves dramatically relative to the full-sequence regressor, reaching \(R^2_{\rm val}=0.784\pm0.032\), while HOM-All reaches \(R^2_{\rm val}=0.818\pm0.016\).  The contrastive encoder therefore appears to smooth or regularize the decision geometry: the resulting low-dimensional classifier remains accurate, and its logits are much more faithfully reconstructed by the abstract feature dictionary than in the full-sequence representation.

We interpret this as a form of semantic filtering.  In the full-sequence and value-index representations, the classifier can route information through many high-dimensional directions, making gradient-based and perturbation-based explanations sensitive to representation-specific artifacts.  By contrast, the semi-hard triplet objective forces samples with the same label to cluster and samples with different labels to separate in a low-dimensional metric space.  Non-discriminative high-variance directions provide little margin benefit but can inject noise into near-boundary triplets.  The encoder is therefore pressured to suppress such directions and retain coordinates that are stable for the class geometry.  This does not prove that the latent classifier uses a single universal feature across all homology tasks; rather, it shows that after contrastive compression, the learned logits are consistently well explained by the ten derived determinant, spectral, and support statistics.

\textbf{Graph Topology Classification (Spheres, Non-Spheres, and Combined).}
The topology tasks show that the latent representation preserves the graph-topology signal even more strongly.  The classifiers achieve BAS \(0.979\pm0.004\) for TOP-Sph, \(0.998\pm0.001\) for TOP-Non, and \(0.992\pm0.001\) for TOP-All.  Thus, even after compression to five latent coordinates, the encoder retains enough information to distinguish star-shaped and \(H\)-shaped plumbing graphs with high accuracy.

The feature-to-logit regressors are also strong.  For TOP-Sph, the regressor achieves
\[
(R^2_{\rm tr},R^2_{\rm val},{\rm PCC}_{\rm tr}, {\rm PCC}_{\rm val})=(0.991\pm0.003,0.944\pm0.004,0.995\pm0.002,0.973\pm0.002).
\]
For TOP-Non, the reconstruction is even stronger:
\[
(R^2_{\rm tr},R^2_{\rm val},{\rm PCC}_{\rm tr}, {\rm PCC}_{\rm val})=(0.997\pm0.002,0.980\pm0.018,0.998\pm0.001,0.990\pm0.009).
\]
The pooled TOP-All task remains highly accurate at the classifier level, although the regressor is less stable across seeds, with \(R^2_{\rm val}=0.838\pm0.332\) and \({\rm PCC}_{\rm val}=0.947\pm0.099\).  This larger variance indicates that some latent TOP-All classifiers learn logit geometries that are less uniformly captured by the ten-feature regressor, even though their hard-label performance remains strong.

These results refine the picture from the full-sequence and value-index experiments.  The contrastive bottleneck does not destroy either the gap-based topology signal visible in TOP-Sph or the determinant/eigenvalue information available in the non-spherical sector.  Instead, it compresses these signals into a five-dimensional metric representation whose logits are usually highly reconstructible from the abstract feature dictionary.  The strongest statement supported by the latent-space results is therefore not that the encoder eliminates all spectral shortcuts, but that it organizes the topology decision into a low-dimensional geometry aligned with the same determinant, spectral, and gap features identified in the higher-dimensional representations.

\begin{table}[ht!]
\centering
\small
\setlength{\tabcolsep}{6pt}
\renewcommand{\arraystretch}{1.15}
\begin{tabular}{c|cc}
\toprule
\textbf{Feature} &
\textbf{TOP coefficient} &
\textbf{HOM coefficient} \\
\midrule
$\det$
& $0.03\,\pm\,0.01$
& $2.17\,\pm\,0.04$ \\
\hline
$\varepsilon_{\rm min}$
& $0.30\,\pm\,0.01$
& $4.43\,\pm\,0.16$ \\
\hline
$\varepsilon_{\rm max}$
& $-0.27\,\pm\,0.04$
& $-0.62\,\pm\,0.03$ \\
\hline
$\mu_{\varepsilon}$
& $0.16\,\pm\,0.02$
& $-0.26\,\pm\,0.02$ \\
\hline
$\sigma_{\varepsilon}$
& $0.19\,\pm\,0.06$
& $0.63\,\pm\,0.04$ \\
\hline
$\Delta$
& $1.13\,\pm\,0.03$
& $-0.35\,\pm\,0.04$ \\
\hline
$\gamma_0$
& $0.06\,\pm\,0.06$
& $0.05\,\pm\,0.02$ \\
\hline
$\mu_{\gamma}$
& $2.23\,\pm\,0.07$
& $0.41\,\pm\,0.16$ \\
\hline
$\sigma_{\gamma}$
& $5.41\,\pm\,0.04$
& $-1.03\,\pm\,0.05$ \\
\hline
$c_{\rm max}$
& $0.21\,\pm\,0.01$
& $0.04\,\pm\,0.01$ \\
\bottomrule
\end{tabular}
\caption{
Multi-seed normalized coefficients \(\kappa_i\) for the linear abstract-feature boundary
approximation \(\sum_i \kappa_i f_i=0\).  The TOP column is obtained from a class-balanced
linear classifier separating graph topology classes, while the HOM column is obtained from a
class-balanced linear classifier separating homology spheres from non-spheres.  For each seed,
the ten abstract features are standardized on the training split, the linear classifier is fit,
and the coefficient vector is rescaled so that
\(\frac{1}{10}\sum_i |\kappa_i|=1\).  Entries report mean \(\pm\) standard deviation over the
five-seed ensemble. Coefficient signs follow the binary label encoding, with positive coefficients pointing
toward the star-shaped class (TOP row) and the non-sphere class (HOM row); coefficient
magnitudes should be interpreted as robust linear feature strengths.
}
\label{tab:linear-boundaries}
\end{table}

\subsubsection{Multi-class Classification}

Finally, we study the geometry of the latent-space decision boundaries in a four-class classification problem.  We use the complete dataset described in Tab.~\ref{tab:dataset-binary}, organizing it into four classes according to both plumbing graph topology and homology type: STAR-Non, STAR-Sph, H-Sph, H-Non. Preliminary experiments showed that the value-index pair representation gave the strongest classifier performance in the binary tasks.  We therefore use the value-index pair format as the input representation for generating the contrastive latent embeddings in this analysis.

For each seed, the encoder is trained on the \(10{,}474\)-dimensional value-index input using semi-hard triplet loss.  The encoder is an MLP with three hidden layers of width \(512\), followed by a \(5\)-dimensional latent output.  Training uses margin \(m=1\), learning rate \(10^{-3}\), batch size \(256\), and early stopping with patience \(10\).  All downstream classifiers and reconstruction models are trained using standard \texttt{scikit-learn} implementations unless otherwise specified.  The analysis is repeated over the seed ensemble
\(
\{42,123,456,789,1337\}.
\)

Although the learned embedding is \(5\)-dimensional, PCA shows that the first two principal components capture approximately \(99.5\%\) of the latent variance across seeds.  The resulting two-dimensional projection is therefore not merely a visualization device: it captures the dominant geometry of the contrastive representation.  This collapse is consistent with the label structure, since the four classes are the intersections of two binary attributes, namely homology class and graph topology.

Working in this two-dimensional PCA subspace, we first train binary logistic regression classifiers for the two induced binary labels.  The topology classifier, separating star-shaped graphs from \(H\)-graphs, achieves near-perfect accuracy across seeds, \(0.998\pm0.002\).  The homology classifier, separating spheres from non-spheres, is also strong but more seed-dependent, with accuracy \(0.943\pm0.050\).  Thus the latent plane is organized by two coarse axes corresponding approximately to graph topology and homology type, with the topology direction more stable across seeds.

As a first linear diagnostic, we fit class-balanced logistic regressions from the ten standardized abstract features directly to the two induced binary labels: graph topology and homology type.  This gives the linear boundary approximation
\[
\sum_i \kappa_i f_i = 0,
\]
with normalized coefficients reported in Tab.~\ref{tab:linear-boundaries}.  The linear topology classifier achieves held-out BAS \(0.988\pm0.001\), while the linear homology classifier achieves held-out BAS \(0.975\pm0.002\).  The coefficient magnitudes show a clean separation of mechanisms: the TOP boundary is dominated by the gap statistics, especially \(\sigma_\gamma\) and \(\mu_\gamma\), with \(\Delta\) also contributing; the HOM boundary is dominated by \(\varepsilon_{\min}\) and \(\det\), with a smaller contribution from \(\sigma_\varepsilon\).  Thus, before analyzing the nonlinear one-vs-rest SVM surfaces, the abstract-feature linear model already recovers the two coarse axes suggested by the contrastive latent geometry.

\begin{figure}[t]
    \centering
    \includegraphics[width=\textwidth]{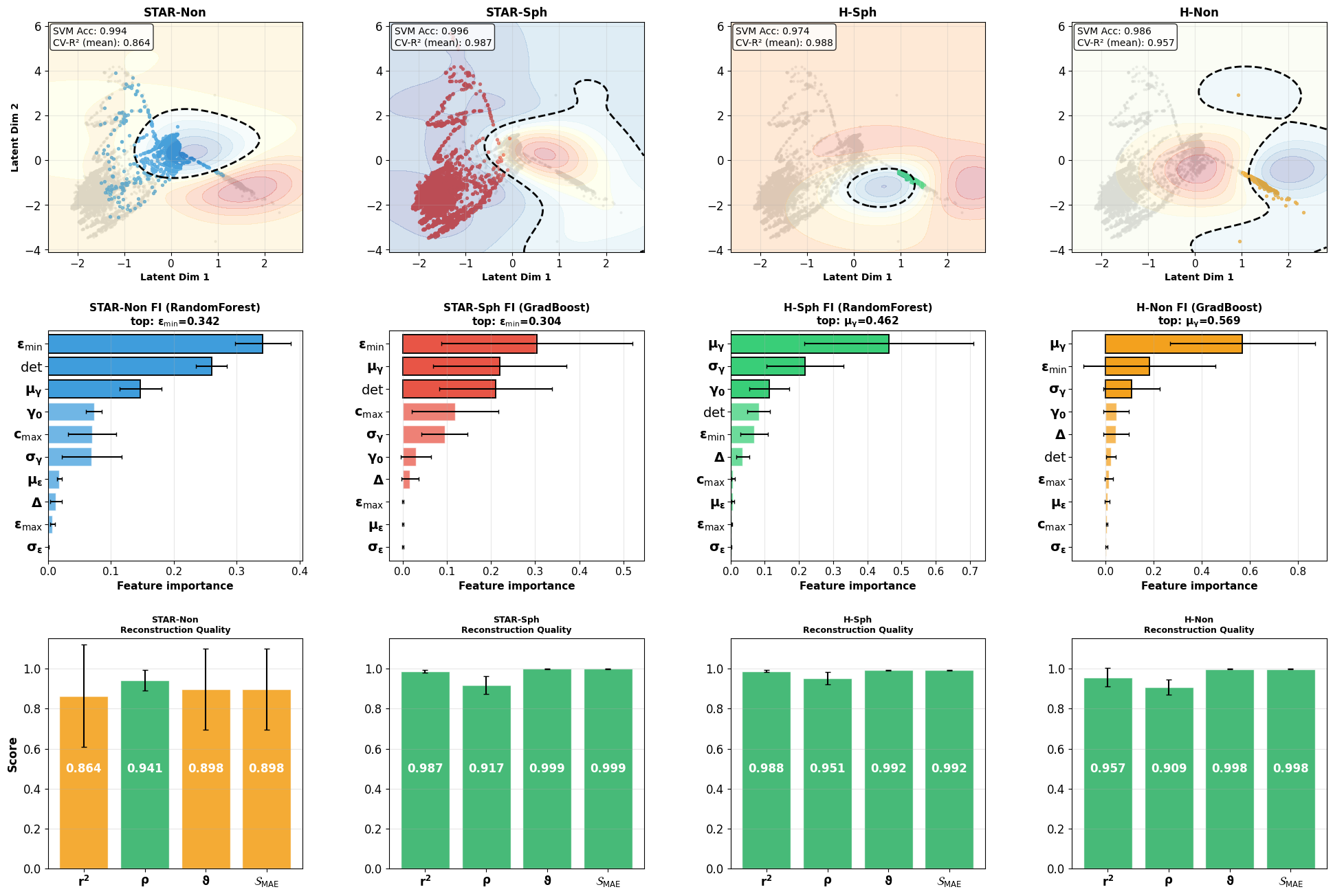}
    \caption{
    One-vs-rest decision-boundary reconstruction in the four-class contrastive latent representation.  Columns correspond to the four classes STAR-Non, STAR-Sph, H-Sph, and H-Non.  The top row shows the \texttt{seed}-\(42\) latent-space projection with an RBF-SVM one-vs-rest decision function: colored points are the positive class, gray points are the complementary classes, filled contours show the SVM decision value, and the dashed black curve is the zero-level decision boundary.  The middle row reports the multi-seed mean feature importances of the cross-validation-selected reconstruction model for the corresponding boundary, with error bars showing seed-to-seed standard deviation.  The bottom row reports multi-seed reconstruction quality using \(R^2\), Spearman correlation \(\rho\), sign accuracy \(\vartheta\), and boundary agreement \(\mathcal{S}_{\rm MAE}\).  Together, the panels show that the latent one-vs-rest boundaries are largely reconstructible from the abstract features, with STAR-side boundaries emphasizing \(\varepsilon_{\rm min}\) and \(\det\), and H-side boundaries emphasizing gap-statistic features such as \(\mu_\gamma\) and \(\sigma_\gamma\).  The STAR-Non boundary is the least stable across seeds and accounts for most of the variance in the reconstruction diagnostics.
    }
    \label{fig:full-latent-RF}
\end{figure}

To capture the nonlinear geometry of the four-class problem, we then train one-vs-rest RBF-SVM classifiers in the two-dimensional latent plane.  The resulting decision functions are shown in Fig.~\ref{fig:full-latent-RF}.  The top row displays the seed-\(42\) latent projection and the corresponding one-vs-rest SVM decision values.  The dashed curves mark the zero-level decision boundaries.  These curves show that the four-class separation is not globally linear: the binary topology and homology axes provide a useful coarse coordinate system, but the individual class boundaries bend around the local cluster geometry.

To relate these nonlinear SVM boundaries to interpretable quantities, we reconstruct each one-vs-rest decision function from the ten abstract features.  For each class and each seed, we scan a small family of reconstruction models, including random forests, gradient boosting, linear regression, ridge regression, and a simple MLP.  Model selection is performed on the validation split.  Reconstruction quality is summarized by the composite score
\( Q_{\mathrm{composite}},\)
which combines the coefficient of determination \(R^2\), Spearman rank correlation \(\rho\), sign accuracy \(\vartheta\), and the normalized MAE-complement boundary score \(S_{\mathrm{MAE}}\), as defined in \eqref{eq:Q-composite}.  The resulting multi-seed reconstruction summary is reported in Tab.~\ref{tab:ovr-boundary-reconstruction}.

\begin{table}[ht!]
\centering
\small
\setlength{\tabcolsep}{4pt}
\renewcommand{\arraystretch}{1}
\resizebox{0.945\textwidth}{!}{%
\begin{tabular}{C{0.2\textwidth}|C{0.2\textwidth}|C{0.30\textwidth}|C{0.2\textwidth}}
\toprule
\textbf{OvR Boundary} &
\textbf{Best Model} &
\textbf{Best FI} &
$\mathbf{Q_{\mathrm{composite}}}$  \\
\midrule
STAR-Non & RandomForest & \rankfeat{\mathbf{\varepsilon_{\rm min}}}{0.342\pm0.049}, \rankfeat{\mathbf{\det}}{0.260\pm0.027}, \rankfeat{\mathbf{\mu_{\gamma}}}{0.147\pm0.037} & $0.898\,\pm\,0.201$\\
\hline
STAR-Sph & GradBoost & \rankfeat{\mathbf{\varepsilon_{\rm min}}}{0.304\pm0.241}, \rankfeat{\mathbf{\mu_{\gamma}}}{0.220\pm0.168}, \rankfeat{\mathbf{\det}}{0.211\pm0.142} & $0.975\,\pm\,0.013$\\
\hline
H-Sph & RandomForest & \rankfeat{\mathbf{\mu_{\gamma}}}{0.462\pm0.276}, \rankfeat{\mathbf{\sigma_{\gamma}}}{0.218\pm0.125}, \rankfeat{\mathbf{\gamma_0}}{0.113\pm0.065} & $0.981\,\pm\,0.009$\\
\hline
H-Non & GradBoost & \rankfeat{\mathbf{\mu_{\gamma}}}{0.569\pm0.337}, \rankfeat{\mathbf{\varepsilon_{\rm min}}}{0.182\pm0.306}, \rankfeat{\mathbf{\sigma_{\gamma}}}{0.108\pm0.131} & $0.963\,\pm\,0.023$\\
\bottomrule
\end{tabular}%
}
\caption{
Multi-seed reconstruction summary for the one-vs-rest decision boundaries in the four-class contrastive latent space.  The top-feature column reports the three largest feature importances of the cross-validation-selected reconstruction model, with mean \(\pm\) standard deviation over seeds displayed below each feature.  The composite score \(Q_{\mathrm{composite}}\) is the weighted combination of the four reconstruction diagnostics defined in Appendix~\ref{app:interpretability_robustness}.  All entries are averaged over the seed ensemble.
}
\label{tab:ovr-boundary-reconstruction}
\end{table}

The reconstruction results show that the one-vs-rest boundaries are strongly, but not uniformly, captured by the ten-feature dictionary.  STAR-Sph, H-Sph, and H-Non have high composite scores,
\[
Q_{\mathrm{composite}}=0.975\pm0.013,\qquad
0.981\pm0.009,\qquad
0.963\pm0.023,
\]
respectively.  STAR-Non is weaker and more seed-dependent, with
\[
Q_{\mathrm{composite}}=0.898\pm0.201.
\]
This is consistent with the bottom row of Fig.~\ref{fig:full-latent-RF}, where the STAR-Non reconstruction diagnostics have the largest error bars.

The feature importances reveal a clear class-dependent structure.  The STAR-side one-vs-rest boundaries emphasize low-end spectral and determinant information: STAR-Non is reconstructed primarily by \(\varepsilon_{\min}\), \(\det\), and \(\mu_\gamma\), while STAR-Sph is reconstructed by \(\varepsilon_{\min}\), \(\mu_\gamma\), and \(\det\).  By contrast, the H-side boundaries emphasize gap statistics.  H-Sph is dominated by \(\mu_\gamma\), followed by \(\sigma_\gamma\) and \(\gamma_0\), while H-Non is dominated by \(\mu_\gamma\), with \(\varepsilon_{\min}\) and \(\sigma_\gamma\) appearing as secondary features.

Synthesizing the binary and one-vs-rest analyses, the multi-class latent space is organized by two coarse physical directions but resolved by class-specific nonlinear boundaries.  At the binary level, topology is separated very cleanly and is primarily associated with gap statistics, while homology is associated with determinant and low-end spectral data.  At the one-vs-rest level, STAR boundaries retain \(\varepsilon_{\min}\) and \(\det\), whereas H-graph boundaries are governed primarily by \(\mu_\gamma\) and related gap statistics.  Thus the contrastive latent representation does not merely memorize the four labels: it arranges the classes in a low-dimensional geometry whose boundaries are largely reconstructible from the same determinant, spectral, and \(q\)-series support features identified throughout the preceding experiments.

\subsubsection{Discussion of Latent Space Results}

Synthesizing the results across the latent space experiments, the most defining characteristic is the convergence of interpretability. By filtering out high-variance noise, the contrastive learning objective aligns gradient sensitivity with global relevance, revealing a "dual-channel" mechanism where the network disentangles the problem into two orthogonal physical axes: a homology axis defined by $\varepsilon_{\rm min}$ and a topology axis defined by $\mu_\gamma$. Crucially, the spontaneous emergence of this clean separation within a universal two-dimensional latent manifold, regardless of the initial bottleneck size, provides strong evidence that these two variables capture the intrinsic, irreducible coordinates of the $\widehat{Z}$-invariant's topological phase space.

\subsection{Control Experiment: Derived Features}
\label{subsec:classification-using-features}

In the preceding experiments, the classifiers received raw or learned representations of the truncated \(\widehat Z\) series, and the interpretability pipeline was used to infer which abstract quantities were reconstructed by the network.  We now perform a direct control experiment: instead of giving the classifier the \(q\)-series, we give it only the ten derived features used throughout the regression and attribution analyses,
\[
(\det,\varepsilon_{\min},\varepsilon_{\max},\mu_\varepsilon,\sigma_\varepsilon,
\Delta,\gamma_0,\mu_\gamma,\sigma_\gamma,c_{\max}).
\]
This asks whether these features are sufficient to solve the same homology and topology tasks, and whether the resulting classifiers rely on the same features identified indirectly in the raw-\(q\)-series experiments.

For each task, we train an MLP classifier with architecture
\((\text{depth},\text{width})=(4,128)\), using the same train/test protocol and early-stopping criteria as in the preceding experiments.  Since the classifier inputs are already the interpretable features, there is no separate feature-to-logit regression stage in this control.  Instead, feature saliency, permutation importance, and SHAP are applied directly to the trained feature classifier.  The multi-seed results are summarized in Tab.~\ref{tab:summary-5}.

\begin{table}[ht!]
\centering
\scriptsize
\setlength{\tabcolsep}{3pt}
\resizebox{0.8\textwidth}{!}{%
\begin{tabular}{c|c|ccc}
\toprule
\textbf{Experiment} &
\textbf{BAS} &
\textbf{FS} &
\textbf{PI} &
\textbf{SHAP} \\
\midrule
HOM-Star & $1.000\,\pm\,0.000$ & \rankfeat{\varepsilon_{\rm min}}{1.0}, \rankfeat{\mu_{\gamma}}{2.2}, \rankfeat{\Delta}{2.8} & \rankfeat{\varepsilon_{\rm min}}{1.0}, \rankfeat{\mu_{\gamma}}{2.0}, \rankfeat{\det}{3.6} & \rankfeat{\varepsilon_{\rm min}}{1.0}, \rankfeat{\mu_{\gamma}}{2.0}, \rankfeat{\sigma_{\varepsilon}}{3.3} \\
\hline
HOM-H & $0.998\,\pm\,0.001$ & \rankfeat{\varepsilon_{\rm min}}{1.2}, \rankfeat{\det}{1.8}, \rankfeat{\sigma_{\gamma}}{3.0} & \rankfeat{\varepsilon_{\rm min}}{1.0}, \rankfeat{\det}{2.0}, \rankfeat{\sigma_{\gamma}}{3.4} & \rankfeat{\varepsilon_{\rm min}}{1.4}, \rankfeat{\det}{1.6}, \rankfeat{\sigma_{\gamma}}{3.2} \\
\hline
HOM-All & $0.995\,\pm\,0.001$ & \rankfeat{\varepsilon_{\rm min}}{1.0}, \rankfeat{\Delta}{2.2}, \rankfeat{\sigma_{\gamma}}{2.8} & \rankfeat{\varepsilon_{\rm min}}{1.0}, \rankfeat{\varepsilon_{\rm max}}{2.5}, \rankfeat{\sigma_{\gamma}}{3.6} & \rankfeat{\varepsilon_{\rm min}}{1.0}, \rankfeat{\sigma_{\gamma}}{2.0}, \rankfeat{\varepsilon_{\rm max}}{4.1} \\
\hline
TOP-Sph & $1.000\,\pm\,0.000$ & \rankfeat{\mu_{\gamma}}{1.0}, \rankfeat{\sigma_{\gamma}}{2.0}, \rankfeat{\Delta}{3.4} & \rankfeat{\mu_{\gamma}}{1.0}, \rankfeat{\sigma_{\gamma}}{2.0}, \rankfeat{c_{\rm max}}{3.4} & \rankfeat{\mu_{\gamma}}{1.0}, \rankfeat{\sigma_{\gamma}}{2.0}, \rankfeat{\varepsilon_{\rm max}}{3.5} \\
\hline
TOP-Non & $1.000\,\pm\,0.000$ & \rankfeat{\varepsilon_{\rm min}}{1.0}, \rankfeat{\gamma_0}{2.0}, \rankfeat{\Delta}{3.4} & \rankfeat{\varepsilon_{\rm min}}{1.0}, \rankfeat{\varepsilon_{\rm max}}{2.6}, \rankfeat{\sigma_{\gamma}}{2.6} & \rankfeat{\varepsilon_{\rm min}}{1.0}, \rankfeat{\sigma_{\varepsilon}}{2.4}, \rankfeat{\sigma_{\gamma}}{2.8} \\
\hline
TOP-All & $1.000\,\pm\,0.000$ & \rankfeat{\sigma_{\gamma}}{1.0}, \rankfeat{\gamma_0}{2.0}, \rankfeat{\mu_{\gamma}}{3.2} & \rankfeat{\sigma_{\gamma}}{1.0}, \rankfeat{\mu_{\gamma}}{2.0}, \rankfeat{\varepsilon_{\rm min}}{3.0} & \rankfeat{\sigma_{\gamma}}{1.0}, \rankfeat{\mu_{\gamma}}{2.1}, \rankfeat{\varepsilon_{\rm min}}{3.5} \\
\bottomrule
\end{tabular}%
}
\caption{
Multi-seed direct feature-classifier results.  The classifier input is the ten-feature vector
\((\det,\varepsilon_{\min},\varepsilon_{\max},\mu_\varepsilon,\sigma_\varepsilon,
\Delta,\gamma_0,\mu_\gamma,\sigma_\gamma,c_{\max})\) rather than the truncated
\(\widehat Z\) sequence.  BAS is reported as mean \(\pm\) standard deviation over
classifier seeds.  The final columns report the three lowest mean feature ranks
for feature saliency (FS), permutation importance (PI), and SHAP, with the
average rank displayed below each feature.
}
\label{tab:summary-5}
\end{table}

The feature controls solve all six tasks with very high accuracy.  All topology controls reach perfect balanced accuracy within numerical precision, and the homology controls achieve BAS between \(0.995\pm0.001\) and \(1.000\pm0.000\).  This confirms that the ten derived quantities are sufficient to recover the class labels in the present datasets.  The more informative question is therefore not whether the features contain enough information, but which features the direct classifier uses once the extraction problem has been removed.

For the homology tasks, the direct classifiers are dominated by the low-end spectrum.  In HOM-Star, HOM-H, and HOM-All, \(\varepsilon_{\min}\) is ranked first by all three attribution methods, up to small seed fluctuations.  The determinant appears as a secondary feature most clearly in HOM-H, while HOM-Star and HOM-All also use gap quantities such as \(\mu_\gamma\), \(\Delta\), and \(\sigma_\gamma\).  Thus the direct-feature homology classifier does not simply implement the exact rule \(|\det M|=1\).  Instead, it learns a low-dimensional spectral proxy in which \(\varepsilon_{\min}\) is the most stable feature, with determinant and support-spacing information providing secondary corrections.

For topology classification, the control experiment separates the homology-sphere and non-sphere sectors.  In TOP-Sph, where the determinant is fixed, the classifier relies almost entirely on the \(q\)-series gap profile: \(\mu_\gamma\) is ranked first and \(\sigma_\gamma\) second by FS, PI, and SHAP.  This directly supports the interpretation that, in the fixed-determinant topology task, star-shaped and \(H\)-shaped graphs are distinguished through the spacing structure of the \(q\)-series.

The TOP-Non control behaves differently.  Although the classifier again reaches perfect BAS, all three attribution methods rank \(\varepsilon_{\min}\) first.  Gap features remain present, with \(\sigma_\gamma\), \(\gamma_0\), or \(\Delta\) appearing among the secondary features, but the dominant control signal is low-end spectral rather than purely gap-based.  This is consistent with the full-sequence TOP-Non analysis, where determinant and eigenvalue information remained important once determinant variation was no longer fixed.

Finally, in TOP-All, the pooled topology control returns to a gap-dominated rule.  All three attribution methods rank \(\sigma_\gamma\) first, and PI and SHAP rank \(\mu_\gamma\) second.  The feature \(\varepsilon_{\min}\) remains visible as a secondary predictor, but the dominant pooled topology signal is the dispersion of the \(q\)-series gaps.  This suggests that, once the derived features are provided explicitly, the most stable topology discriminator across the full dataset is the gap profile, even though the non-spherical sector alone can be solved by a low-end spectral shortcut.

\paragraph{Comparison with \(q\)-series models.}
The direct-feature controls clarify the role of the interpretability pipeline.  The raw \(q\)-series classifiers must first extract relevant determinant, spectral, and support-spacing information from the truncated series.  Depending on representation and sector, they sometimes emphasize high-variance spectral proxies, especially in the non-spherical topology tasks.  When the ten derived features are supplied directly, however, the learned rules become sharper and more task-specific: homology classification is organized primarily by \(\varepsilon_{\min}\), with determinant information secondary, while topology classification is organized by gap statistics in the homology-sphere and pooled sectors, with a low-end spectral shortcut persisting in TOP-Non.  The control experiment therefore supports the main conclusion of Sec.~\ref{sec:denseRep}: the classifiers are not using opaque artifacts of the raw input representation, but are largely driven by a small set of determinant, spectral, and \(q\)-series support features.

\section{Cobordism Experiments}
\label{sec:cobordism-experiments}

\subsection[\texorpdfstring{Learning $d$ from $\widehat{Z}$}{Learning d from Zhat}]{\texorpdfstring{Learning $\boldsymbol{d}$ from $\boldsymbol{\widehat{Z}}$}{Learning d from Zhat}}

\subsubsection{Theoretical Background and Estimates}\label{subsec:theoretical-background-d}

To set the stage for our results, we first review the known connections between $\widehat{Z}$ and the Heegaard Floer correction term $d$. One implicit relation discovered in \cite{cobordism_invariants_bps} links the radial limit of $\widehat{Z}$ at $q \to i$ to the invariant $d  \pmod 4$:
\begin{equation}\label{eq:zhat-d}
    \frac{1}{2\pi} \operatorname{arg} \widehat{Z}\big|_{q \to   i} \equiv \frac{1}{4} + \frac{3}{4}d \pmod 1.
\end{equation}
Among the connections between $\widehat{Z}$ and $d$ studied so far, the relation between $\Delta$ and $d$ has received the most attention. For integer homology spheres \cite{cobordism_invariants_bps} showed that: \[\Delta \equiv \frac{1}{2} + d \pmod 1.\]
The interested reader may also find a $\operatorname{Spin}^c$-dependent version of this relation for rational homology spheres in \cite{cobordism_invariants_bps}. Thereafter, \cite{harichurn2024deltaainvariantsnonperturbativecomplex} gave an explicit formula for $\Delta$ for Brieskorn spheres and compared $\Delta$ and $d$ across various families of Brieskorn spheres, including some homology cobordant to $S^3$. Most recently \cite{delta_plumbed_manifolds} pushed this comparison further, showing that for Seifert homology spheres $Y =\Sigma(b_1, \dots, b_r)$
\begin{equation}\label{delta}
    \Delta(Y) = \frac{1}{2} - \frac{\kinv(Y)}{4}\end{equation}
where $\kinv(Y)$, written $K^2 + s$ in the prior literature \cite{Borodzik2013}, is the invariant recalled below. Additionally \cite{delta_plumbed_manifolds} further developed the comparisons between $d$ and $\Delta$. We build on this story in this section.

We first introduce Dedekind sums following the notation in \cite{Borodzik2013} as they will be useful in the formulas to appear. The generalized Dedekind sum is defined for $p \in \mathbb{Z} \setminus \{0\}$, $q \in \mathbb{Z}_{\geq 1}$, and $x, y \in \mathbb{R}$ by:
\begin{equation}
    s\left(p, q;x, y\right) = \sum_{i=0}^{q-1} \left\langle\frac{i+y}{q}\right\rangle \left\langle\frac{p(i+y)}{q} + x\right\rangle
\end{equation}
with the classical Dedekind sum being the specialization $s(p, q) := s(p, q;0, 0)$. In the above $\langle \cdot \rangle$ is the sawtooth function defined for $x \in \mathbb{R}$ by $\langle x\rangle = x - \lfloor x \rfloor - \frac{1}{2}$ if $x \not\in \mathbb{Z}$ and $0$ otherwise. Expressing the manifold $Y$ as the Seifert manifold $Y=M(b; \frac{a_1}{b_1}, \dots, \frac{a_r}{b_r})$, we have the following expression for $\kinv$: 
\begin{equation}\label{eq:kinv}
    \kinv(Y) = \left(2-r + \sum_{i=1}^r \frac{1}{b_i}\right)^2\frac{1}{e} + e + 5 - 12\sum_{i=1}^rs(a_i, b_i)
\end{equation}
where $e$ is the orbifold Euler number of $Y$. In \cite{Borodzik2013} Borodzik and N\'emethi showed: 
\begin{equation}\label{d}
    d(Y) = \frac{\kinv(Y)}{4} - 2 \min_{k \geq 0} \tau(k)
\end{equation}
which together with 
 \eqref{delta} implies:
\begin{equation}\label{approx-ii}
    d = - \Delta + \frac{1}{2} - 2\min_{k \geq 0} \tau(k).
\end{equation}
In the above, 
 $\tau$ is defined as 
\begin{equation}\label{eq:tau-base}
    \tau(k) = \sum_{j=0}^{k-1} \delta_j \quad  \text{and} \quad  \delta_j := 1-jb -\sum_{i=1}^r \bigg\lceil \frac{j a_i}{b_i} \bigg\rceil.
\end{equation}
Let $P = b_1\cdots b_r$ and $\hat{b}_i = P/b_i$; there is an alternative formula for $\tau$ (cf. \cite[eq. 3.6]{Borodzik2013}) that we find instructive:
\begin{align}\label{eq:tau-alt}
\begin{split}
   \tau(k) = &\frac{k(k-1)}{2P} + k\left(1 - \frac{r}{2}\right) + \frac{r}{2} \\ + &\sum_{i=1}^r \left(-\frac{1}{2} \left\langle\frac{k a_i}{b_i}\right\rangle + \frac{1}{2}\left\lfloor \frac{k-1}{b_i} \right\rfloor - s\left(\hat{b}_i, b_i;\frac{k}{b_i}, 0\right)+s\left(\hat{b}_i, b_i\right)\right). 
\end{split}    
\end{align}
We now fix $r = 3$ for the remainder of this subsubsection. By inspecting \eqref{eq:tau-alt} one can show, as we do in Appendix \ref{appendix:bound-min-tau}, that:
\begin{equation}\label{eq:tau-approx}
    \bigg|\min_{k\geq 0} \tau(k) + \frac{P}{8}\left(1 - \sum_{i=1}^3 \frac{1}{b_i}\right)^2\bigg| \leq \sum_{i=1}^3 b_i
\end{equation}
Combining \eqref{approx-ii} with the behavior of $\min \tau$ we see:
\begin{equation}\label{approx}
    \bigg | d +   \Delta - \frac{1}{2} - \frac{P}{4}\left(1 - \sum_{i=1}^3 \frac{1}{b_i}\right)^2\bigg| \leq 2\sum_{i=1}^3 b_i,
\end{equation}
which, as far as we know, is a new observation in the $\widehat{Z}$ literature. Interestingly, by comparing the expression for $\Delta$ in \cite[eq. 7.1]{delta_plumbed_manifolds} with \eqref{eq:tau-approx}, one sees that $\Delta$ and $|2\min_{k \geq 0} \tau(k)|$ are similar in magnitude. In comparison, $d$ is much smaller than either quantity (cf. \eqref{eq:d-estimate}) and varies erratically as a function of the $b_i$ due to the presence of Dedekind sums as seen in \eqref{eq:tau-alt}.

\subsubsection{Regression of $d$ from $\widehat Z$ Exponents}

\subsubsection*{The Experiment}
Given the relation \eqref{approx-ii}, we design the regression task around $d+\Delta-\frac12=-2\min_{k\geq0}\tau(k)$. Writing
\begin{equation}
    \widehat{Z}(Y;q) = q^{\Delta}\left(1 + c_{n_1}q^{n_1} + c_{n_2}q^{n_2} + \cdots\right)
\end{equation}
where $c_{n_i} \in \mathbb{Z}$ are all nonzero and $0<n_1<n_2<\cdots$ with $n_i \in \mathbb{N}$, we ask whether the leading exponent $\Delta$ and the ordered nonzero exponent offsets $n_i$ contain enough information to predict the $\tau$-function contributions.  This formulation is mathematically equivalent to predicting $d$, since $d$ is recovered from \eqref{approx-ii}. We emphasize, however, that this is not the same as asking a generic smooth regressor to learn the full correction term formula directly from the raw Seifert parameters $(b_1,b_2,b_3)$: the latter involves floor, ceiling, and Dedekind-sum corrections, and is not well approximated by the simple smooth models tested here.

Two families of datasets enter this section. The main three-fiber dataset $\mathcal{D}_3^{20}$ starts from the
pairwise-coprime triples $2 \le b_1 < b_2 < b_3 \le 44$ and keeps the $685$
manifolds whose first twenty nonzero $\widehat{Z}$-coefficients occur at
exponents at most $1.28\times10^4$, yielding the leading exponent $\Delta$
and nineteen subsequent offsets per manifold; in effect this restricts
$b_1 b_2 b_3 \le 2442$. The per-fiber-count sets $\mathcal{D}_r$, $r = 3, 4, 5$, consist of the Seifert
fibered integer homology spheres with pairwise-coprime multiplicities
$2 \le b_1 < \dots < b_r \le 45,\ 30,\ 20$ for $r = 3, 4, 5$ respectively,
oriented so that the orbifold Euler number is negative, generated independently
under a deeper extraction (first one hundred nonzero coefficients at exponents below $3.3\times10^{6}$; see the reproduction notebook \cite{harichurn_2026_18734903}), giving $n = 3{,}020$, $1{,}743$, and $306$
manifolds respectively. 

All results in this section and the following subsection use $\mathcal{D}_3^{20}$, except where a $\mathcal{D}_r$ is named explicitly: Table~\ref{tab:poly2-fiber-scaling}, the Kernel-Ridge $\Delta$-augmentation, the cross-$r$ transfer, and the prefix-scaling analysis. Throughout, a model's input is a prefix of this offset sequence, the first $L$ offsets $(n_1, \ldots, n_L)$, optionally augmented by $\Delta$ and further features as stated; we call $L$ the prefix length.

\subsubsection*{Results}
We began with a sweep over regressor classes and input dimensions.  Throughout, each reported \(R^2\) is the mean \(\pm\) standard deviation over ten random seeds; for each seed we average the held-out \(R^2\) over a freshly shuffled five-fold partition.  For the deterministic models reported here, the fold partition is the only seed-dependent quantity. Exact-integer accuracies follow the same scheme, scored once per seed on the pooled out-of-fold predictions (one held-out prediction per manifold). 

On the three-fiber Brieskorn dataset, using the first four exponent offsets \((n_1,\ldots,n_4)\), the strongest models were a degree-three polynomial regression and Kernel Ridge regression with an RBF kernel, achieving respectively \(R^2=0.9995 \pm 0.0000\) and \(R^2=0.9993 \pm 0.0002\), followed closely by degree-two polynomial regression with \(R^2=0.9990\pm0.0000\).  Linear regression remained lower, at \(R^2=0.9453 \pm 0.0004\).  Adding more exponent offsets did not systematically improve these scores, and in some cases degraded performance, indicating that the smooth, large-scale variation of \(-2\min_{k\geq0}\tau(k)\) is already captured by a short exponent prefix.  The performance of the degree-two model suggests that the dominant trend is essentially quadratic in these leading offsets, although, as discussed below, the exact integer-valued correction contains finer number-theoretic structure not captured by such a smooth approximation alone. 

Feature-importance analysis with three attribution methods (permutation importance, mean $|$SHAP$|$, and MLP gradient saliency (FS), each recomputed over ten re-drawn train/test splits) yields the consensus hierarchy
$n_3 > n_1 > n_4 > n_2$ (mean ranks $1.50$, $2.07$, $2.43$, $4.00$). Permutation importance and $|$SHAP$|$ reproduce this ordering on every split, while gradient saliency instead ranks $n_4$ first with $n_1$ and
$n_3$ statistically indistinguishable, as the FS bars of
Figure~\ref{fig:learning-d-from-zhat} show (per-split rank tables in \cite{harichurn_2026_18734903}). The dominance of $n_3$ is consistent with the arithmetic structure of the Brieskorn false-theta expansion: among the early offsets, $n_3$ typically mixes several Seifert parameters, heuristically $n_3 \approx (b_1-1)(b_2+b_3)$ on much of the dataset. Separately, degree-two polynomial regression reconstructs $\kinv(Y)/4$ from
$(n_1,\ldots,n_4)$ with $R^2 = 0.9979 \pm 0.0000$ and $-2\min_k \tau(k)$ with $0.9990 \pm 0.0000$.

To test whether this phenomenon persists beyond three singular fibers, we expanded the dataset to Seifert fibered integer homology spheres with \(r=4\) and \(r=5\) exceptional fibers.  The same qualitative pattern persists: low-degree polynomial and kernel models achieve very high \(R^2\) from short exponent prefixes, and adding \(\Delta\) further improves the reconstruction.  The number of exponent offsets required to reach a fixed high-accuracy threshold grows with \(r\). Under the Kernel Ridge RBF protocol, the smallest prefix length
reaching \(R^2 \ge 0.999\) is \(L_{99.9\%} = 4, 8, 15\) for
\(r = 3, 4, 5\), identical across all ten seeds. This is consistent with the
quadratic count \(r^2 - 2r\), which matches exactly at \(r = 4, 5\)
and is off by one at \(r = 3\) (empirically \(4\), not \(3\)).  We treat this as an empirical scaling law rather than a theorem: the higher-\(r\) regimes have fewer samples and larger candidate feature spaces, and the threshold estimates depend on the model class and available prefix length.

Importantly, the polynomial degree need not grow with \(r\) in these experiments.  A degree-two polynomial remains highly effective across all tested fiber counts, especially when \(\Delta\) is included as an additional feature.  Higher-degree polynomials, by contrast, are more prone to overfitting because the number of cross-terms grows rapidly relative to the available sample size.

\begin{table}[ht]
  \centering
  \begin{tabular}{l|c|c|c}
    \toprule
     & $r=3$ & $r=4$ & $r=5$ \\
    \midrule
    Poly(2), offsets only (best over $L$) & $1.0000 \pm 0.0000$ & $1.0000 \pm 0.0000$ & $0.9950 \pm 0.0019$ \\
    Poly(2), offsets $+\,\Delta$          & $1.0000 \pm 0.0000$ & $1.0000 \pm 0.0000$ & $1.0000 \pm 0.0000$ \\
    \bottomrule
  \end{tabular}
  \caption{Cross-validated $R^2$ for a degree-2 polynomial regressing
  $d+\Delta-\tfrac{1}{2}$ on the $\widehat{Z}$ exponent offsets of Seifert fibered
  integer homology spheres with $r$ exceptional fibers ($n=3020,1743,306$ for
  $r=3,4,5$). The best-over-$L$ scores in the
offsets-only row are attained at $L = 6$, $12$, and $16$ for
$r = 3, 4, 5$ respectively. The $+\Delta$ entries are evaluated at fixed base prefix lengths
$L = 4, 7, 10$ for $r = 3, 4, 5$: the empirical Poly(2)
$R^2 \ge 0.99$ crossings for $r = 3, 4$, with $L = 10$ chosen for
$r = 5$, whose crossing is unstable at $n = 306$. These evaluation
defaults are distinct from the $99.9\%$ thresholds $L_{99.9\%}$
discussed in the text.}
  \label{tab:poly2-fiber-scaling}
\end{table}

The leading exponent \(\Delta\) carries additional nontrivial
information. On the \(\mathcal{D}_r\) datasets, at the base prefix
lengths of Table~\ref{tab:poly2-fiber-scaling}, adding \(\Delta\) to the offset array improves the Kernel Ridge reconstruction of
\(d+\Delta-\frac12\): in the \(r=3\) case the score increases from \(0.9998\) using offsets alone to \(1.0000\) at the reported precision, while for \(r=4\) it increases from \(0.9943\) to \(0.9999\), and for \(r=5\) from \(0.9928\) to \(0.9994\). On \(\mathcal{D}_3^{20}\) at \(L=4\), adding \(\Delta\) also
raises the degree-two polynomial reconstruction of \(\kinv(Y)/4\) to \(1.0000\pm0.0000\), and the reconstruction of \(-2\min_k\tau(k)\) to \(0.9997\pm0.0000\).

The exact coefficients of the quadratic map are fiber-count-dependent. A polynomial trained on three-fiber
manifolds does not transfer directly to four-fiber manifolds, giving \(R^2<0\). However, transfer between
adjacent higher fiber counts can succeed: an \(r=4\) model evaluated on \(r=5\) gives \(R^2=0.6485\).  This asymmetry is consistent with the possibility that lower-fiber polynomial relations arise as specializations of a higher-dimensional arithmetic structure, although we regard this as suggestive rather than conclusive.

The exponent offsets encode more than the scalar correction term alone.  In the three-fiber experiments, from the first four offsets, \(\min_k\tau(k)\) is reconstructed with \(R^2=0.9990\), the range of \(\tau\) with \(R^2=0.9998\), and the location of the minimum \(k^*=\arg\min_{k\geq0}\tau(k)\) with \(R^2=0.9951\).  Even after projecting out the variance linearly correlated with the \(d\)-invariant, the exponent offsets still explain \(96\%\) of the variance in \(k^*\), indicating that the exponent spectrum carries information about the shape of the \(\tau\)-function beyond the value of \(d\) itself.

In summary, for the Seifert fibered integer homology spheres in our datasets, the macroscopic shape of \(d+\Delta-\frac12\) is predominantly captured by a degree-two polynomial in a short prefix of the exponent support, supplemented by the leading exponent \(\Delta\).  The exponent spectrum of \(\widehat Z\) also encodes several related geometric quantities, including the minimum, argmin, and range of the \(\tau\)-function, as well as \(\kinv(Y)/4\).

These high-\(R^2\) results should be interpreted as smooth approximation results rather than exact arithmetic reconstruction.  The success of the exponent-offset features is nevertheless consistent with a more rigid arithmetic mechanism.  In the three-fiber Brieskorn case, the false-theta sector shifts contain pairwise combinations of the Seifert multiplicities, such as
\[
    b_i b_j-b_i-b_j,
\]
and once the three pairwise offsets are identified one can recover \((b_1,b_2,b_3)\) by elementary square-root formulas.  The ordered exponent offsets are not labeled by sector, and the pairwise offsets need not coincide with the first three ordered offsets because other sector contributions can interleave.  However, direct enumeration on the finite three-fiber dataset verifies that the prefix \((n_1,n_2,n_3,n_4)\) is injective in the Brieskorn triple.  For \(r>3\), we treat the analogous injectivity statement as empirical: the observed scaling in the number of required offsets suggests that a small prefix of the exponent support captures enough Seifert arithmetic to approximate the \(\tau\)-function contribution to \(d\), but we do not claim a closed-form inverse in general.

\subsubsection{Towards Exact Integer Predictions}

Despite the near-perfect \(R^2\) scores, the polynomial predictions are continuous, whereas the target
\[
    t=-2\min_{k\geq0}\tau(k)
\]
is integer-valued, and in this setting lies in \(2\mathbb Z\). Directly rounding the degree-two polynomial prediction for \(t\) from the four offsets alone gives exact agreement on only \(17.9\%\pm0.7\%\) of the dataset. The residuals are tightly bounded, with in-sample standard deviation
approximately \(2.30\), maximum \(8.5\), and cross-validated MAE
\(1.86\), but they are not well explained by the tested polynomial
functions of the exponent offsets. This is consistent with the structure of \(\tau(k)\), whose floor and ceiling terms, together with the Dedekind-sum contributions in the Seifert formula, introduce arithmetic corrections beyond the smooth quadratic trend. Appending $\Delta$ as a fifth input feature doubles the exact-recovery rate to
$36.1\% \pm 0.8\%$, the leftmost bar of
Figure~\ref{fig:exact_error_profile}.

To isolate the finer arithmetic correction, we modified the target and augmented the feature set.  First, we trained the regressor on \(t/2\in\mathbb Z\) rather than \(t\in2\mathbb Z\), so that rounding errors are measured on the natural integer scale of the half-target, lifting exact recovery to \(61.5\%\pm1.0\%\) (the ``\(\mathbin{\mathchar"2204}2\)'' bar).
Separately, widening the input from the first four offsets to all nineteen lifts this further to \(73.8\%\pm0.5\%\) (``+all offsets''), recovering nearly half of what the Dedekind augmentation below achieves from the first four alone.
Second, returning to the first four offsets, we supplemented the exponent features with symmetrized Dedekind-sum features derived from the Seifert data: writing $P=b_1b_2b_3$, these
are $D_i = s(P/b_i,\,b_i)$ for $i=1,2,3$ (each a symmetric function of the two
complementary multiplicities; up to sign, the sums $s(a_i,b_i)$ of
\eqref{eq:kinv}) together with their total $D_1+D_2+D_3$.  The symmetrization is important: raw, unsymmetrized Dedekind sums do not respect the permutation symmetry of the Brieskorn triple and empirically degrade the exact recovery rate.  Using the nine-dimensional feature set consisting of \(\Delta\), four exponent offsets \((n_1,\ldots,n_4)\), and four symmetrized Dedekind-sum features, a degree-two polynomial expansion followed by linear regression gives exact integer recovery of \(d\) on \(88.6\%\pm0.6\%\) of the Brieskorn dataset.

The remaining failures are not catastrophic.  In the failed cases, the predicted \(t/2\) differs from the true value by exactly \(\pm1\).  These errors concentrate among manifolds with large Seifert multiplicities, large product \(b_1b_2b_3\), and large leading exponent \(\Delta\).

We also tested whether the Dedekind-sum features used in this arithmetic augmentation are themselves recoverable from the \(\widehat Z\)-exponent data.  This auxiliary experiment separates three notions: finite identifiability, smooth regression, and arithmetic reconstruction.  First, the short prefix \((n_1,n_2,n_3,n_4)\) has no collisions in the finite three-fiber Brieskorn dataset, either in the Brieskorn triple \((b_1,b_2,b_3)\) or in the associated Dedekind-feature vector.  Thus, on this dataset, the Dedekind features are not independent external information; they are determined by the leading exponent support.  Second, we trained regressors from longer prefixes \((\Delta,n_1,\ldots,n_L)\) to the Dedekind features.  Smooth recovery improves with \(L\), but remains far from exact for the full Dedekind vector: the aggregate \(R^2\) rises from approximately \(0.19\) for \(L=4\) to roughly \(0.41\) for \(L\geq 12\).  The total Dedekind contribution is much more predictable than the individual symmetrized components, indicating that the arithmetic combination relevant for \(d\) is smoother than the full vector of Dedekind data.  Third, nonparametric and parameter-recovery tests show that injectivity does not automatically imply train/test interpolation: nearest-neighbor recovery of held-out Brieskorn triples is poor even though full-catalog collisions vanish, reflecting the arithmetic, non-Euclidean nature of the map.  We therefore interpret the Dedekind features as structured Seifert arithmetic latent in the \(\widehat Z\)-exponent support, but not as information that the tested low-degree regressors extract reliably from short prefixes alone.  Efficient recovery appears to require either longer exponent prefixes together with richer models, or an explicit arithmetic decoding step from exponent support to Seifert data.

\section{Other Experiments}\label{sec:otherexp}

\subsection{Multi-class classification: Singular fibers}

We first extend the binary classification framework to a genuinely multi-class task.  The goal is to determine the number of singular fibers in a Seifert manifold, equivalently the degree of the central node in the corresponding star-shaped plumbing graph.  We generated a balanced dataset of \(30{,}000\) samples, with \(6{,}000\) examples in each of the five classes corresponding to \(3,4,5,6,\) and \(7\) singular fibers.  The classifier was trained on the value-index pair representation and achieved balanced accuracy \(0.9426\) on the test set.

This task is substantially more geometrically complex than the binary experiments in Sec.~\ref{sec:denseRep}.  In the binary setting, the classifier logits can often be interpreted through a single separating direction, and the feature-to-logit regression problem reduces to reconstructing a comparatively simple decision geometry.  Here, the classifier output is five-dimensional, and the logit cloud is organized by multiple pairwise class separations.  Visualizing the logits reveals a piecewise structure resembling intersecting affine regions rather than a single dominant separating direction.  This increased nonlinearity is reflected in the interpretability pipeline: the regression network reconstructs the logits with reduced quality, achieving \(R^2=0.8006\).  Thus the ten-feature dictionary still captures a substantial part of the learned representation, but the multi-class geometry is not as cleanly reducible to a single linear feature mechanism.

Despite this reduction in regression quality, the feature analysis remains informative.  The most stable signals are spectral and support-spacing quantities that vary systematically with the number of arms in the star-shaped graph.  In particular, the eigenvalue dispersion of the plumbing matrix and the gap statistics of the \(q\)-series provide natural summaries of the increasing graph complexity.  As the number of singular fibers changes, both the spectrum of \(M\) and the distribution of exponents in the \(\widehat Z\) series are altered.  The classifier appears to exploit this combined information: spectral-spread features capture changes in the quadratic form determined by \(M^{-1}\), while gap statistics capture the induced changes in the support structure of the truncated \(q\)-series.

The singular-fiber experiment therefore supports two conclusions.  First, the value-index representation remains effective beyond binary tasks, achieving high accuracy even when the output space contains five ordered graph-complexity classes.  Second, the interpretability picture becomes less one-dimensional.  Unlike the binary homology and topology tasks, where determinant, low-end spectral data, or gap statistics often dominate a single separating direction, the singular-fiber task requires a collection of class-dependent boundaries.  The resulting feature analysis should therefore be interpreted as identifying the dominant coordinates of a multi-class decision geometry, not as isolating one universal scalar invariant for the number of singular fibers.

\subsection[\texorpdfstring{Learning $SL_2(\mathbb{Z})$ orbit}{Learning SL(2,Z) orbit}]{\texorpdfstring{Learning $\boldsymbol{SL_2(\mathbb{Z})}$ orbit}{Learning SL(2,Z) orbit}}
\label{subsec:LearningOrbit}

In the final experiment, we move from classification to a regression task motivated by the modular properties of \(\widehat Z\).  The goal is to test whether a network can predict the size of the \(SL_2(\mathbb{Z})\) orbit associated with a \(\widehat Z\)-invariant directly from truncated \(q\)-series data.  See \cite{Cheng:2018vpl,Cheng:2024vou} for discussions of the \(SL_2(\mathbb{Z})\) orbits underlying these invariants.

For a general plumbed \(3\)-manifold there is no simple closed formula for the orbit size.  A useful controlled family is provided by Brieskorn spheres \(\Sigma(b_1,b_2,b_3)\), where the fiber orders \(b_i\) are pairwise coprime and square-free.  In this case, the orbit size \(\alpha\) is determined by the Milnor number:
\begin{align}
    \alpha\left(\Sigma(b_1,b_2,b_3)\right)
    =
    \frac{(b_1-1)(b_2-1)(b_3-1)}{4}
    =
    \frac{\mu}{4}.
\end{align}
We use this family as an in-distribution benchmark for learning the orbit size from the corresponding truncated \(q\)-series.

The dataset consists of \(q\)-series expansions for Brieskorn spheres satisfying the coprime and square-free conditions, ranging from \(\Sigma(2,3,5)\) to \(\Sigma(19,67,115)\).  Since these \(q\)-series are sparse, we computed expansions to order \(O(10^8)\), yielding samples with between \(64\) and \(4619\) nonzero coefficients.  We fixed the input length to \(64\) terms.  The input is a serialized value-index representation in which the coefficient and exponent arrays are concatenated:
\[
[c_0,\ldots,c_{63},e_0,\ldots,e_{63}].
\]
This gives a \(128\)-dimensional input vector.  We focus on this ablated representation, without appending the leading exponent \(\Delta\), so that the network must infer the orbit size from the coefficient--exponent sequence itself rather than from an explicitly supplied leading shift.

We trained a transformer encoder on this flat-block representation using a projection width of \(32\), two encoder layers, four attention heads, and one-dimensional adaptive average pooling.  The input is ordered as
\[
[c_0,\ldots,c_{63},e_0,\ldots,e_{63}],
\]
with scalar token shape \((128,1)\).  The regression target was standardized during training and inverse-transformed for reporting on the original orbit-size scale.  The experiment was repeated over a \(30\)-seed ensemble.  On held-out in-distribution Brieskorn-sphere samples, the transformer achieves
\[
R^2_{\rm pred}=0.9929\pm0.0083,
\]
where the mean and standard deviation are taken over seeds.  Aggregating held-out predictions across the ensemble gives an ensemble-mean prediction score
\[
R^2_{\rm pred}=0.9967,
\]
with linear calibration
\[
R^2_{\rm lin}=0.9968,\qquad
\widehat{\alpha}=0.9883\,\alpha+700.53 .
\]
Thus the flat-block transformer learns a highly accurate in-distribution interpolation rule for this Brieskorn-sphere family.  The fit is not exact, but the seed-to-seed variation is small: the per-seed calibration slopes are \(0.9891\pm0.0183\), and the per-seed prediction scores concentrate around \(R^2_{\rm pred}=0.9929\pm0.0083\).

Averaging predictions over the unique held-out samples appearing across the seed ensemble yields the calibration shown in Fig.~\ref{fig:sl2-indistribution-linear-fit}.  The ensemble-mean prediction achieves
\[
R^2_{\rm seed} =0.9929\pm0.008 ,\quad
R^2_{\rm pred}=0.9967,\quad
R^2_{\rm lin}=0.9968,\quad
\widehat{\alpha}=0.9883\alpha+700.53,
\]
over \(5329\) unique held-out samples.  The calibration slope remains close to one, confirming that the model captures the global scale of the orbit size, while the small positive intercept and slope below one indicate mild regression toward the mean at the largest orbit sizes.  The scatter in Fig.~\ref{fig:sl2-indistribution-linear-fit} shows that the largest residuals are concentrated among a small number of high-orbit examples and lower-\(\alpha\) outliers, but the overall ensemble calibration is substantially tighter than in the earlier robustness run.  We therefore interpret the flat-block transformer as learning a robust in-distribution interpolation rule within the square-free Brieskorn-sphere family, rather than an exact closed-form formula for the \(SL_2(\mathbb{Z})\)-orbit size.

\begin{figure}[t]
\centering
    \includegraphics[width=\textwidth]{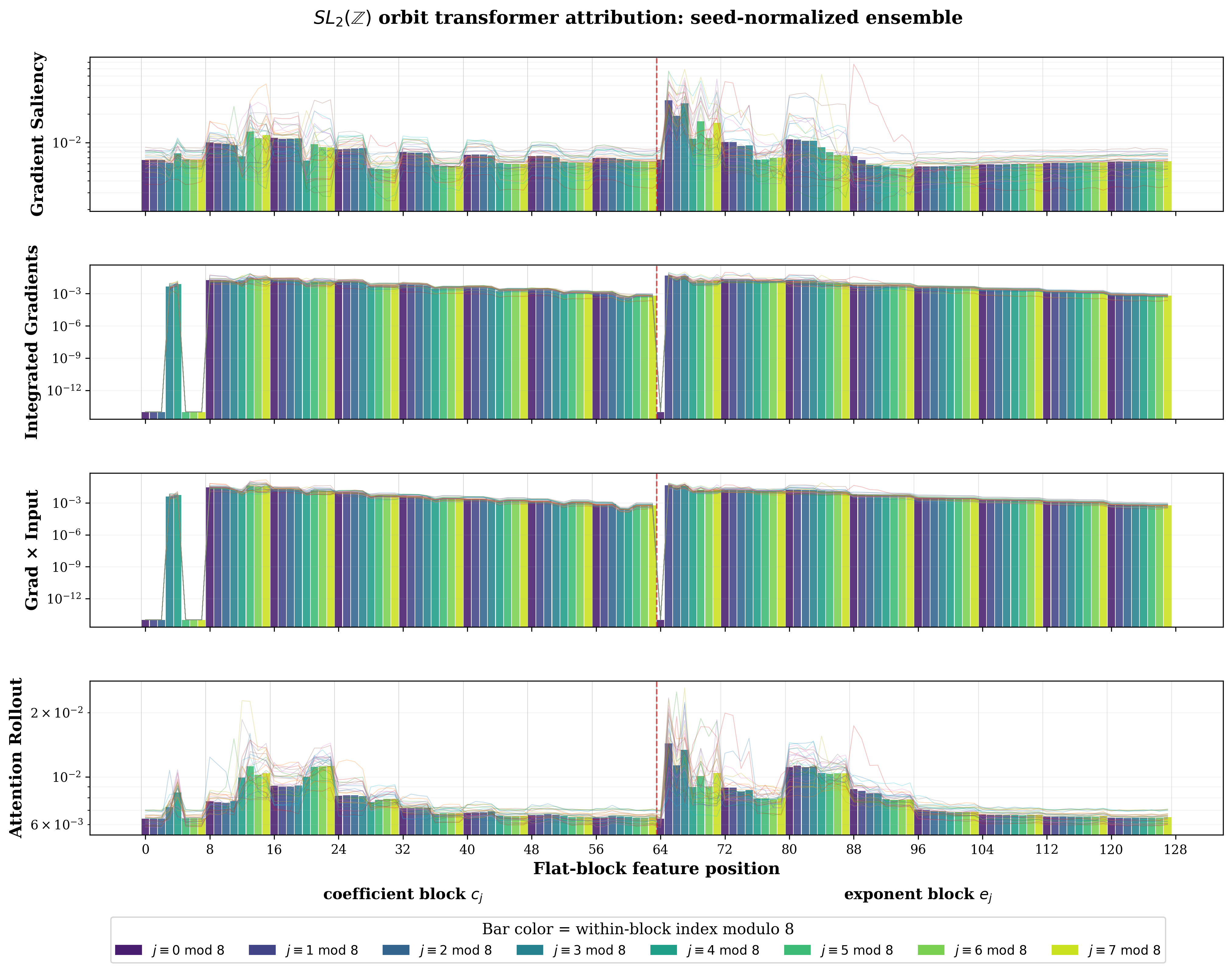}
    \caption{
    \(SL_2(\mathbb{Z})\)-orbit regression: attribution diagnostics for the flat-block transformer.
    The input is ordered as
    \([c_0,\ldots,c_{63},e_0,\ldots,e_{63}]\).
    The four panels show gradient saliency, integrated gradients, gradient-times-input, and attention rollout, averaged over the \(30\)-seed robustness ensemble after normalizing each seed by its total attribution mass.
    The dashed red line separates the coefficient block from the exponent block.
    Bar color denotes the within-block index modulo \(8\), reflecting positional organization associated with the eight interleaved partial-theta sectors in the Brieskorn-sphere expression.
    }
    \label{fig:sl2-transformer-attribution}
\end{figure}

To probe what the transformer uses in the serialized input, we apply gradient-based attribution to the trained models: saliency, integrated gradients, gradient\(\times\)input, and attention rollout. Because the input dimension is large and the task is a regression problem, we restrict the diagnostics to these attribution methods rather than running the full FS/PI/SHAP pipeline. The attributions concentrate on the active coefficient--exponent region and displays an approximately periodic modulation across the serialized terms, as shown in Fig.~\ref{fig:sl2-transformer-attribution}.  This structure is compatible with the partial-theta decomposition of the Brieskorn-sphere \(\widehat Z\)-series discussed below.  Importantly, this should be interpreted as evidence for a structured in-family interpolation mechanism, not as evidence that the transformer has learned a general algorithm for \(SL_2(\mathbb{Z})\)-orbit size.

To understand this, note that the $\widehat Z$-invariants for Brieskorn spheres $\Sigma(b_1,b_2,b_3)$ take the following form\footnote{This equation has to be modified with the addition of a monomial term if and only if $(b_1,b_2,b_3)=(2,3,5)$. }: 
\begin{gather}\begin{split}\label{eq:z_hat_brieskorn_sphere}
    \widehat{Z}(\Sigma(b_1, b_2, b_3);\tau) &= cq^{\tilde{\Delta}} 
\sum_{(\epsilon_{1},\epsilon_{2},\epsilon_{3})\in ({\mathbb Z}/2)^3} (-1)^{\sum_i\epsilon_i} \widetilde{P}_{b_1b_2b_3,\chi_{\epsilon_{1},\epsilon_{2},\epsilon_3}}(q)\\
\chi_{\epsilon_{1},\epsilon_{2},\epsilon_{3}}&:= b_1b_2b_3-(-1)^{\epsilon_{1}}  b_1b_3 - (-1)^{\epsilon_{2}} b_1b_2- (-1)^{\epsilon_{3}} b_2b_3
\end{split}\end{gather}
for some $\tilde\Delta \in \mathbb Q$ and $c\in \mathbb C$. In the above, $\tilde P_{m,r}(q)$ denotes the partial theta function 
\[
\tilde P_{m,r}(q)= \sum_{k\geq 0}q^{\frac{(2mk+r)^2}{4m}}.
\]
From this we see that the ordered set $\{e_i\}$ is naturally split into the following set: 
$$
e_{8n+i}= \frac{(2b_1b_2b_3 n+r_i)^2-r_1^2}{4b_1b_2b_3} 
$$
for $i\in\{1,2,\dots,8\}$. The quasi-periodicity simply indicates that the attribution profiles assign comparable weight to terms belonging to a specific partial theta function $\tilde P_{b_1b_2b_3,r_i}$. 

To test generalization, we evaluated the trained networks on out-of-distribution \(q\)-series whose \(SL_2(\mathbb{Z})\) orbit sizes were computed in \cite[Tab.~14]{Cheng:2018vpl}.  These examples lie outside the square-free Brieskorn-sphere family used for training.  Both the MLP and transformer fail this test: neither architecture correctly predicts the OOD orbit sizes, and the OOD regression quality is worse than a baseline expectation, with \(R^2<0\).  This failure is important.  It shows that the high in-distribution performance should not be interpreted as learning the general \(SL_2(\mathbb{Z})\)-orbit problem.  Rather, the networks learn a highly accurate interpolation rule within a restricted Brieskorn-sphere family, where the orbit size is tightly correlated with the growth pattern of the exponent sequence.

\section{Conclusions and outlook}
\label{sec:conclusion}

In this work, we have demonstrated that finite truncations of $\widehat{Z}$-invariants for weakly negative plumbed 3-manifolds carry significant topological information that can be effectively extracted and interpreted using machine learning. Below we discuss the key conclusions. 

\begin{itemize}
\item {\bf Generalizable Data Pipeline.}
While we focused on the $\widehat{Z}$-invariants of plumbed 3-manifolds in this work, we have established a systematic pipeline for handling mathematical data structured as (truncated) infinite $q$-series. This framework is readily applicable to series originating from diverse sources, such as Nahm sums, $q$-expansions of modular forms, partition functions of other types of quantum field theories, or other quantum invariants.

\item {\bf Novel Connections to Cobordism.}
Our experiments revealed a relationship between the \(q\)-series data of the \(\widehat Z\)-invariant and three-dimensional homology cobordism invariants: regression models fit \(d+\Delta-\frac12\), equivalently the \(\tau\)-function contribution in \(d=-\Delta+\frac12-2\min_k\tau(k)\), with \(>99.9\%\) explained variance from a small number of nonzero exponent offsets of \(\widehat Z\). Supplementing these exponent features with symmetrized Dedekind sums yields a learned quadratic approximation to \(-2\min_k\tau(k)\) giving an accurate (\(\sim90\%\)) integer prediction for \(d\).  For three-fiber Brieskorn spheres, this success is partially explained by the arithmetic structure of the false-theta expansion: the partial-theta sector shifts contain low-order combinations of the Seifert multiplicities \(b_i\), and in the finite three-fiber dataset the leading exponent offsets are injective in the Brieskorn triple \((b_1,b_2,b_3)\); once these multiplicities are known, the normalized Seifert numerators \(a_i\) are fixed by the homology-sphere condition, so the quantities entering the known formula for \(d\), including the Dedekind sums and the \(\tau\)-function, become arithmetic functions of the same underlying Seifert data.  Thus the high regression accuracy should be interpreted as evidence that, in this Brieskorn family, the exponent support carries enough Seifert arithmetic for the network to approximate the correction-term formula; the related \(SL_2(\mathbb Z)\)-orbit experiment supports the same sector-arithmetic interpretation, while also cautioning against assuming out-of-family generalization.

\item {\bf Interpretability Beyond ``Black Boxes''.}
Finally, our methodology---specifically the use of an interpretability regression network to map mathematical features to classifier logits---provides a window into the network's predictive mechanism. By identifying the specific arithmetic and geometric data used by the model, we highlight the fact that machine learning methods do not necessarily lead only to black-box results.  Instead, suitably designed, explainable learning tasks serve to elucidate complex mathematical structures, guiding researchers toward the discovery of human-understandable conjectures.
A concrete illustration is the network's use of $\varepsilon_{\rm min}$ as a proxy for the determinant condition in the homology classification tasks. The spectral bound of Lemma~\ref{prop:spectral_bound} provides a justification for this strategy, and at the same time offers a precise structural account of the classifier's residual generalization error (see \S\ref{subsubsec:FS-homology-classification}).

\end{itemize}

\textbf{Acknowledgements.} We thank Giorgi Butbaia, Mrunmay Jagadale, Davide Passaro and Josef Svoboda for useful conversations. SG was supported by the Simons Collaboration Grant on "New Structures in Low-Dimensional Topology", by the NSF grant DMS-2245099, and by the U.S. Department of Energy, Office of Science, Office of High Energy Physics, under Award No. DE-SC0011632. R.-K. S. is supported by an Outstanding Young Scientist Grant (RS-2025-00516583) of the National Research Foundation of Korea (NRF). He is also partly supported by the BK21 Program ("Next Generation Education Program for Mathematical Sciences", 4299990414089) funded by the Ministry of Education in Korea and the National Research Foundation of Korea (NRF). SH was supported by the FirstRand FNB 2020 Fund Education Scholarship and the University of KwaZulu-Natal's 2024 Vincent Maphai Scholarship Award. The research of MC and BR is supported by the Vici grant (number VI.C.232.117) from the Dutch Research Council (NWO). MC is also supported by the AS-IAIA-114 grant from Academia Sinica Taiwan. The work of FR is supported by the NSF grants PHY-2210333, PHY-2609835, and PHY-2019786 (The NSF AI Institute for Artificial Intelligence and Fundamental Interactions), as well as by startup funding from Northeastern University.

\appendix
\section{Data Analysis Tools and Concepts}
\label{sec:data-analysis}
For the document to be self-contained, in the appendix we introduce and review various data analysis tools and concepts relevant for the experiments in the paper. 

\subsection{Model Performance Metrics} \label{sec:performance-metrics}

In this appendix, we briefly review the methods used to assess network performance across each of the classification and regression tasks.  

\noindent{\textbf{Balanced Accuracy Score (BAS)}}\\

To evaluate the performance of the trained network on the classification tasks, we utilize the Balanced Accuracy Score (BAS). We selected this metric to account for the significant population imbalances between labels in the dataset (see Tab.~\ref{tab:dataset-binary}). Instead of truncating the dataset to enforce parity, we compute the average recall across all classes. For $n_c$ classes, the BAS is defined as:
\begin{align}
 {\rm recall} =  \frac{TP}{TP+FN}~,~~~
     {\rm BAS} = \frac{1}{n_c}\sum_{i=1}^{n_c} {\rm recall}_i
\end{align}
where $TP_i$ is the number of true positives and $FN_i$ is the number of false negatives for the $i^{\rm th}$ class.

\noindent{\textbf{{Pearson Correlation Coefficient (PCC)}}}

The first metric we use to evaluate network performance on the regression task is the Pearson Correlation Coefficient (PCC). The PCC measures the strength and direction of the linear relationship between the ground truth and the predicted values. It is normalized such that PCC $\in [-1,1]$, where $\text{PCC} = +1$ ($-1$) indicates perfect correlation (anti-correlation). Consider the set of $n$ ground truth samples $\{y_i\}$ and predicted values $\{\hat{y}_i\}$. The PCC is defined as:
\begin{align}
    {\rm PCC} = \frac{\sum_{i=1}^n(y_i - \bar{y})(\hat{y}_i - \bar{\hat{y}})}{\sqrt{\sum_{i=1}^n(y_i-\bar{y})^2}\sqrt{\sum_{i=1}^n(\hat{y}_i-\bar{\hat{y}})^2}}
\end{align}
where $\bar{y}$ and $\bar{\hat{y}}$ denote the sample means of the ground truth and predictions, respectively. In our use case, $y$ corresponds to the logits of the classifier network and $\hat{y}$ to the outputs of the regression network. Thus, the more accurate the regression network, the closer the resulting PCC score will be to $1$.

\noindent{\textbf{Coefficient of Determination ($R^2$ Score)}}\\\newline
The second metric used to assess the performance of the regression network is the coefficient of determination, or $R^2$ score. Complementary to the PCC, the $R^2$ score evaluates the goodness of fit by quantifying the proportion of variance in the observed data explained by the model. For a set of $n$ ground truth samples $\{y_i\}$ with mean $\bar{y}$ and $n$ predicted values $\{\hat{y}_i\}$, the score is defined as:
\begin{align}
    R^2 = 1 - \frac{\sum_{i=1}^n (y_i-\hat{y}_i)^2}{\sum_{i=1}^n(y_i - \bar{y})^2}~.
\end{align}
Values of $R^2 < 0$ indicate that the model performs worse than a constant baseline predicting the sample mean. An $R^2=0$ implies the model explains none of the variance around the mean. As $R^2$ approaches $1$, the approximation improves; $R^2=1$ indicates that the predictions perfectly match the ground truth.

\noindent{\textbf{Spearman Rank Correlation ($\rho$)}}

We employ a regressor, such as an RF regressor, to characterize the non-linear decision boundaries learned by the SVM, in the latent space learned using a multi-class contrastive learning network. In this context, the standard $R^2$ score is insufficient because it primarily measures how well the model captures variance and linear trends. However, the decision boundaries in the latent space  are often complex and non-linear. As a remedy, we use a complementary metric--the Spearman rank correlation ($\rho$)--that assesses the topological quality of the fit rather than just the magnitude of the error.

Unlike the standard PCC score, which assesses linearity, Spearman rank correlation $\rho$ evaluates the \textit{monotonic} relationship between the ground truth and the predictions. A relationship is monotonic if an increase in one variable consistently corresponds to an increase (or decrease) in the other, even if that rate of change is not constant.

Let $y$ and $\hat{y}$ denote the vectors of ground truth and predicted values, respectively. We convert these values into rank vectors $\mathcal{R}$ and $\hat{\mathcal{R}}$, where $\mathcal{R}_i$ represents the rank of the $i$-th sample within the set. The Spearman rank correlation is defined as the PCC of these rank vectors:
\begin{align}
    \rho = \frac{\mathrm{cov}(\mathcal{R},\hat{\mathcal{R}})}{\sigma_{\mathcal{R}}\sigma_{\hat{\mathcal{R}}}}~
\end{align}
where $\sigma_\mathcal{R}$ and $\sigma_{\hat{\mathcal{R}}}$ represent the standard deviations of the rank variables. Note that,  for a sample size $n$,  in the absence of tied ranks we have \[\sigma_\mathcal{R} = \sigma_{\hat{\mathcal{R}}}=\sqrt{\tfrac{n^2-1}{12}},\] allowing the formula to simplify to: 
\begin{align}
    \rho = 1 - \frac{6\sum_{i=1}^n (\mathcal{R}_i- \hat{\mathcal{R}}_i)^2}{n(n^2-1)}~.
\end{align}
The coefficient ranges over $\rho \in [-1,1]$. A value of $+1$ indicates a strictly increasing monotonic relationship (perfect ordering), while $-1$ indicates a strictly decreasing one. 

The utility of this metric is that it verifies whether the predicted boundary preserves the \textit{ordinal structure} of the true decision function. That is, $\rho = 1$ implies that if the ground truth implies $y_i > y_j$, the model correctly predicts $\hat{y}_i > \hat{y}_j$, regardless of the exact magnitude of the values. This makes $\rho$ more robust to outliers and non-linearities than the $R^2$ score. However, $\rho$ should not be used in isolation; a high $\rho$ score guarantees correct ranking, but the model might still suffer from significant scaling errors (poor $R^2$ score), as $\rho$ is insensitive to even large magnitude errors.

\noindent{\textbf{Sign Accuracy ($\vartheta$)}}

While the PCC and $R^2$ metrics evaluate the continuous regression performance, we also require a metric to assess the implied classification performance. In the context of SVM decision boundaries, the hyperplane is defined at $y=0$. Thus, we employ the Sign Accuracy ($\vartheta$), which effectively treats the regression output as a binary classifier. It is defined as the fraction of samples where the predicted decision value falls on the same side of the decision boundary as the ground truth:
\begin{align}
    \vartheta = \frac{1}{n}\sum_{i=1}^n \mathbb{I}\left[\text{sgn}(y_i)=\text{sgn}(\hat{y}_i)\right]
\end{align}
where $y$ and $\hat{y}$ denote the true and predicted decision values (logits), and $\mathbb{I}[\cdot]$ is the indicator function. This is equivalent to the standard classification accuracy metric:
\begin{align}
    \vartheta = \frac{TP + TN}{TP + TN + FP + FN}
\end{align}
where $TP$ and $TN$ represent the true positives and true negatives, respectively.

This metric is crucial for determining classification consistency near the boundary. Notably, because $\vartheta$ relies solely on the sign, it is magnitude agnostic. Unlike $R^2$, it does not penalize the model for predicting a value close to the boundary when the truth is far, provided both are on the correct side.

\noindent{\textbf{Complement Normalized MAE Score ($S_{\mathrm{MAE}}$)}}\\
\newline
The final individual metric used to evaluate the regression fit is derived from the Mean Absolute Error (MAE). To ensure consistency with PCC and $R^2$ (where higher values indicate better performance), we define a complement score $S_{\mathrm{MAE}}$. Given the ground truth and the predicted vector $y$ and $\hat y$, the score is defined as:
\begin{align}
    S_{\mathrm{MAE}} = 1 - \frac{1}{n  (\max(y) - \min(y))}\sum_{i=1}^n \lvert y_i -\hat{y}_i\rvert~.
\end{align}
For typical models, this yields a score $S_{\mathrm{MAE}} \in [0,1]$. The primary advantage of this metric over squared-error metrics (like $R^2$) is that it imposes a linear penalty on errors, making it less sensitive to moderate outliers. Furthermore, normalizing by the range renders the metric scale-invariant, allowing us to incorporate it into the composite score described below.

\noindent{\textbf{Composite Score ($Q_{\mathrm{composite}}$)}}

We now assess the performance of the various regressors trained to approximate the decision boundaries around the clusters labeled by the four classes (STAR-Sph, STAR-Non, H-Sph, H-Non) that the multi-class SVM classifier learned in the latent space.  That is, we focus on a single cluster, extract the decision values that define the boundary region from the SVM classifier, and train a regressor network from the input space of $q$-series and plumbing matrix features to the decision values.  To assess the performance of the regressor network, we construct a composite score using the four metrics described above. A conservative approach would be to utilize the minimum score, $Q_{\rm min} =\min\{R^2, \rho, \vartheta, S_{\mathrm{MAE}}\}$. While this ensures no single metric is poor, it does not leverage the specific complementary strengths of each measure. Instead, we define a weighted composite score to emphasize the most descriptive metrics for this topology:
\begin{align}\label{eq:Q-composite}
    Q_{\mathrm{composite}} = 0.3 ~R^2 + 0.25 ~\rho + 0.3 ~\vartheta + 0.15~S_{\mathrm{MAE}}~.
\end{align}

We justify this specific weighting by examining the complementary nature of the chosen metrics:
\begin{itemize}
    \item {\textbf{$R^2$ vs. $\rho$:}} These metrics capture distinct aspects of the fit. A perfect rank correlation ($\rho=1$) can be achieved even with a poor $R^2$ score if the predictions preserve the ordinal structure but suffer from systematic scaling errors. $\rho$ validates the topology, while $R^2$ validates the magnitude.
    \item {\textbf{$R^2$ vs. $\vartheta$:}} These metrics emphasize different performance regimes. The $R^2$ score is dominated by outliers, measuring global regression fidelity. In contrast, $\vartheta$ is magnitude-invariant and focuses solely on the consistency of the classification. 
    \item {\textbf{$R^2$ vs. $S_{\mathrm{MAE}}$:}} Both metrics measure absolute errors but apply different penalties. $R^2$ scales quadratically with error, making it highly sensitive to outliers. $S_{\mathrm{MAE}}$ applies a linear penalty, providing a measure of the "typical" error that is less skewed by extreme outliers.
    \item {\textbf{$\vartheta$ vs. $S_{\mathrm{MAE}}$:}} These metrics are largely independent. It is possible to achieve $\vartheta=1$ (perfect classification) while having a poor $S_{\mathrm{MAE}}$ if the model places points on the correct side of the boundary but with incorrect magnitudes (e.g., predicting a value close to $0$ when the truth is large).
\end{itemize}

Tab.~\ref{tab:model-composite-scores} compares reconstruction models for the one-vs-rest SVM decision functions in the four-class contrastive latent space.  Random Forest and Gradient Boosting are consistently the strongest methods, with nearly identical composite scores on all four boundaries.  Random Forest is nevertheless the top model for each individual boundary, achieving \(0.966\pm0.020\) on STAR-Sph, \(0.977\pm0.009\) on STAR-Non, \(0.970\pm0.015\) on H-Sph, and \(0.952\pm0.035\) on H-Non, with the best overall average score \(0.966\pm0.015\).  Gradient Boosting is only marginally lower, with average score \(0.964\pm0.015\), while the linear baselines are substantially weaker and the neural network reconstruction, although competitive, does not match the tree ensembles.  The per-boundary reconstructions and feature-importance summaries of Tab.~\ref{tab:ovr-boundary-reconstruction} use the cross-validation-selected model for each boundary and seed; given the near-degeneracy of the two
tree ensembles, this selection alternates between Random Forest and Gradient Boosting across boundaries,
consistent with the out-of-fold comparison here.

\begin{table}[ht!]
\centering
\scriptsize
\setlength{\tabcolsep}{4pt}
\resizebox{\textwidth}{!}{%
\begin{tabular}{c|ccccc}
\toprule
\textbf{Model} &
\textbf{STAR-Sph} &
\textbf{STAR-Non} &
\textbf{H-Sph} &
\textbf{H-Non} &
\textbf{Average} \\
\midrule
Random Forest
& \(\mathbf{0.966\,\pm\,0.020}\)
& \(\mathbf{0.977\,\pm\,0.009}\)
& \(\mathbf{0.970\,\pm\,0.015}\)
& \(\mathbf{0.952\,\pm\,0.035}\)
& \(\mathbf{0.966\,\pm\,0.015}\) \\
\hline
Gradient Boost
& \(0.964\,\pm\,0.021\)
& \(0.975\,\pm\,0.010\)
& \(0.968\,\pm\,0.016\)
& \(0.949\,\pm\,0.036\)
& \(0.964\,\pm\,0.015\) \\
\hline
Ridge Regression
& \(0.521\,\pm\,0.055\)
& \(0.789\,\pm\,0.017\)
& \(0.663\,\pm\,0.074\)
& \(0.588\,\pm\,0.099\)
& \(0.640\,\pm\,0.032\) \\
\hline
Linear Regression
& \(0.521\,\pm\,0.055\)
& \(0.789\,\pm\,0.017\)
& \(0.663\,\pm\,0.074\)
& \(0.588\,\pm\,0.099\)
& \(0.640\,\pm\,0.032\) \\
\hline
Neural Network
& \(0.951\,\pm\,0.026\)
& \(0.965\,\pm\,0.011\)
& \(0.943\,\pm\,0.019\)
& \(0.908\,\pm\,0.038\)
& \(0.942\,\pm\,0.016\) \\
\bottomrule
\end{tabular}%
}
\caption{
Multi-seed composite model-selection scores for reconstructing the one-vs-rest SVM decision functions in the four-class contrastive latent space.  Each entry reports the mean $\pm$ standard deviation of \(Q_{\mathrm{composite}}\) over seeds, computed from out-of-fold predictions of the corresponding reconstruction model.  The composite score combines calibrated reconstruction \(R^2\), Spearman rank correlation \(\rho\), sign accuracy \(\vartheta\), and the range-normalized MAE-complement score \(S_{\mathrm{MAE}}\).  Larger values indicate better reconstruction.  Bold entries mark the best model for each boundary and the best average across boundaries.
}
\label{tab:model-composite-scores}
\end{table}

\subsection{Model Explainability Techniques}
This appendix outlines the interpretability methods used to determine feature importance. We categorize these tools into two distinct classes based on the nature of the information they extract from the trained model:

\begin{enumerate}
    \item \textbf{Local Sensitivity (Feature Saliency):} These gradient-based methods measure the rate of change of the network output with respect to the input. They answer the question: \textit{``Which feature change causes the steepest immediate change in the output?''} High sensitivity implies the network is locally responsive to a feature, but does not necessarily imply the feature is critical for global accuracy.
    
    \item \textbf{Global Relevance (PI and SHAP):} These perturbation-based and game-theoretic methods measure the contribution of a feature to the model's predictive performance. They answer the question: \textit{``How much does the model rely on this feature to correctly classify samples?''}
\end{enumerate}

Distinguishing between these two perspectives is crucial for analyzing the results in \S4, as features with high local sensitivity  often differ from those with high global relevance.

\paragraph{Feature Saliency (FS)}
To interpret the regression model, we employ a gradient-based saliency method inspired by Class Activation Mapping (CAM) techniques \cite{Selvaraju_2019}. While originally designed for Convolutional Neural Networks (CNNs) to identify spatial regions of interest, the underlying principle -- weighting feature activations by their gradient contribution to the output -- can be adapted to the hidden layers of an MLP.

We define the ``importance weight'' of a hidden unit $h_j$ for a given class logit $c$ (or regression target) as the gradient of the output $y^c$ with respect to that unit:
\begin{align}
    \alpha_j^c = \frac{\partial y^c}{\partial h_j}~.
\end{align}
This weight quantifies how strongly the $j$-th hidden unit influences the final prediction. Analogous to the Grad-CAM localization map, we compute a weighted activation map for the hidden layer:
\begin{align}
    L_{\text{MLP}}^c = \text{ReLU}\left(\sum_j \alpha_j^c h_j \right)~.
\end{align}
The ReLU function ensures we only consider features that have a positive influence on the class of interest. To project this importance back to the input space (feature attribution), we estimate the contribution of the $i$-th input feature $x_i$ as:
\begin{align}
    S_i^c = \sum_j \alpha_j^c h_j W_{ji}~,
\end{align}
where $W_{ji}$ are the weights connecting input $i$ to hidden unit $j$. This formulation effectively decomposes the network's decision, highlighting which input variables maximize the activation of the most critical hidden neurons.

\noindent{\textbf{Permutation Importance (PI)}}\\
To complement the gradient-based internal analysis, we employ Permutation Importance (PI), a model-agnostic statistical method that quantifies feature contribution based on predictive performance. The fundamental premise is that if a feature is important, corrupting its information content should significantly degrade the model's score. Conversely, if a feature is irrelevant, the model's performance should remain unchanged.

To implement this, we take the dataset and randomly shuffle (permute) the values of a single feature $i$ while keeping all other feature columns fixed. This breaks the relationship between feature $i$ and the target while preserving the marginal distribution of the features. For a trained model $f$ and dataset $\mathcal{D}$, the importance of feature $i$, denoted $I_i$, is defined as the expected drop in the scoring metric $S$:
\begin{align}
    I_i = S\big(f,\mathcal{D}\big) - \mathbb{E}_{\pi}\!\left[S\big(f,\mathcal{D}^{\pi(i)}\big)\right],
\end{align}
where $S$ is a chosen utility metric (where higher values indicate better performance, such as $R^2$ or accuracy) and $\mathcal{D}^{\pi(i)}$ is the dataset with the $i$-th feature randomly permuted. The expectation value $\mathbb{E}_\pi$ is estimated by repeating the shuffling process numerous times to ensure statistical robustness. For the regression tasks in this analysis, we utilize the $R^2$ score as the metric $S$.

This approach provides an \emph{external, performance-based measure} of reliance on features, independent of the model's internal weights or gradients. Therefore, it serves as a valuable cross-check to the gradient-based saliency maps described above. However, a crucial caveat is that PI can underestimate the importance of correlated features; if two features are highly collinear, permuting one may not severely impact performance because the model can extract similar information from the other. We implement this analysis using the built-in function available in \texttt{sklearn.inspection}.

\noindent{\textbf{Shapley Additive Explanations (SHAP)}}

We employ SHAP (SHapley Additive exPlanations) \cite{lundberg2017unified} to interpret the model's predictions. This framework adapts Shapley values \cite{shapley:book1952} from cooperative game theory, where they are used to fairly distribute a ``payout'' among a set of ``players'' based on their marginal contributions. In the context of machine learning, the game is the prediction task, the players are the input features, and the payout is the model's output relative to the baseline expectation. This approach unifies several explanation methods (e.g., LIME \cite{ribeiro2016should}, DeepLIFT \cite{shrikumar2016not}) under the class of additive feature attribution methods. We implement this analysis using the standard \texttt{shap} Python package.

Formally, let $F$ be the set of all features with cardinality $|F|=M$. For a given datapoint $x$, the Shapley value $\phi_i$ for a specific feature $i \in F$ is calculated as the weighted average of its marginal contribution across all possible feature coalitions $S \subseteq F \setminus \{i\}$:
\begin{align}
    \phi_i(x) = \sum_{S\subseteq F \setminus \{i\}}\frac{|S|!(M-|S|-1)!}{M!}\left[ f_{S\cup \{i\}}(x_{S\cup \{i\}}) - f_S(x_S)\right]~,
\end{align}
where $f_S(x_S)$ represents the model's prediction given $x_S$, the subset of features $S$, marginalized over the absent features with the whole dataset. A key property of SHAP is local accuracy, or additivity, meaning the sum of the feature attributions equals the deviation of the prediction from the dataset average:
\begin{align}
    f(\boldsymbol{x}) = \overline{f}+ \sum_{i=1}^M \phi_i(x)~.
\end{align}

\begin{figure}
    \centering
\includegraphics[width=1\linewidth]{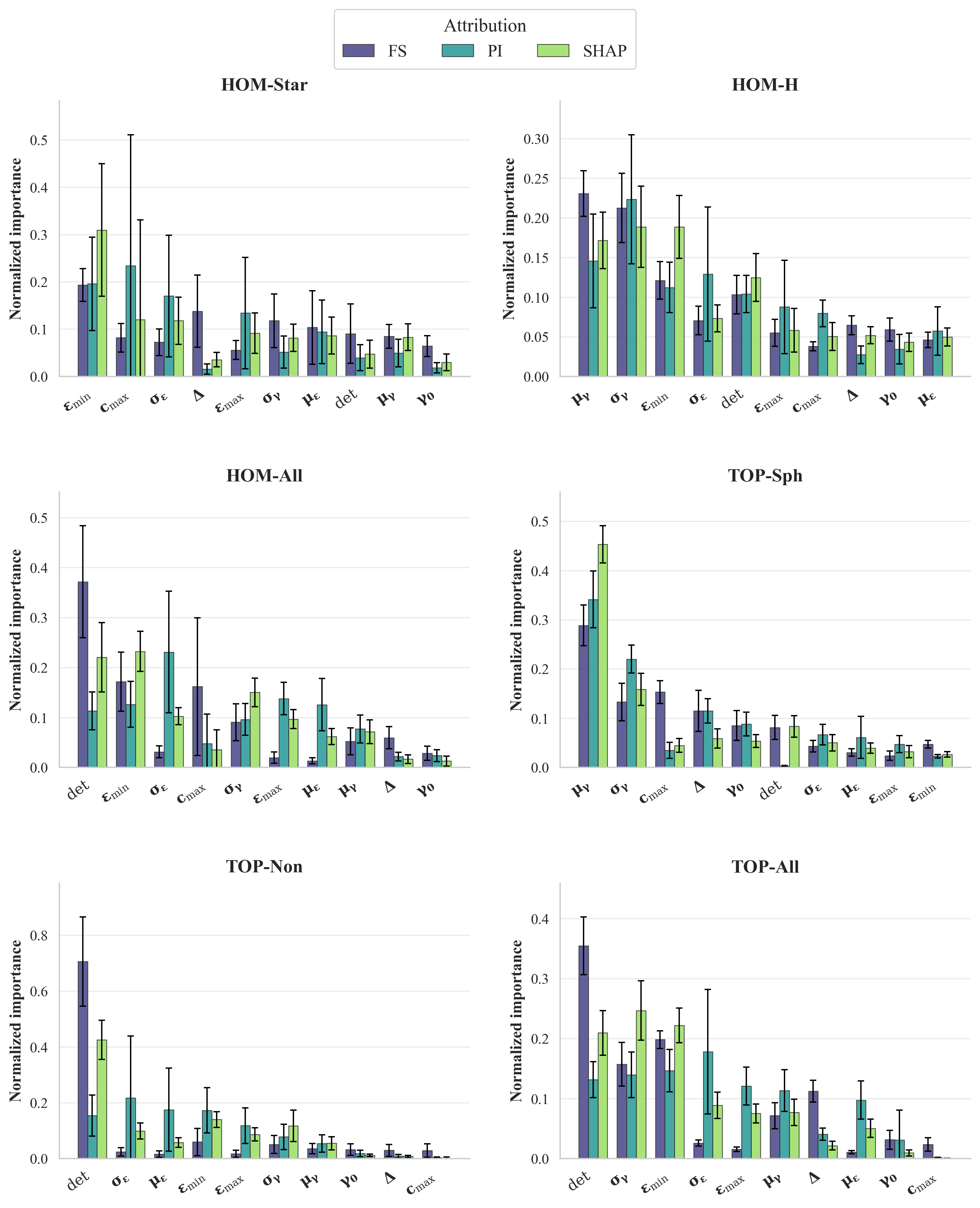}
    \caption{\textbf{Full Sequence.} Mean normalized feature importance scores for feature saliency, permutation importance, and SHAP; averaged over all five seeds for all HOM and TOP experiments.}
    \label{fig:FS_importance}
\end{figure}

\paragraph{Normalized Feature Importance}
For visualization and comparison across attribution methods, we convert raw feature scores into normalized feature-importance profiles.  Given an attribution vector \(S=(S_1,\ldots,S_M)\), we define
\begin{align}
    \widehat S_i
    =
    \frac{S_i}{\sum_{j=1}^M |S_j|+\varepsilon_0},
\end{align}
where \(\varepsilon_0>0\) is a small numerical stabilizer.  Thus the denominator is the total absolute attribution mass.  This normalization removes arbitrary scale differences between seeds, models, and attribution methods while preserving the sign of methods that can produce signed effects, such as permutation importance.  In figures such as Fig.~\ref{fig:FS_importance}, we plot these normalized scores over the ordered feature coordinates; for coefficient--exponent representations, the horizontal axis records the retained coefficient positions followed by the corresponding exponent positions.  Peaks in absolute normalized score indicate coordinates on which the trained model concentrates a disproportionate fraction of its attribution mass.

\paragraph{Other Gradient-Based and Attention-Based Diagnostics}
In addition to the feature-saliency score above, we compute three input-level diagnostics for the \(SL_2(\mathbb Z)\)-orbit regression experiment: gradient-times-input, integrated gradients, and attention rollout.  These diagnostics are applied to the scalar regression output \(f(x)\), where the input is the flat coefficient--exponent block
\[
    x=[c_0,\ldots,c_{L_{\rm eff}-1},e_0,\ldots,e_{L_{\rm eff}-1}] .
\]
Thus each attribution coordinate corresponds either to a retained coefficient position or to the corresponding retained exponent position.

The gradient-times-input score is defined by
\[
    S_i^{\rm gx}(x)
    =
    \left|
    x_i\,\frac{\partial f(x)}{\partial x_i}
    \right| .
\]
Unlike raw gradient saliency, this quantity weights local sensitivity by the actual input value.  It therefore emphasizes coordinates that are both locally influential and active in the represented sample.

We also compute integrated gradients.  Given a baseline input \(x'\), taken to be the zero input in standardized coordinates, the integrated-gradient attribution is
\[
    S_i^{\rm IG}(x)
    =
    \left|
    (x_i-x_i')
    \int_0^1
    \frac{\partial f\!\left(x' + \alpha(x-x')\right)}
         {\partial x_i}
    \,d\alpha
    \right| .
\]
Numerically, this integral is approximated by a finite Riemann sum along the straight-line interpolation from \(x'\) to \(x\).  Integrated gradients probe sensitivity accumulated along this path rather than only at the endpoint \(x\), and hence provide a complementary diagnostic to both raw feature saliency and gradient-times-input.

For transformer regressors, we further compute attention rollout.  Let \(A^{(\ell,h)}\) denote the attention matrix for head \(h\) in layer \(\ell\).  We first average over heads,
\[
    \bar A^{(\ell)}=\frac{1}{H}\sum_{h=1}^{H} A^{(\ell,h)},
\]
and include the residual connection by forming
\[
    \tilde A^{(\ell)}
    =
    \frac{\bar A^{(\ell)}+I}
         {\operatorname{row\text{-}sum}(\bar A^{(\ell)}+I)} .
\]
The rollout matrix is then the layerwise product
\[
    R=\tilde A^{(L)}\tilde A^{(L-1)}\cdots \tilde A^{(1)} .
\]
We use the resulting token-level mass as an attention-based attribution profile over the coefficient and exponent positions.  Unlike the gradient-based diagnostics, attention rollout does not measure sensitivity of the output with respect to infinitesimal input perturbations; rather, it records how information is routed through the self-attention layers.  It is therefore used as a complementary structural diagnostic rather than as a substitute for gradient attribution.

For the multi-seed robustness analysis, all attribution profiles are computed separately for each trained seed.  Before averaging over seeds, each profile is normalized by its total attribution mass on the represented feature block.  This prevents a small number of high-magnitude seeds from dominating the ensemble attribution profile.  We interpret the resulting averaged diagnostics as evidence for which coefficient and exponent positions are used or routed through by the trained regression model, not as a closed-form symbolic rule for the \(SL_2(\mathbb Z)\)-orbit size.

\subsection{Contrastive Learning}
\label{appendix:contrastive_learning}

We use a contrastive learning algorithm based on \emph{semi-hard triplet loss} (SHTL) to perform clustering-based data compression. The rationale for this choice is the ability of the network to learn meaningful, clustered embeddings in a low-dimensional space even in the presence of limited or imbalanced data, which is often the case in our analysis. Compared to other contrastive losses, SHTL offers a good trade-off between computational efficiency and representational robustness.

Contrastive learning originated in computer vision as a method for image verification and facial recognition. Given a labeled dataset $P$ with class labels $c_i$, the goal is to learn a mapping $f: P \rightarrow M$ from the input space to a latent space $M$ (typically $M \subset \mathbb{R}^d$) such that similar samples are embedded close together and dissimilar samples are far apart. A commonly used contrastive loss for binary-labeled pairs $(x_{1,i}, x_{2,i})$ with a pair relationship label $y^i \in \{0,1\}$ is given by \cite{1467314}:
\begin{align*}
    \mathcal{L}_{\rm con} = \frac{1}{2}\sum_{\text{pairs}~i} \left((1 - y^i)\Vert f(x_{1,i}) - f(x_{2,i}) \Vert^2 + y^i \max\left[0, \kappa - \Vert f(x_{1,i}) - f(x_{2,i}) \Vert \right]^2\right)
\end{align*}
Here, $\kappa$ defines a margin in the latent space, and $\Vert \cdot \Vert$ denotes the $L_2$-norm. We adopt the convention where $y=0$ denotes a similar pair (same class) and $y=1$ denotes a dissimilar pair. Consequently, pairs with $y = 0$ contribute to the first term and are pulled together. Negative pairs ($y = 1$) contribute to the second term only if they fall within the margin $\kappa$, effectively pushing them apart.

Triplet loss generalizes this approach by evaluating triplets $(A, P, N)$ consisting of an anchor $A$, a positive $P$ (same class), and a negative $N$ (different class) example. The loss encourages the distance between the anchor and positive to be smaller than the distance between the anchor and negative by at least a margin $\kappa$:
\begin{align*}
    \mathcal{L}_{\rm trip} &= \sum_i \max\left[0, D(A_i, P_i, N_i)\right] \\
    D(A_i, P_i, N_i) &= \Vert f(A_i) - f(P_i) \Vert^2 - \Vert f(A_i) - f(N_i) \Vert^2 + \kappa
\end{align*}
Training with all possible triplets is computationally expensive and often inefficient, as many triplets satisfy the margin condition (i.e., they are already correctly clustered) and contribute zero gradient. Conversely, ``hard'' negatives (where $N$ is very close to $A$) can destabilize training in the early stages.

To address these issues, \emph{semi-hard triplet loss} (SHTL) restricts training to \emph{semi-hard} triplets--those where the negative is farther from the anchor than the positive, but still within the margin:
\begin{align*}
    \Vert f(A_i) - f(P_i) \Vert^2 < \Vert f(A_i) - f(N_i) \Vert^2 < \Vert f(A_i) - f(P_i) \Vert^2 + \kappa
\end{align*}
This constraint defines a \emph{shell} in the latent space: the inner radius is determined by the distance to the positive, and the shell thickness is fixed by $\kappa$. Negatives within this shell are considered \emph{semi-hard} and contribute to the loss. Negatives within the inner radius are labeled \emph{hard}, while those outside the shell are \emph{easy}. By focusing training on semi-hard cases, SHTL improves convergence speed and embedding quality while avoiding the instability associated with hard negatives.

While more complex formulations such as $N$-tuple contrastive losses exist \cite{NIPS2016_6b180037}, we find that given the relatively small number of classes in our experimental datasets, SHTL strikes an effective balance between simplicity and performance. It encourages well-separated clusters in the latent space without requiring extensive sampling or computational overhead.\\

\section{Interpretability Robustness}\label{app:interpretability_robustness}
In this appendix, we detail the protocol of the multi-seed robustness runs that underlie the binary full-
sequence and value-index classification experiments of Sec.~\ref{sec:topExp}. The single-seed results
reported for those experiments in the main body are taken from this robustness experiment. The goal is to
establish whether the results are robust to the choice of random seed, with all other pipeline stages held
fixed.

\subsection{Multi-seed robustness protocol}
\label{app:multiseed_protocol}

This subsection describes the multi-seed robustness experiment underlying the Sec.~\ref{sec:topExp} classifier, logit-regression, and interpretability results.  The purpose of this experiment is to separate stable structural effects from artifacts of a particular random initialization, train/validation split, or attribution pass.  All seed-averaged Sec.~\ref{sec:topExp} quantities are computed from this experiment, and the normalized Margin Reconstruction analysis in the next subsection uses the same source teachers and stored splits.

\paragraph{Seed bank and stages.}
All stages draw from the fixed seed bank
\[
    \mathcal S=\{42,123,456,789,1337\}.
\]
For each task and representation, we train one classifier for each seed in \(\mathcal S\).  For each trained classifier, we then train logit regressors over the same seed bank, and interpretability passes are likewise repeated over the same seed set.  Thus the robustness experiment is organized into three nested components: source classifiers, abstract-feature logit regressors, and attribution diagnostics.

\paragraph{Tasks.}
The six binary tasks are
\[
\mathrm{HOM\text{-}Star},\quad
\mathrm{HOM\text{-}H},\quad
\mathrm{HOM\text{-}All},\quad
\mathrm{TOP\text{-}Sph},\quad
\mathrm{TOP\text{-}Non},\quad
\mathrm{TOP\text{-}All}.
\]
The HOM tasks distinguish homology-sphere from non-homology-sphere examples on different graph sectors, while the TOP tasks distinguish \(H\)-graph from STAR examples on different topological sectors.  Each task is evaluated as a binary classification problem with balanced accuracy as the primary classifier metric.

\paragraph{Representations.}
For each task we evaluate three input representations.  The \emph{full-sequence} or sparse representation embeds each normalized \(\widehat Z\) coefficient sequence into a fixed \(10{,}000\)-dimensional sparse vector indexed by exponent.  The dense representation serializes each sequence as a value-index pair vector, with coefficient entries followed by exponent entries, and applies an asinh scaling.  The contrastive-learning representation starts from the full-sequence input and passes it through a learned encoder trained with a semi-hard triplet objective.  The main full-sequence results in Sec.~\ref{sec:topExp} use the sparse representation; the dense and contrastive-learning representations are reported as robustness comparisons.

\paragraph{Abstract feature dictionary.}
The logit-regression and interpretability stages use the ten-dimensional abstract feature vector
\[
\Phi(x)=
\bigl(
\det,\,
\varepsilon_{\min},\,
\varepsilon_{\max},\,
\mu_\varepsilon,\,
\sigma_\varepsilon,\,
\Delta,\,
\gamma_0,\,
\mu_\gamma,\,
\sigma_\gamma,\,
c_{\max}
\bigr).
\]
Here \((\varepsilon_{\min},\varepsilon_{\max},\mu_\varepsilon,\sigma_\varepsilon)\) are eigenvalue summary statistics, while \((\gamma_0,\mu_\gamma,\sigma_\gamma)\) are gap summary statistics.  The abstract features are standardized before being used by the regression and interpretability models.

\paragraph{Classifier training.}
For each task, representation, and classifier seed \(s\in\mathcal S\), the data are split into train and held-out validation sets using the stored split associated to that seed.  A shallow ReLU classifier is then trained with cross-entropy loss, Adam optimization, early stopping on held-out balanced accuracy, hidden width \(64\), learning rate \(10^{-3}\), batch size \(128\), and patience \(20\).  The best validation checkpoint is retained.  The classifier records store the trained model, the fitted scaler, the label encoder, the train/validation indices, the validation inputs, and the train/validation labels.  The training design matrix is intentionally omitted from the stored classifier records to avoid unnecessary memory growth; when needed, it is reconstructed from the stored split indices and fitted scaler.

The primary classifier metric is validation balanced accuracy. As a reproducibility check, all reported
classifier means and standard deviations are recomputed directly from the released models and validation
inputs, rather than read from cached summary tables; the recomputed balanced accuracies agree with the
recorded checkpoint values, with a largest observed deviation of \(2.2\times10^{-4}\) (one HOM-Star seed). In
particular, the full-sequence HOM-Star result used in the main text is \(0.928\pm0.003\).

\paragraph{Logit-regression stage.}
After training the source classifiers, the next stage asks how much of the teacher's two-logit geometry can be reconstructed from the abstract feature dictionary.  For each source classifier, the classifier logits are extracted on the train and validation splits.  A regressor is then trained from the standardized abstract features to the two-dimensional teacher logits.  This is a deliberately stricter test than reproducing the hard class label: it requires the abstract features to reconstruct the continuous pre-softmax geometry of the teacher.  Regressor performance is reported by train and validation \(R^2\) and Pearson correlation coefficient.  The checkpoint selected by validation performance is used for attribution.

\paragraph{Attribution diagnostics.}
For each trained logit regressor, we compute feature-level attribution diagnostics on the abstract feature dictionary.  The reported diagnostics include gradient saliency, grouped permutation importance, and SHAP-style attribution.  These methods are not interpreted as causal interventions on the original \(\widehat Z\) sequence.  Rather, they quantify which abstract features are most responsible for reconstructing the trained teacher's logit geometry within the abstract-feature regression model.  Rankings are aggregated over the seed ensemble and reported as mean ranks or seed-averaged attribution summaries.

\paragraph{Reporting and figure generation.}
The final Sec.~\ref{sec:topExp} tables and figures are generated directly from the trained models and stored
data of this multi-seed experiment.  Classifier confusion matrices and logit-space plots are recomputed directly from the stored full-sequence classifiers and held-out inputs.  The logit-space visualization standardizes logits within each panel for display and applies central quantile filtering so that a small number of extreme logits does not determine the plotting range.  These display transformations are used only for visualization and do not affect the reported balanced accuracy values.

\paragraph{Interpretation.}
The multi-seed experiment distinguishes three levels of robustness.  First, classifier robustness asks whether the full-sequence, dense, and contrastive-learning teachers achieve stable validation balanced accuracy across seeds.  Second, logit-regression robustness asks whether the abstract feature dictionary reconstructs the continuous teacher logits in a stable way.  Third, attribution robustness asks whether the same abstract features remain important across seeds and attribution methods.  The normalized Margin Reconstruction experiment in the next subsection complements this pipeline by relaxing the logit-regression target: instead of asking whether abstract features reconstruct calibrated logits, it asks whether they reconstruct the source teacher's hard decision boundary.

\subsection{Normalized Margin Reconstruction protocol}
\label{app:boundaryfirst}

This appendix describes the normalized Margin Reconstruction (MR) protocol used in Table~\ref{tab:boundaryfirst_zseed_margin_profile_attribution}.  The purpose of the protocol is to test whether low-dimensional abstract features reproduce the \emph{hard decision boundary} of the full-sequence classifiers, rather than their calibrated logits.  This distinction matters because the logit-regression experiment tests reconstruction of the continuous two-logit geometry, while MR tests reconstruction of the teacher's induced classification rule.

\paragraph{Source teachers.}
For each task \(T\in\{\mathrm{HOM\text{-}Star},\mathrm{HOM\text{-}H},\mathrm{HOM\text{-}All},\mathrm{TOP\text
{-}Sph},\mathrm{TOP\text{-}Non},\mathrm{TOP\text{-}All}\}\), the source teacher is the full-sequence
classifier from the multi-seed robustness experiment; no separate MR teacher is trained. For each teacher
seed \(s\in\{42,123,456,789,1337\}\), we use that experiment's stored train/validation split, fitted input
scaler, trained full-sequence classifier, and validation labels. The training inputs are reconstructed from
the stored split and scaler, and the reconstructed teacher predictions are checked against the recorded
validation BAS for consistency.

\paragraph{Student targets.}
Let \(f_{T,s}\) denote the source full-sequence teacher for task \(T\) and seed \(s\), and let
\[
    \ell_{T,s}(x)=(\ell_0(x),\ell_1(x))
\]
be its two-logit output on a validation sample \(x\).  The MR student does not regress \(\ell_{T,s}(x)\).  Instead, it is trained to reproduce the hard teacher label
\[
    \hat y_{T,s}(x)=\arg\max_{a\in\{0,1\}}\ell_a(x).
\]
Thus the primary target is balanced agreement with \(\hat y_{T,s}\), not balanced accuracy against the original ground-truth label.  True-label BAS is reported only as a reference.

\paragraph{Abstract feature dictionary.}
Students receive only subsets of the abstract feature vector
\[
\Phi(x)=
\bigl(
\det,\,
\varepsilon_{\min},\,
\varepsilon_{\max},\,
\mu_\varepsilon,\,
\sigma_\varepsilon,\,
\Delta,\,
\gamma_0,\,
\mu_\gamma,\,
\sigma_\gamma,\,
c_{\max}
\bigr).
\]
Here \((\varepsilon_{\min},\varepsilon_{\max},\mu_\varepsilon,\sigma_\varepsilon)\) are the eigenvalue statistics, and \((\gamma_0,\mu_\gamma,\sigma_\gamma)\) are the gap statistics.  The candidate feature subsets include \(\det\)-only, eigenvalue statistics, gap statistics, \(c_{\max}\)-only, and several combined sets such as \(\det+\gamma\)-statistics and \(\det+\varepsilon\)-statistics.

\paragraph{Student model families.}
For each task, teacher seed, feature subset, and student seed, we train students from three simple model families:
\[
    \text{logistic regression},\qquad
    \text{random forest},\qquad
    \text{small ReLU MLP}.
\]
Logistic regression and the MLP use standardized abstract features.  The random forest is trained on the corresponding raw abstract-feature subset.  The MLP has a single hidden layer matching the small classifier family used elsewhere in the robustness pipeline and is selected by an internal validation split within the teacher-training split.  The final evaluation is always performed on the held-out split inherited from the source teacher.

\paragraph{Margin slices.}
MR evaluates not only full held-out teacher agreement but also agreement on margin-defined subsets.  For each validation sample \(x\), define the source-teacher margin
\[
    m(x)=|\ell_1(x)-\ell_0(x)|.
\]
The boundary slices \(B_5\), \(B_{10}\), and \(B_{20}\) are the 5\%, 10\%, and 20\% of validation samples with smallest \(m(x)\), respectively.  The easy slice \(E_{50}\) is the 50\% of validation samples with largest \(m(x)\).  Balanced teacher agreement on these slices distinguishes true boundary reconstruction from easy-region agreement.

\paragraph{Selection criterion.}
For each task, a single MR student is selected by a margin-aware score combining full teacher agreement, boundary-slice teacher agreement, easy-region teacher agreement, true-label BAS, and a small feature-count penalty.  The score used for the selected rows is
\[
\begin{aligned}
S
={}&0.30\,A_{\mathrm{Full}}
+0.20\,A_{B_5}
+0.20\,A_{B_{10}}
+0.15\,A_{B_{20}}
+0.05\,A_{E_{50}}
+0.10\,\mathrm{BAS}_{\mathrm{true}}
-0.002\,d,
\end{aligned}
\]
where \(A_{\mathrm{Full}}\) is balanced agreement with the source teacher on the full held-out split, \(A_{B_p}\) is balanced agreement on the \(p\%\) smallest-margin slice, \(A_{E_{50}}\) is balanced agreement on the 50\% largest-margin slice, \(\mathrm{BAS}_{\mathrm{true}}\) is balanced accuracy against the original labels, and \(d\) is the number of abstract features used by the student.  The selected row for each task is the candidate with largest mean score over the teacher/student seed ensemble.

\paragraph{Feature attribution.}
For the selected MR student in each task, we compute feature-level attributions on the \(B_{10}\) slice.  Three diagnostics are used: saliency, SHAP attribution, and permutation drop in teacher agreement.  For logistic students, saliency is the absolute standardized coefficient; for random forests, saliency is the impurity-based feature importance; and for MLP students, saliency is the mean absolute input gradient of the class-margin score.  SHAP values are computed with model-appropriate explainers when available, with the logistic case using the standard linear coefficient-times-input contribution.  Permutation importance is computed by permuting one selected abstract feature at a time within the evaluation slice and recording the drop in balanced agreement with the source teacher.  The ``Dominant \(B_{10}\) features'' column in Table~\ref{tab:boundaryfirst_zseed_margin_profile_attribution} reports the leading one or two features under the average of normalized saliency, SHAP, and permutation-drop scores.

\paragraph{Interpretation.}
MR is not intended to replace the full-sequence classifier.  Instead, it asks whether a low-dimensional abstract feature subset gives a faithful surrogate for the teacher's hard decision boundary.  High true-label BAS alone is insufficient: a student can classify the original labels well while failing to reproduce the teacher near its boundary.  The margin-slice profile is therefore essential.  In the present experiments, determinant and gap statistics explain most of the HOM-H/HOM-All and topology teacher geometry, respectively, while HOM-Star remains poorly reconstructed by the current abstract feature dictionary.

\section{Spectral bound on \texorpdfstring{$|\varepsilon_{\mathrm{min}}|$}{|epsilon_min|}}
\label{app:spectral_bound}

In this appendix we give the full derivation of Lemma~\ref{prop:spectral_bound}. Let $M$ be the plumbing matrix of a three-legged star-shaped graph associated to a Seifert manifold $M(b;\, a_1/b_1,\, a_2/b_2,\, a_3/b_3)$ with singular fiber orders $b_1, b_2, b_3$. Being an intersection form, $M$ is a nonsingular symmetric integer matrix; we write $\varepsilon_{\rm min}$
for its eigenvalue of smallest absolute value.

Because $M$ is symmetric, its inverse $M^{-1}$ is also symmetric. The spectral radius of $M^{-1}$ is $1/|\varepsilon_{\rm min}|$, and by the Rayleigh quotient (taking $x = e_v$), the absolute value of any diagonal entry is bounded by the spectral radius:
\begin{equation}\label{eq:rayleigh_app}
    \frac{1}{|\varepsilon_{\rm min}|} \ge |(M^{-1})_{vv}| \quad \text{for all } v \in V.
\end{equation}
By Cramer's rule, the diagonal entries of the inverse are $(M^{-1})_{vv} = \det M_v / \det M$, where $M_v$ is the submatrix obtained by deleting vertex $v$.

Consider the central node $v_0$. Deleting $v_0$ disconnects the star into three independent paths, so $M_{v_0}$ is block-diagonal and $|\det M_{v_0}| = \prod_{i=1}^{3} |\det L_i|$, where $L_i$ is the tridiagonal intersection matrix of the $i$-th leg. The absolute determinant of each leg satisfies a recurrence relation
\[\Delta_k = w_k \Delta_{k-1} - \Delta_{k-2},\quad \text{with}~\Delta_0 = 1,\; \Delta_{-1} = 0,\]
which coincides with the numerator recurrence for the Hirzebruch--Jung continued fraction $[w_1, \dots, w_n] = b_i / a_i$. Hence $|\det L_i| = b_i$, giving $|\det M_{v_0}| = b_1 b_2 b_3$. See Appendix~A of \cite{Cheng:2022rqr}, in particular equations (A.24),  for more details.

Combining with Cramer's rule, $|(M^{-1})_{v_0 v_0}| = b_1 b_2 b_3 / |\det M|$, and the spectral bound \eqref{eq:rayleigh_app} becomes:
\begin{equation}
    |\varepsilon_{\rm min}| \le \frac{|\det M|}{b_1 b_2 b_3},
\end{equation}
completing the proof of Lemma~\ref{prop:spectral_bound}.

The bound reveals two independent mechanisms that conspire to make $|\varepsilon_{\rm min}|$ small for Brieskorn spheres. First, the homology sphere condition $|\det M| = 1$ fixes the numerator to its minimal possible value, whereas for non-spheres $|\det M|$ can be arbitrarily large, directly loosening the bound. Second, the pairwise coprimality of the $b_i$ forces a large denominator: since the smallest pairwise coprime integers $\ge 2$ are $2, 3, 5$, one has $b_1 b_2 b_3 \ge 30$, and typically much larger for generic Seifert homology spheres. Together, these two factors guarantee that a Brieskorn sphere must possess a small $|\varepsilon_{\rm min}|$, bounded by $1/30$ from above, for all three-fiber Brieskorn homology spheres.

\section{Bound involving \texorpdfstring{$\min\tau$}{min tau}}
\label{appendix:bound-min-tau}
In this section we give a proof that for every pairwise coprime triple $2 \leq b_1 < b_2 < b_3$,
\[
\Big|\min_{k\ge 0}\tau(k) +
\frac{P}{8}\Big(1-\sum_{i=1}^{3}\frac{1}{b_i}\Big)^{\!2} \Big|
\leq  \sum_{i=1}^{3} b_i .
\]
In what follows we use the notation of Section \ref{subsec:theoretical-background-d}.  Additionally we introduce the notation $\{x\}$ to denote $x - \lfloor x \rfloor$ for $x \in \mathbb{R}$. It is clear that $\{x\} \in [0, 1)$. Similarly, note the sawtooth function satisfies $|\langle x \rangle | \leq \frac{1}{2}$ for every $x \in \mathbb{R}$. Recall the definition of the generalized and classical Dedekind sums defined respectively as
\[
  s(p,q;x,y) \;=\; \sum_{l=0}^{q-1}
  \saw{\frac{l+y}{q}}\,\saw{\frac{p(l+y)}{q}+x},
  \qquad s(p,q) := s(p,q;0,0).
\]
We will use the result (cf. \cite{RademacherGrosswald}) that, for $\gcd(p,q) = 1$, 
\begin{equation}\label{eq:classical-dedekind-bound}
  |s(p,q)| \;\leq\; s(1,q) \;=\; \frac{(q-1)(q-2)}{12\,q}.  
\end{equation}
We begin by writing $S = 1- \sum_{i=1}^3 \frac{1}{b_i}$. Apart from the case $(b_1, b_2, b_3) = (2, 3, 5)$ where $S < 0$, in all other cases one has $0 < S < 1$.
Substituting
$\lfloor (k-1)/b_i \rfloor = (k-1)/b_i - \{(k-1)/b_i\}$ into
\eqref{eq:tau-alt}, allows us to write for every integer $k \ge 0$,
\begin{equation}\label{eq:decomp}
\tau(k) = g(k) + R(k), \qquad
g(k) := \frac{k(k-1)}{2P} - \frac{kS}{2},
\end{equation}
with
\begin{equation}
 R(k) := \frac{3}{2} - \frac{1}{2}\sum_{i=1}^3 \left[\frac{1}{b_i}
+  \Big\{\frac{k-1}{b_i}\Big\}
+  \bigg\langle \frac{k a_i}{b_i}\bigg\rangle
- 2\Big( s(\hat b_i, b_i)
  - s\Big(\hat b_i, b_i; \frac{k}{b_i}, 0\Big)\Big)\right].   
\end{equation}
Let us now establish a bound for $R(k)$. Each generalized Dedekind sum appearing in $R$ has $y = 0$, so its $l=0$ summand contains the factor $\saw{0} = 0$ and vanishes; the remaining
$b_i - 1$ summands are each a product of two sawtooth values, thus for every $x$ we find
$|s(\hat b_i, b_i; x, 0)| \le (b_i-1)/4$. Finally,
pairwise coprimality gives
$\sum_{i=1}^3 1/b_i \le \tfrac12+\tfrac13+\tfrac15 = \tfrac{31}{30}$.
Therefore, after noting that $s(\hat b_i, b_i)$ satisfies \eqref{eq:classical-dedekind-bound}, for every $k \ge 0$ we obtain,
\begin{equation}\label{eq:R-bound}
|R(k)| \;\leq\; \frac32 + \frac{31}{60} + \frac32 + \frac34 + {\sum_{i=1}^{3} \frac{(b_i - 1)(b_i - 2)}{12\, b_i}
    + \frac14 \sum_{i=1}^{3} (b_i - 1)
    \;\leq\; \frac{529}{180} + \frac{1}{3}\sum_{i=1}^3 b_i} .
\end{equation}
Since $g(k)$ grows quadratically and $|R|$ is bounded we see $\tau(k)\to \infty$ as $k \to \infty$, so $\min_{k\ge0}\tau(k)$ exists. Define $B:=\frac{529}{180} + \frac{1}{3}\sum_{i=1}^3 b_i$. We then, moreover, see from \eqref{eq:decomp} that $|\tau(k) - g(k)| \leq B$ for every integer $k \geq 0$.

We now turn to analyzing $g$. Extended to a real argument, $g$ is a parabola with leading
coefficient $\tfrac{1}{2P}$, vertex
$x^{*} = \tfrac{1+PS}{2}$, and minimal value,
\[
g(x^{*}) \;=\; -\frac{(1+PS)^2}{8P}
\;=\; -\left[\frac{PS^2}{8} + \frac{S}{4} + \frac{1}{8P}\right].
\]
Let $k_0$ denote a nearest integer to $x^*$, then $\min_{k \in \mathbb{Z}_{\geq 0}}g(k) = g(k_0)$. If $PS$ is odd, then $x^* \in \mathbb{Z}$ and $k_0 = x^*$. On the other hand if $PS$ is even then $k_0 = x^* \pm \frac{1}{2}$. Note that $g(x^* \pm \frac{1}{2}) = g(x^*) + \frac{1}{8P}$. In all cases we thus obtain
\begin{equation}
    0 \leq \min_{k \in \mathbb{Z}_{\geq 0}}g(k) - g(x^*) \leq \frac{1}{8P}.
\end{equation}
This can be in turn used to show 
\begin{equation}\label{eq:min-g-k}
    \bigg|\min_{k \in \mathbb{Z}_{\geq 0}}g(k) + \frac{PS^2}{8}\bigg| \leq \frac{S}{4} + \frac{1}{8} \leq \frac{3}{8}
\end{equation}
where in the last inequality we use the assumption that $0 < S < 1$. However \eqref{eq:min-g-k} holds even when $(b_1, b_2, b_3) = (2, 3, 5)$ and thus it holds for all cases of $S$. Since $|\min_{k \geq 0} \tau(k) - \min_{k \in \mathbb{Z}_{\geq 0}}g(k)| \leq B$ a further calculation, after noting that $\sum_{i=1}^3 b_i \geq 2+3+5=10$, shows 
\begin{equation}\label{eq:final-estimate-tau}
   \bigg|\min_{k \geq 0}\tau(k) + \frac{PS^2}{8}\bigg| \leq B+\frac{3}{8} = \frac{1193}{360} + \frac{1}{3}\sum_{i=1}^3 b_i \leq \sum_{i=1}^3 b_i.
\end{equation}
This completes the proof. This result also implies
\begin{equation}\label{eq:d-estimate}
    d(\Sigma(b_1, b_2, b_3)) \leq 2(b_1 + b_2 + b_3).
\end{equation}
To prove this, first note that the orbifold Euler number of $\Sigma(b_1, b_2, b_3)$ is given by $e = -1/P$. After substituting this into \eqref{eq:kinv} and comparing with \eqref{d} one obtains
\begin{equation}\label{eq:master-2Sb}
    d(\Sigma(b_1,b_2,b_3))
    \;=\; \frac54 - \frac{1}{4P} - 3 \sum_{i=1}^{3} s(a_i, b_i)
     -2 \min_{k \geq 0} \tau(k) - \frac{P S^2}{4}
\end{equation}
The middle expression in \eqref{eq:final-estimate-tau} implies $2\big| \min_{k \geq 0} \tau(k)
    + \frac{P S^2}{8} \big|
    \;\leq\; \frac{1193}{180} + \frac{2}{3}\sum_{i=1}^3 b_i$. 
An application of \eqref{eq:classical-dedekind-bound} together with $\sum_{i=1}^3 1/b_i \leq \frac{31}{30}$ yields the estimate
\[
   3 \sum_{i=1}^{3} |s(a_i, b_i)|
    \;\leq\; \frac{1}{4}\sum_{i=1}^3 b_i - \frac94 + \frac{31}{60}.
\]
With these facts at hand, applying the triangle
inequality to \eqref{eq:master-2Sb} gives
\[
    d(\Sigma(b_1, b_2, b_3)) \;\leq\; \frac{11}{12}\sum_{i=1}^3 b_i + \frac{553}{90} + \frac{1}{4P}.
\]
and since $\sum_{i=1}^3 b_i \geq 10$ and $P \geq 30$ we see that $\frac{553}{90} + \frac{1}{4P} \leq \frac{13}{12}\,\sum_{i=1}^3 b_i$. From this \eqref{eq:d-estimate} follows. It is possible to obtain tighter estimates of $d$, but we do not pursue this here.

\section{Splice diagrams and integral homology Spheres}

\label{appendix:splicing_diagrams}

In this section we give a self-contained overview of splice diagrams following \cite{neumann2005complex}. The main goal will be to outline an algorithm from \cite[Section 9]{neumann2005complex} which allows us to produce integral homology spheres from a given splice diagram.

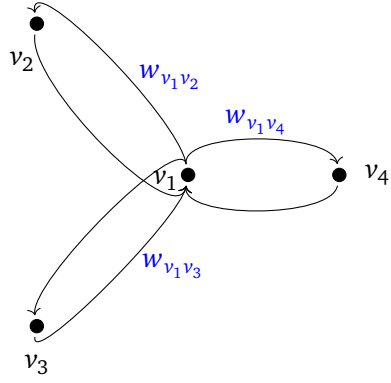
\begin{figure}[ht]
    \centering
    \begin{tikzpicture}
    \begin{scope}[every node/.style={circle, draw, fill=black, inner sep = 0pt, minimum size=5pt}]
        \node (1) at (0, 0) {};
        \node (2) at (2, 0) {};
        \node (3) at (-2, 2) {};
        \node (4) at (-2, -2) {};
    \end{scope}
    \draw[->, shorten <=0.05cm, shorten >= 0.05cm] (1) to[out=100, in=100, distance=
    0.5cm] (2);
    \draw[->, shorten <=0.05cm, shorten >= 0.05cm] (1) to[out=100, in=100, distance=
    0.7cm] (3);
    \draw[->, shorten <=0.05cm, shorten >= 0.05cm] (1) to[out=100, in=100, distance=
    0.5cm] (4);

    \draw[->, shorten <=0.05cm, shorten >= 0.05cm] (2) to[out=-100, in=-100, distance=
    0.5cm] (1);
    \draw[->, shorten <=0.05cm, shorten >= 0.05cm] (3) to[out=-100, in=-100, distance=
    0.7cm] (1);
    \draw[->, shorten <=0.05cm, shorten >= 0.05cm] (4) to[out=-100, in=-100, distance=
    0.5cm] (1);
% vertex labels
\node (1A) at (-0.3, -0.07) {$v_1$};
\node (2A) at (2.5, 0) {$v_4$};
\node (3A) at (-2.2, 1.5) {$v_2$};
\node (4A) at (-2, -2.5) {$v_3$};
% weight labels
\node (1B) at (0.9, 0.7) [text=blue]{$w_{v_1v_4}$};
\node (1B) at (-0.24, 1.3) [text=blue]{$w_{v_1v_2}$};
\node (1A) at (-0.2, -1.2) [text=blue]{$w_{v_1v_3}$};
\end{tikzpicture}
    \caption{A splice diagram represented by a digraph. The weights on outgoing edges from nodes are depicted as $w_{v_iv_j}$ and the weights on outgoing edges from leaves are omitted.}    \label{fig:digraph_splicing}
\end{figure}

Recall that earlier in the paper we defined a splice diagram as a tree satisfying some conditions. For clarity, it is helpful to instead define a \textit{splice diagram} to be a digraph (see Figure \ref{fig:digraph_splicing}), whose underlying graph structure has finitely many vertices which have $\deg = 1$ or $\deg \geq 3$. For each of the outgoing edges of a node an integer weight is assigned arbitrarily. On the other hand, for each of the outgoing edges of a leaf, a weight is assigned as per the \textit{"Splicing Leaf Repair"} procedure (see \cite[Theorem 9.1]{neumann2005complex})  outlined below.

Suppose $v_i$ is a leaf of a splice diagram $\widetilde{\Gamma}$ and $E =(v_i, v_j)$ is an outgoing edge connecting $v_i$ to the node $v_j$. To assign the weight $w_{v_iv_j}$ on $E$: we first let $b = w_{v_jv_i}$ be the weight on the edge $(v_j, v_i)$. Let $a$ be the product of the weights on edges $(v_j, v_k)$ where $v_k \neq v_i$ is a vertex adjacent to the node $v_j$, then define \begin{equation}
    w_{v_iv_j} := \bigg\lceil \frac{a}{b} \bigg\rceil.
\end{equation}  
We call this procedure "Splicing Leaf Repair" as it assigns a weight to each of the outgoing edges from a leaf. It is summarized in Algorithm \ref{alg:splicing_repair}.

In what follows we will always assume that a splice diagram $\widetilde{\Gamma}$ has the Splicing Leaf Repair procedure applied to it and satisfies the edge-determinant and pairwise-coprimality hypotheses of the splice-diagram definition; the constructions below are stated under, and must preserve, these hypotheses.

\subsection{An algorithm to produce integral homology Spheres}\label{sec:algorithm_zhs_overarching}

We now turn to the main goal of this section, which is to outline the algorithm to produce negative-definite plumbed manifolds which are integral homology spheres from splice diagrams.

At a high level the algorithm is as follows: one starts off with a splice diagram $\widetilde{\Gamma}$ and then constructs what is known as a \textit{maximal splice diagram} $\widetilde{\Gamma}_{\operatorname{max}}$ from it. The maximal splice diagram has the same underlying graph structure as a plumbing graph $\Gamma$. From $\widetilde{\Gamma}_{\operatorname{max}}$ one can then construct the plumbing matrix of an associated plumbed manifold and by \cite[Thm 1.1]{neumann2005complex} one sees that this plumbed manifold is an integral homology sphere. The reader can find the relevant pseudocode in Algorithm \ref{alg:splice_to_plumbing}.

\subsubsection*{Maximal Splice Diagram from Splice Diagram}

We first outline Algorithm \ref{alg:splice_to_max_splice} which generates the maximal splice diagram $\widetilde{\Gamma}_{\operatorname{max}}$ from a splice diagram $\widetilde{\Gamma}$.

\begin{figure}[ht]
    \centering
    \begin{tikzpicture}
    \begin{scope}[every node/.style={circle, draw, fill=black, inner sep = 0pt, minimum size=5pt}]
        \node (1) at (-2, 0) {};
        \node (2) at (2, 0) {};
    \end{scope}
    \node (3) at (-4, 2) {};
    \node (4) at (-4, -2) {};
    \node (7) at (-4, 0.7) {};
    \node (8) at (-3.5, 0) {$\vdots$};
    \node (9) at (-3.5, -0.5) {$\vdots$};
    \node (5) at (4, 2) {};
    \node (6) at (4, -2) {};
\node (10) at (4, 0.7) {};
    \node (11) at (3.5, 0) {$\vdots$};
    \node (12) at (3.5, -0.5) {$\vdots$};
    
    \draw (1)--(2);
    \draw (1)--(3);
    \draw (1)--(4);
    \draw (1)--(7);
    \draw (2)--(5);
    \draw (2)--(6);
    \draw (2)--(10);
% vertex labels
\node (1B) at (-2, -0.5) {$v_i$};
\node (2B) at (2, -0.5) {$v_j$};
% weight labels
\node (1B) at (-1.6, 0.3) [text=blue]{$b$};
\node (2B) at (1.6, 0.3) [text=blue]{$c$};
\node (1B) at (-2.5, 0.9) [text=blue]{$a_1$};
\node (1B) at (-3.15, 0.6) [text=blue]{$a_2$};
\node (1B) at (2.5, 0.9) [text=blue]{$d_1$};
\node (1A) at (-2.5, -0.9) [text=blue]{$a_r$};
\node (1B) at (3.05, 0.6) [text=blue]{$d_2$};
\node (1A) at (2.5, -0.9) [text=blue]{$d_s$};
\end{tikzpicture}
\caption{Starting point for algorithm}    \label{fig:alg_start}
\end{figure}
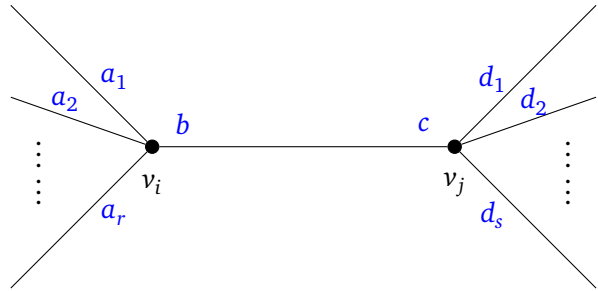

It works as follows: let $E = (v_i, v_j)$ be an edge between two nodes $v_i$ and $v_j$ (see Figure \ref{fig:alg_start}).

\begin{figure}[ht]
    \centering
    \begin{tikzpicture}
    \begin{scope}[every node/.style={circle, draw, fill=black, inner sep = 0pt, minimum size=5pt}]
        \node (1) at (0, 0) {};
        \node (2) at (2, 0) {};
        \node (3) at (4, 0) {};
        \node (4) at (-2, 0) {};
        \node (5) at (-4, 0) {};
    \end{scope} 
    \node (6) at (5.5, 0) {};
    \node (7) at (-5.5, 0) {};
    \node (8) at (6.5, 0) {};
    \node (9) at (-6.5, 0) {};
    \draw (1)--(2);
    \draw (1)--(4);
    \draw (2)--(3);
    \draw (3)--(6);
    \draw (4)--(7);
    \draw[dashed] (6)--(8);
    \draw[dashed] (7)--(9);
% weight labels
\node (1B) at (-4.4, 0.3) [text=blue]{$1$};
\node (1B) at (-3.6, 0.3) [text=blue]{$3$};
\node (1B) at (-2.4, 0.3) [text=blue]{$1$};
\node (1B) at (-1.6, 0.3) [text=blue]{$2$};
\node (1B) at (-0.4, 0.3) [text=blue]{$1$};
\node (2B) at (0.4, 0.3) [text=blue]{$1$};
\node (2B) at (1.6, 0.3) [text=blue]{$2$};
\node (2B) at (2.4, 0.3) [text=blue]{$1$};
\node (2B) at (3.6, 0.3) [text=blue]{$3$};
\node (2B) at (4.4, 0.3) [text=blue]{$1$};
\end{tikzpicture}
\caption{The infinite linear graph used as a building block within the algorithm.}    \label{fig:infinite_graph}
\end{figure}
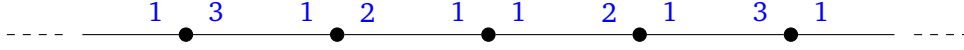

Set $a = \prod_{i=1}^r a_i$ and $d = \prod_{i=1}^s d_i$. In the case that either $v_i$ or $v_j$ is a leaf instead of a node, then we set $a = 1$ or $d=1$ respectively. Consider the infinite linear graph in Figure \ref{fig:infinite_graph}.

\begin{figure}[ht]
    \centering
    \begin{tikzpicture}
    \begin{scope}[every node/.style={circle, draw, fill=black, inner sep = 0pt, minimum size=5pt}]
        \node (1) at (2, 0) {};
        \node (2) at (4, 0) {};
        \node (3) at (6, 0) {};
        \node (4) at (-2, 0) {};
        \node (5) at (-4, 0) {};
    \end{scope} 
    \node (10) at (-0.8, 0) {};
    \node (12) at (0, 0) {$\mapsto$};
    \node (11) at (1, 0) {};
    \node (6) at (7.5, 0) {};
    \node (7) at (-5.5, 0) {};
    \draw (1)--(2);
    \draw (2)--(3);
    \draw (2)--(11);
    \draw (3)--(6);
    \draw (4)--(7);
    \draw (4)--(10);
% weight labels
\node (1B) at (-4.4, 0.3) [text=blue]{$\alpha$};
\node (1B) at (-3.6, 0.3) [text=blue]{$\beta$};
\node (1B) at (-2.4, 0.3) [text=blue]{$\gamma$};
\node (1B) at (-1.6, 0.3) [text=blue]{$\delta$};
\node (1B) at (1.8, 0.3) [text=blue]{$\alpha$};
\node (2B) at (2.3, 0.3) [text=blue]{$\beta$};
\node (2B) at (3.4, 0.3) [text=blue]{$\alpha + \gamma$};
\node (2B) at (4.6, 0.3) [text=blue]{$\beta + \delta$};
\node (2B) at (5.8, 0.3) [text=blue]{$\gamma$};
\node (2B) at (6.3, 0.3) [text=blue]{$\delta$};
\end{tikzpicture}
\caption{Graph Refinement}    \label{fig:subdivision}
\end{figure}
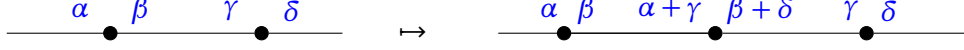

On the infinite linear graph in Figure \ref{fig:infinite_graph}, vertices     \begin{tikzpicture}
    \begin{scope}[every node/.style={circle, draw, fill=black, inner sep = 0pt, minimum size=5pt}]
        \node (1) at (0, 0) {};
    \end{scope} 
    \node (3) at (1, 0) {};
    \node (4) at (-1, 0) {};
    \draw (1)--(3);
    \draw (1)--(4);
% weight labels
\node (1B) at (-0.3, 0.3) [text=blue]{$x$};
\node (1B) at (0.3, 0.3) [text=blue]{$y$};
\end{tikzpicture} are ordered by the size of $x/y$. The algorithm proceeds by modifying the infinite graph via the process outlined in Figure \ref{fig:subdivision} until both the vertices \begin{tikzpicture}
    \begin{scope}[every node/.style={circle, draw, fill=black, inner sep = 0pt, minimum size=5pt}]
        \node (1) at (0, 0) {};
    \end{scope} 
    \node (3) at (1, 0) {};
    \node (4) at (-1, 0) {};
    \draw (1)--(3);
    \draw (1)--(4);
% weight labels
\node (1B) at (-0.3, 0.3) [text=blue]{$a$};
\node (1B) at (0.3, 0.3) [text=blue]{$b$};
\end{tikzpicture} and \begin{tikzpicture}
    \begin{scope}[every node/.style={circle, draw, fill=black, inner sep = 0pt, minimum size=5pt}]
        \node (1) at (0, 0) {};
    \end{scope} 
    \node (3) at (1, 0) {};
    \node (4) at (-1, 0) {};
    \draw (1)--(3);
    \draw (1)--(4);
% weight labels
\node (1B) at (-0.3, 0.3) [text=blue]{$c$};
\node (1B) at (0.3, 0.3) [text=blue]{$d$};
\end{tikzpicture} appear on the graph.

The algorithm then replaces the edge $(v_i, v_j)$ by the extracted portion of the modified linear graph: the two endpoint vertices are identified with $v_i$ and $v_j$ (whose weights along this edge are unchanged), and the interior vertices subdivide the edge. This process is repeated for every edge of $\widetilde{\Gamma}$ and we call the resulting graph $\widetilde{\Gamma}_{\operatorname{max}}$ the \textit{maximal splice diagram}. 

\subsubsection*{Plumbing Graph from Maximal Splice Diagram}

To retrieve the plumbing graph $\Gamma$ from a maximal splice diagram $\widetilde{\Gamma}_{\operatorname{max}}$, recall that the underlying graph structure of $\widetilde{\Gamma}_{\operatorname{max}}$ (i.e. the vertices and undirected edges) is the same as the graph structure of $\Gamma$. In particular this allows one to read off the adjacency matrix of $\Gamma$ from  the underlying graph structure of $\widetilde{\Gamma}_{\operatorname{max}}$. To find the plumbing matrix $M$ associated with $\Gamma$, we are further required to find the weights $w_i$ associated with the vertices $v_i$ of $\Gamma$. In what follows, (similar to \cite{neumann2005complex}), we use the notation $v_i-v_j$ to mean the vertices $v_i$ and $v_j$ are connected by an edge. If $v_k$ is any vertex adjacent to $v_i$ we obtain:  
\begin{equation}\label{eq:finding_weights}
    w_i = \frac{-1}{\ell_{v_iv_k}}\left(\sum_{\{v_j \in \operatorname{Vert}(\widetilde{\Gamma}_{\operatorname{max}}) \mid v_i-v_j\}}\ell_{ v_jv_k}\right)
\end{equation}
where the notation $\ell_{v_iv_j}$ is used to denote the product of the weights adjacent to, but not on, the shortest path from $v_i$ to $v_j$ (with $\ell_{v_iv_i}$ denoting the product of the weights adjacent to $v_i$) in $\widetilde{\Gamma}_{\operatorname{max}}$.

\subsection{Algorithms and Pseudocode}

In this subsection we list the algorithms in the form of pseudocode for the interested reader.

\begin{algorithm}[H]
    \caption{Splicing Leaf Repair}
    \label{alg:splicing_repair}
    \KwData{splice diagram $\widetilde{\Gamma}$
        }
    \KwResult{splice diagram $\widetilde{\Gamma}_r$ with weights on edges adjacent to leaves } 
set $\widetilde{\Gamma}_r = \widetilde{\Gamma}$\\
\For{vertices $v_i$ in $\widetilde{\Gamma}$ with $\deg(v_i)=1$}
{
let $v_j$ be the node adjacent to the vertex/leaf $v_i$ \\
set $b = w_{v_jv_i}$ the weight on the edge $(v_j, v_i)$\\
set $a=1$\\
\For{vertices $v_k$ adjacent to $v_j$}{
\If{$v_k \neq v_i$}{
$a = a \times w_{v_jv_k}$ the weight on the edge $(v_j, v_k)$
}
}
set the weight on the edge $(v_i, v_j)$ to be equal to $\lceil \frac{a}{b} \rceil$
}
return $\widetilde{\Gamma}_r$
\end{algorithm}

\begin{algorithm}[H]
    \caption{Generate maximal splice diagram from splice diagram}
\label{alg:splice_to_max_splice}
    \KwData{splice diagram $\widetilde{\Gamma}$
        }
    \KwResult{Maximal splice diagram $\widetilde{\Gamma}_{\operatorname{max}}$ } 
set $\widetilde{\Gamma}_r$ equal to Splicing Leaf Repair of $\widetilde{\Gamma}$ by Algorithm \ref{alg:splicing_repair}\\
set $\widetilde{\Gamma}_{\operatorname{max}} = \widetilde{\Gamma}_r$\\
\For{all edges $(v_i, v_j)$ of $\widetilde{\Gamma}_r$}
{
set $b = w_{v_iv_j}$ the weight on the edge $(v_i, v_j)$\\
set $c = w_{v_jv_i}$ the weight on the edge $(v_j, v_i)$\\
set $a = 1$\\
Set $d = 1$\\
\For{$v \in \{v_i, v_j\}$}{
\If{$v$ is a node}{
\For{vertices $v_k$ adjacent to $v$}{
\If{edge $(v, v_k) \not\in \{(v_i, v_j), (v_j, v_i)\}$}{
set $w_{vv_k}$ equal to the weight on the edge $(v, v_k)$\\
\uIf{$v = v_i$}{
$a = a \times w_{vv_k}$
}
\Else{
$d = d \times w_{vv_k}$
}
}
}
}
}
Let $I$ be the infinite graph in Figure \ref{fig:infinite_graph}\\
Add edges to $I$ by following the procedure in Figure \ref{fig:subdivision} until  vertices with     \begin{tikzpicture}
    \begin{scope}[every node/.style={circle, draw, fill=black, inner sep = 0pt, minimum size=5pt}]
        \node (1) at (0, 0) {};
    \end{scope} 
    \node (3) at (1, 0) {};
    \node (4) at (-1, 0) {};
    \draw (1)--(3);
    \draw (1)--(4);
% weight labels
\node (1B) at (-0.3, 0.3) [text=blue]{$a$};
\node (1B) at (0.3, 0.3) [text=blue]{$b$};
\end{tikzpicture} and     \begin{tikzpicture}
    \begin{scope}[every node/.style={circle, draw, fill=black, inner sep = 0pt, minimum size=5pt}]
        \node (1) at (0, 0) {};
    \end{scope} 
    \node (3) at (1, 0) {};
    \node (4) at (-1, 0) {};
    \draw (1)--(3);
    \draw (1)--(4);
% weight labels
\node (1B) at (-0.3, 0.3) [text=blue]{$c$};
\node (1B) at (0.3, 0.3) [text=blue]{$d$};
\end{tikzpicture} appear in $I$\\
Replace the edge $(v_i, v_j)$ of $\widetilde{\Gamma}_{\operatorname{max}}$ by the portion of $I$ between \begin{tikzpicture}
    \begin{scope}[every node/.style={circle, draw, fill=black, inner sep = 0pt, minimum size=5pt}]
        \node (1) at (0, 0) {};
    \end{scope} 
    \node (3) at (1, 0) {};
    \node (4) at (-1, 0) {};
    \draw (1)--(3);
    \draw (1)--(4);
% weight labels
\node (1B) at (-0.3, 0.3) [text=blue]{$a$};
\node (1B) at (0.3, 0.3) [text=blue]{$b$};
\end{tikzpicture} and     \begin{tikzpicture}
    \begin{scope}[every node/.style={circle, draw, fill=black, inner sep = 0pt, minimum size=5pt}]
        \node (1) at (0, 0) {};
    \end{scope} 
    \node (3) at (1, 0) {};
    \node (4) at (-1, 0) {};
    \draw (1)--(3);
    \draw (1)--(4);
% weight labels
\node (1B) at (-0.3, 0.3) [text=blue]{$c$};
\node (1B) at (0.3, 0.3) [text=blue]{$d$};
\end{tikzpicture}, identifying its endpoint vertices with $v_i$ and $v_j$
}
return $\widetilde{\Gamma}_{\operatorname{max}}$
\end{algorithm}

\begin{algorithm}[H]
    \caption{Convert a splice diagram to a plumbing graph}\label{alg:splice_to_plumbing}
    \KwData{splice diagram $\widetilde{\Gamma}$
        }
    \KwResult{Plumbing matrix $M$ of the plumbing graph $\Gamma$ corresponding to $\widetilde{\Gamma}$ } 
convert splice diagram $\widetilde{\Gamma}$ to maximal splice diagram $\widetilde{\Gamma}_{\operatorname{max}}$ using Algorithm \ref{alg:splice_to_max_splice}\\
set $M$ equal to the adjacency matrix of underlying graph of $\widetilde{\Gamma}_{\operatorname{max}}$\\
\For{each vertex $v_i$ in $\widetilde{\Gamma}_{\operatorname{max}}$}{
pick a vertex $v_k$ adjacent to $v_i$\\
compute $w_i$ via \eqref{eq:finding_weights}\\
set $M_{ii} = w_i$
}
return $M$
\end{algorithm}

As mentioned earlier, the resulting plumbing graph $\Gamma$ produced from the splice diagram $\widetilde{\Gamma}$ is an integral homology sphere by \cite[Thm 1.1]{neumann2005complex}.

\bibliographystyle{utphys}
\bibliography{ZhatTDA}

\newpage

\section*{Figures}
\begin{figure}[ht!]
    \centering
    \includegraphics[width = 0.92\textwidth]{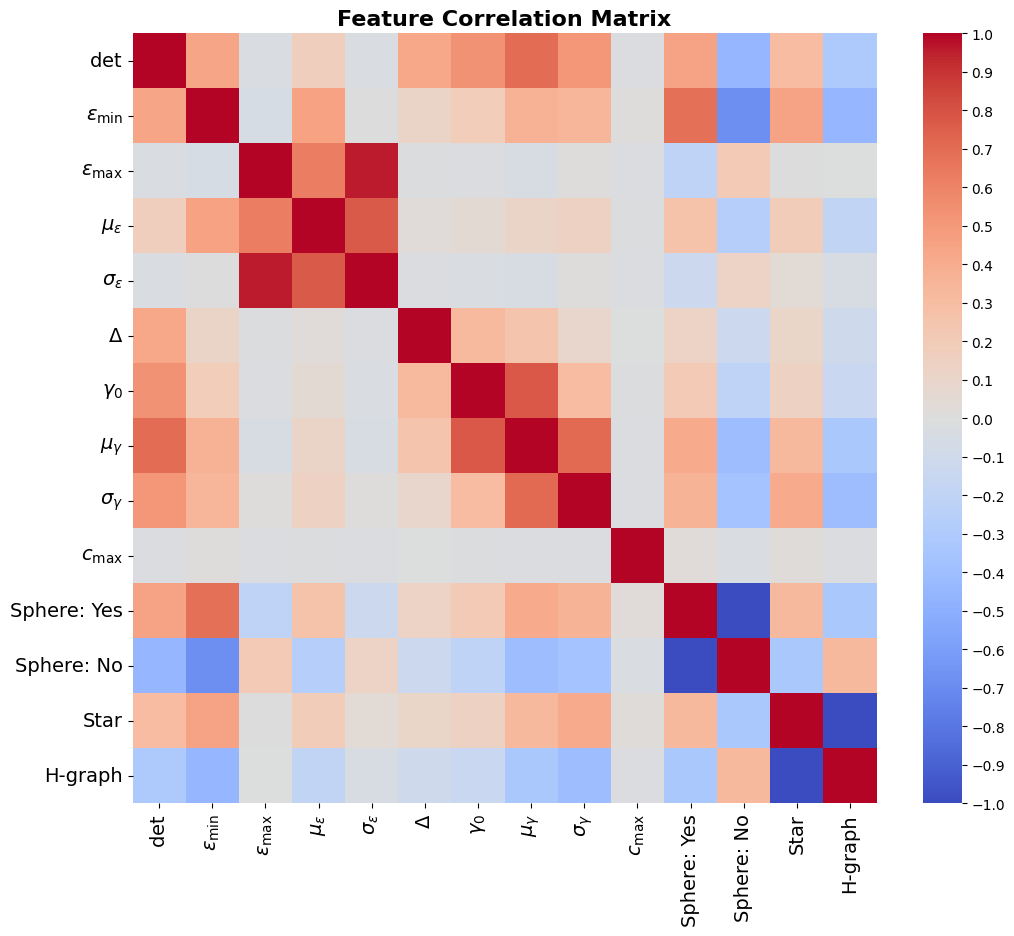}
    \caption{Heatmap of correlation matrix of the features used as input to regression networks and their binary labels.}\label{fig:feature-correlation-heatmap}
\end{figure} 

\begin{figure}
    \centering
\includegraphics[width=1\linewidth]{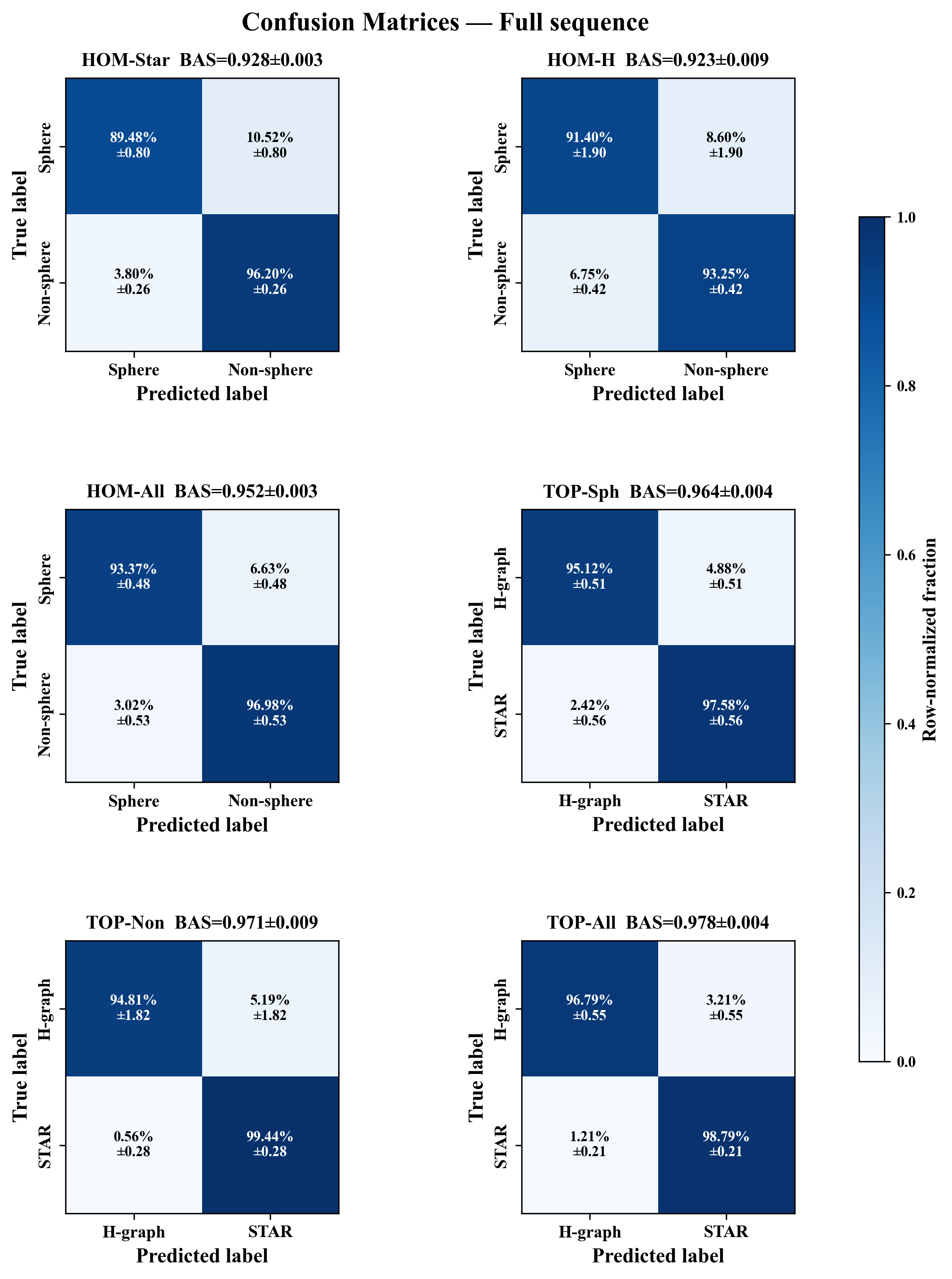}
    \caption{\textbf{Full Sequence.} Mean confusion matrices across five seeds for all HOM and TOP experiments. }
    \label{fig:FS_confusion}
\end{figure}

\begin{figure}
    \centering
    \includegraphics[width=1\linewidth]{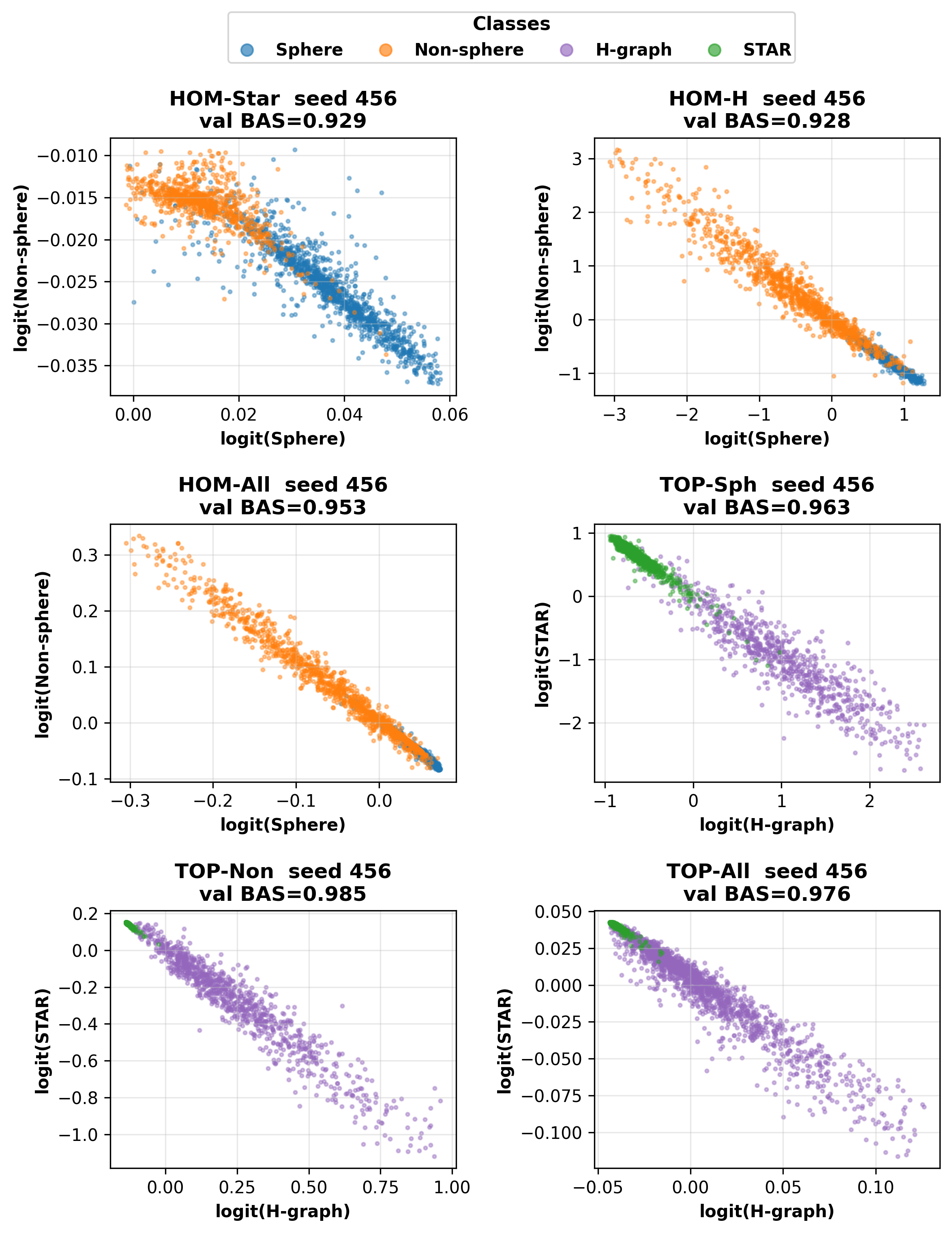}
    \caption{\textbf{Full Sequence.} Logit space visualization for a representative seed (\texttt{seed}-456) for all HOM and TOP experiments. For display only, the two logits are standardized within each panel and central quantile filtering is used to avoid axis ranges being determined by a small number of extreme logits.}
    \label{fig:FS_logits}
\end{figure}

\begin{figure}
    \centering
\includegraphics[width=1\linewidth]{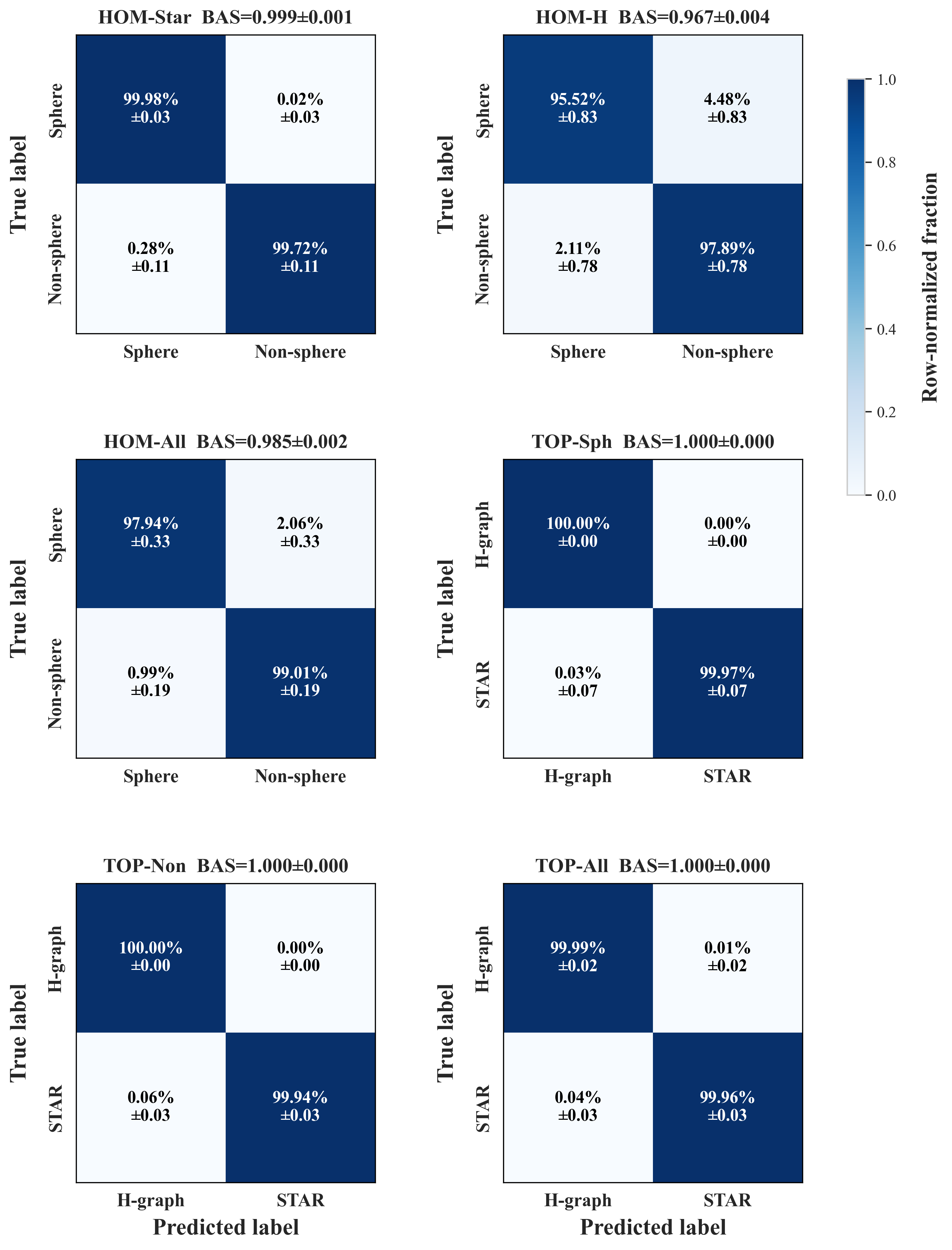}
    \caption{\textbf{Value-Index.} Mean confusion matrices across five seeds for all HOM and TOP experiments. }
    \label{fig:VI_confusion}
\end{figure}

\begin{figure}
    \centering
    \includegraphics[width=1\linewidth]{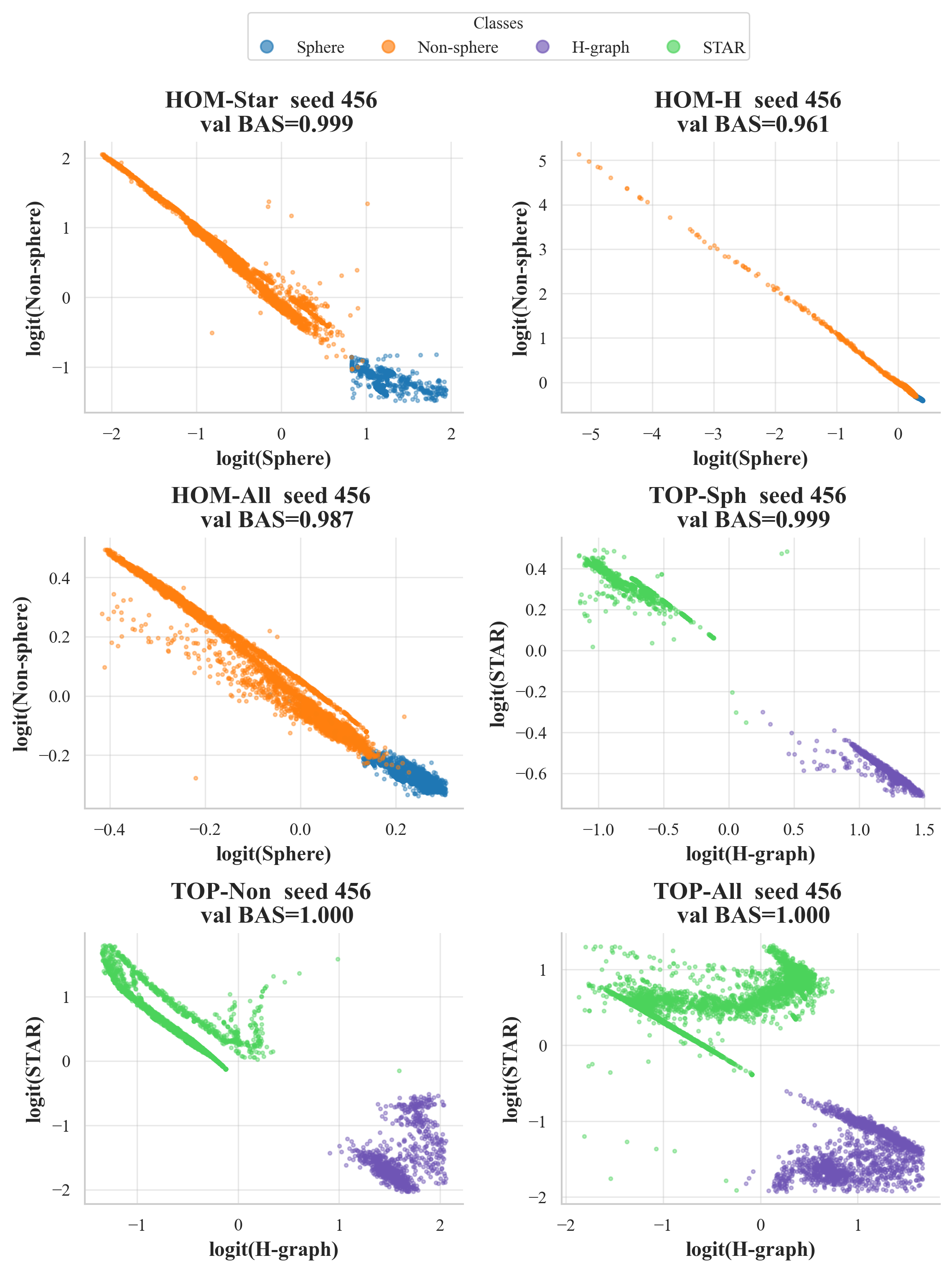}
    \caption{\textbf{Value-Index.} Logit space visualization for a representative seed (\texttt{seed}-42) for all HOM and TOP experiments.}
    \label{fig:VI_logits}
\end{figure}

\begin{figure}
    \centering
\includegraphics[width=1\linewidth]{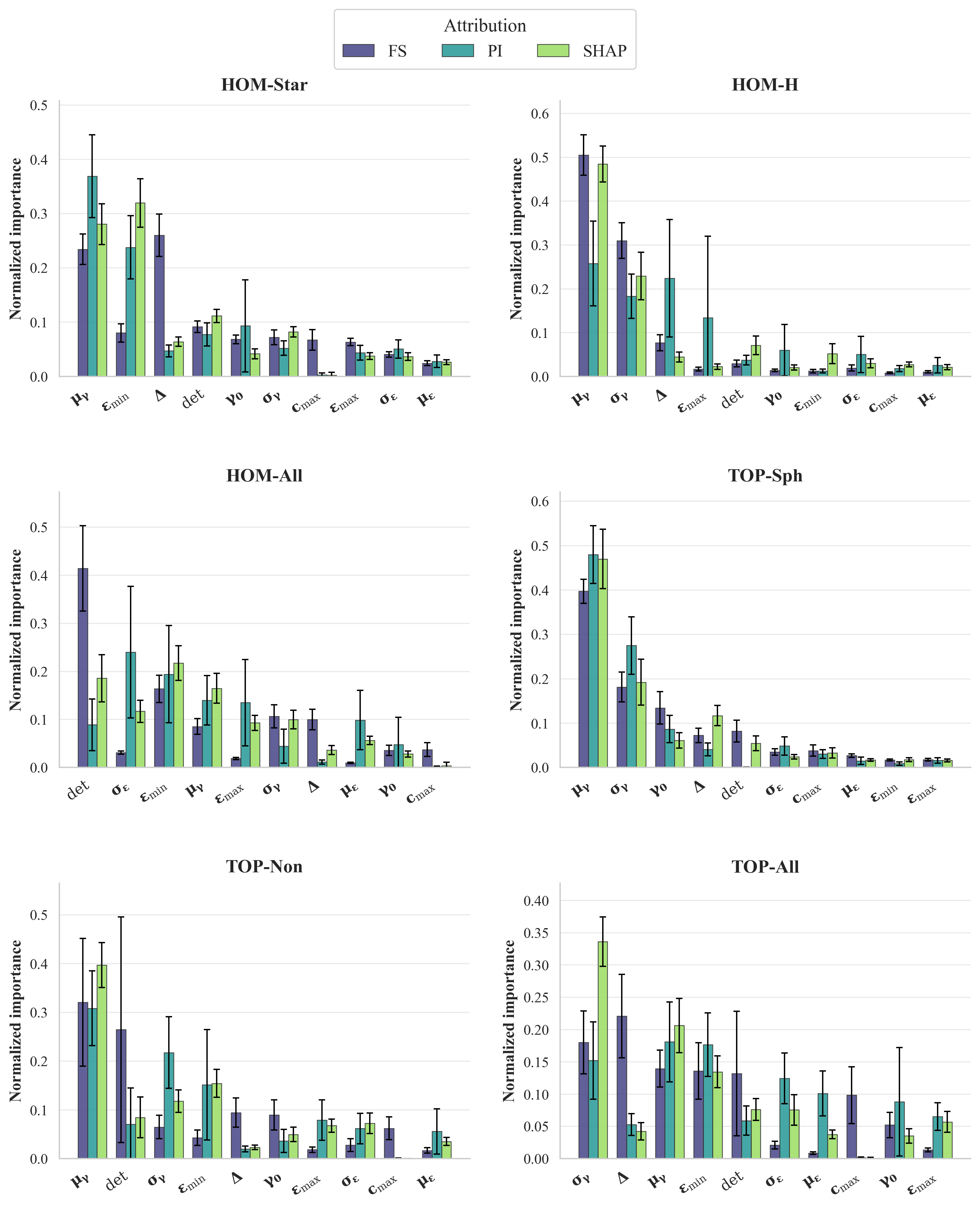}
    \caption{\textbf{Value-Index.} Mean normalized feature importance scores for feature saliency, permutation importance, and SHAP; averaged over all five seeds for all HOM and TOP experiments.}
    \label{fig:VI_importance}
\end{figure}

\begin{figure}
    \centering
    \includegraphics[width=1\linewidth]{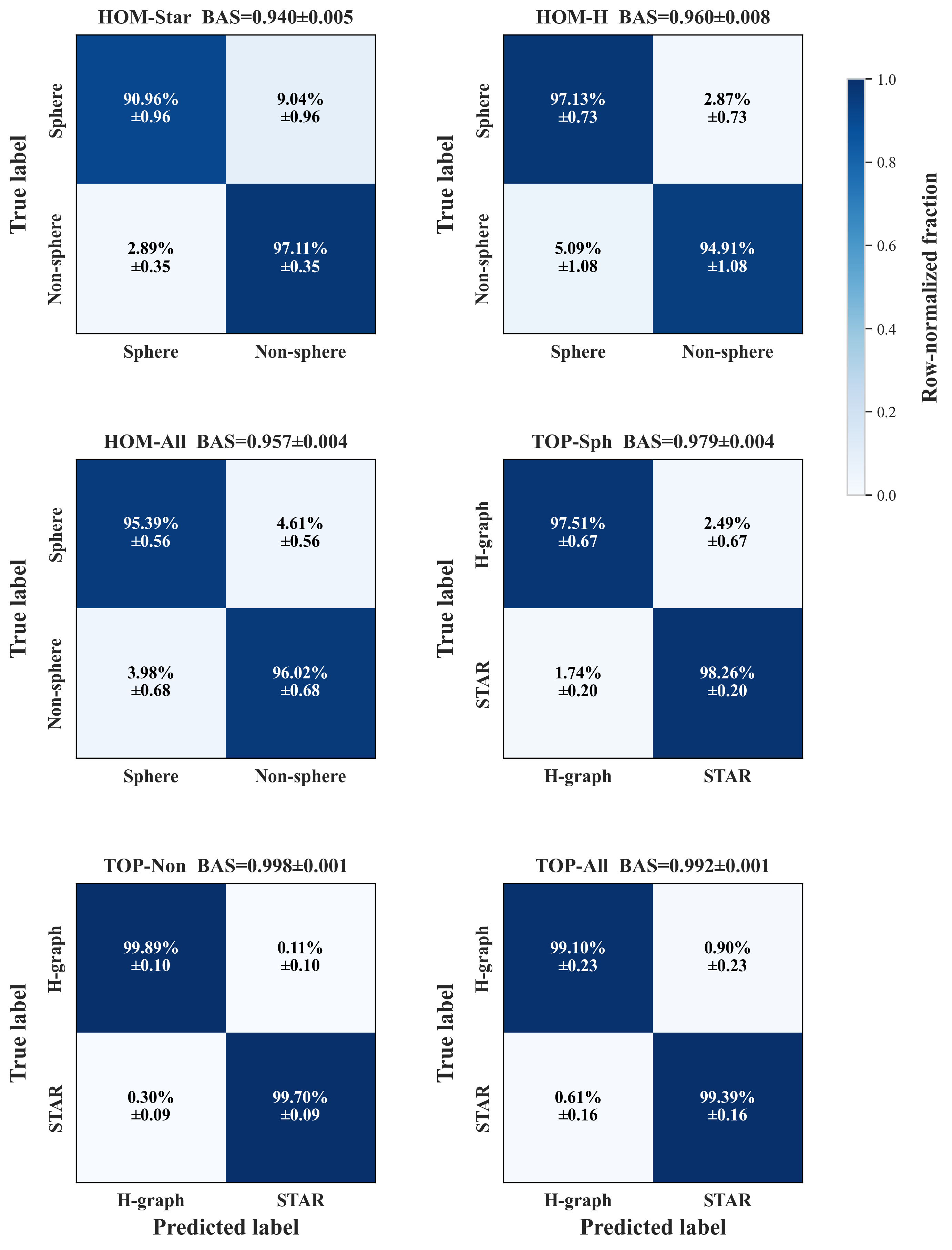}
    \caption{\textbf{Latent Space.} Mean confusion matrices across five seeds for all HOM and TOP experiments. }
    \label{fig:LS_confusion}
\end{figure}
\begin{figure}
    \centering
    \includegraphics[width=1\linewidth]{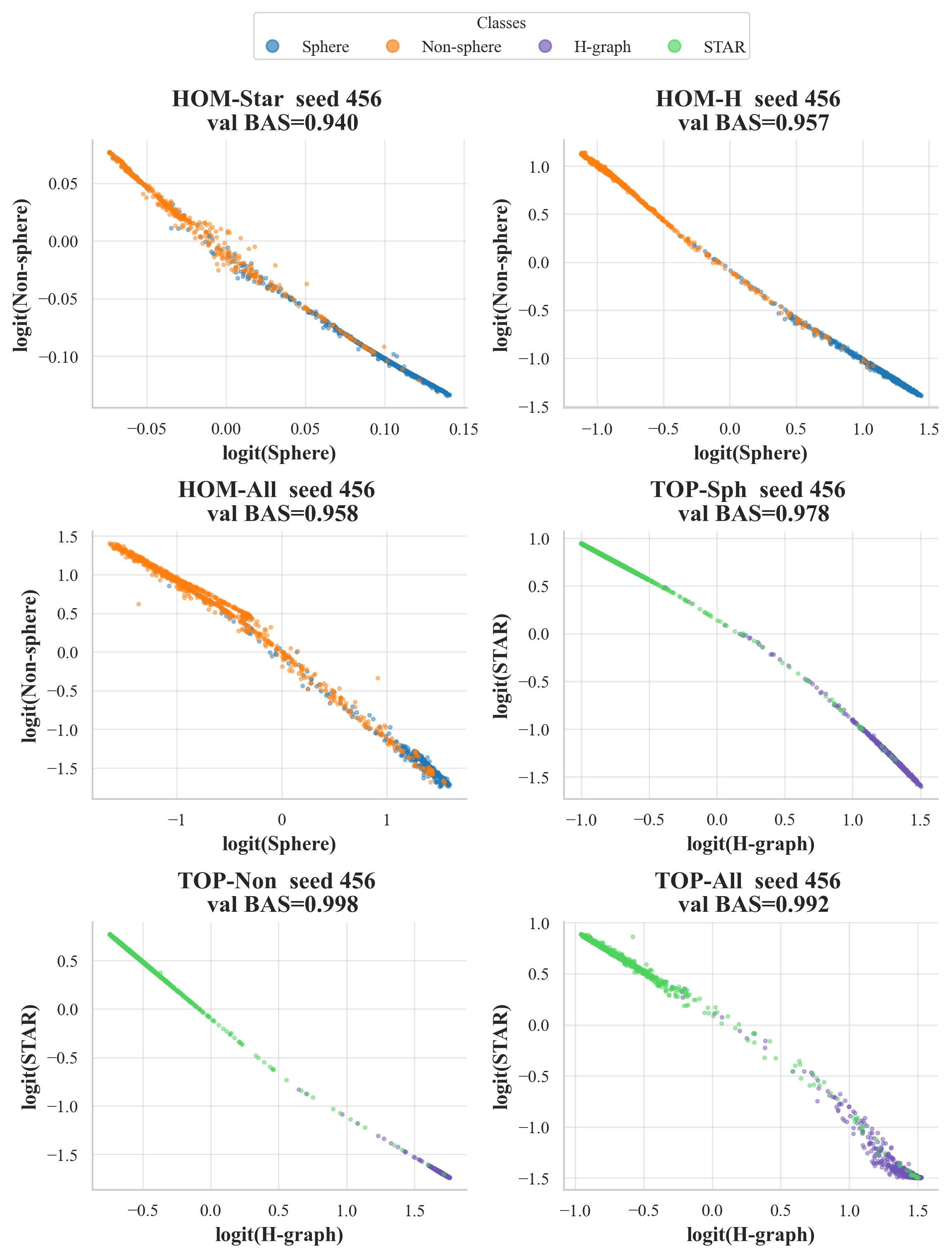}
    \caption{\textbf{Latent Space.} Logit space visualization for a representative seed (\texttt{seed}-42) for all HOM and TOP experiments.}
    \label{fig:LS_logits}
\end{figure}

\begin{figure}
    \centering
\includegraphics[width=1\linewidth]{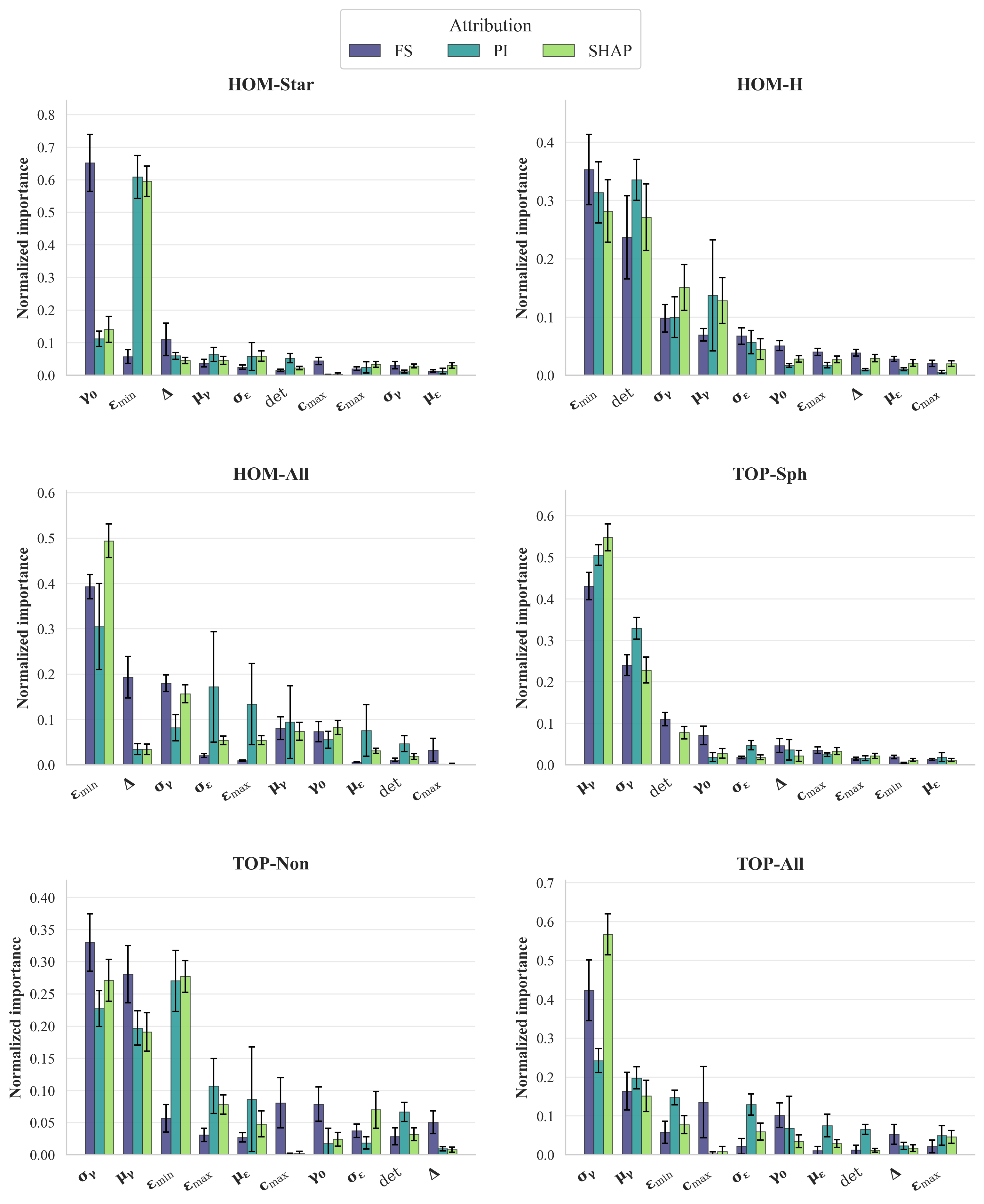}
    \caption{\textbf{Latent Space.} Mean normalized feature importance scores for feature saliency, permutation importance, and SHAP; averaged over all five seeds for all HOM and TOP experiments.}
    \label{fig:LS_importance}
\end{figure}

\begin{figure}
    \centering
    \includegraphics[width=1\linewidth]{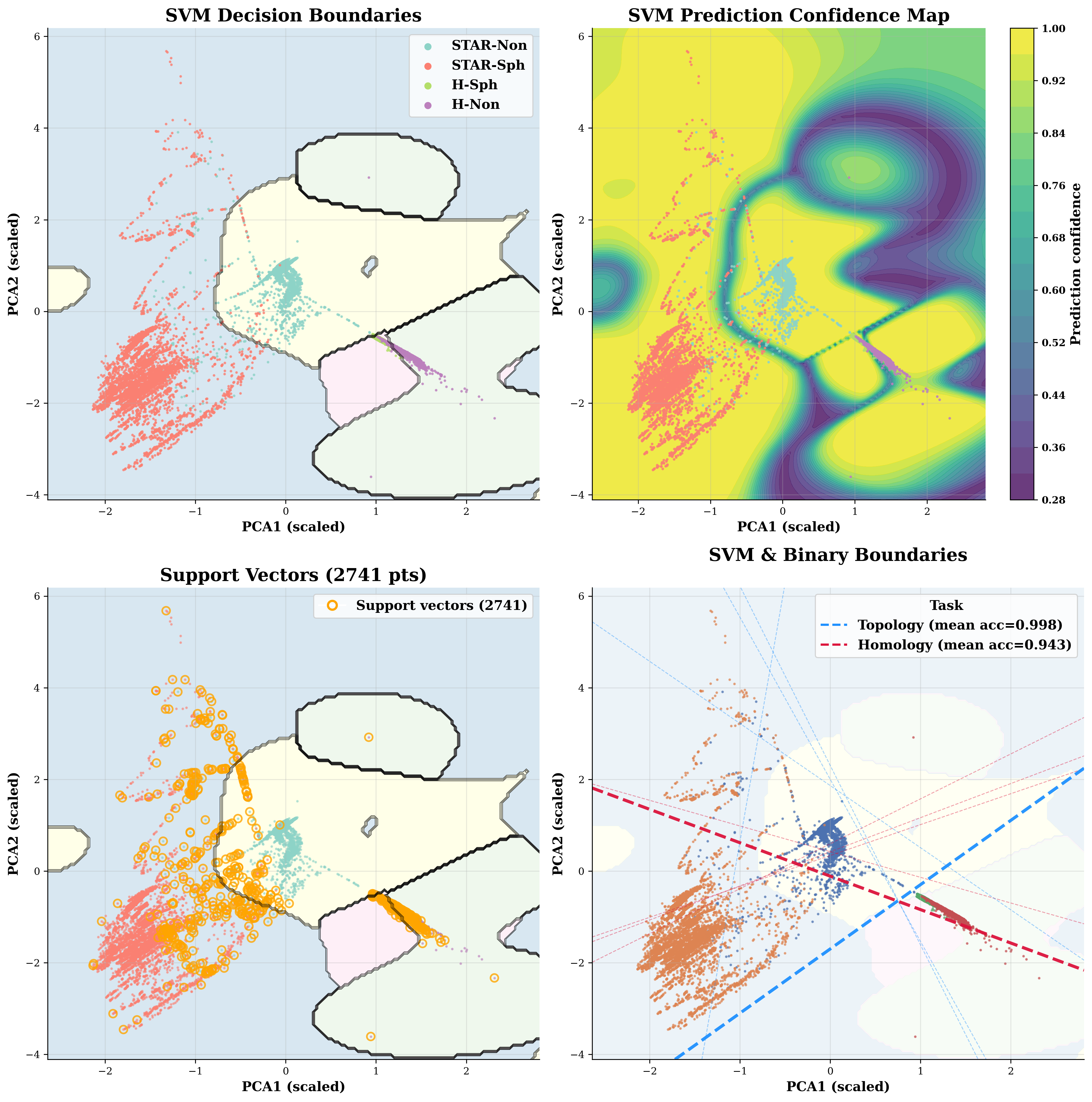}
    \caption{\textbf{Multi-class Latent Space; \texttt{seed}-42.} Top Left: SVM decision boundaries in latent space separating STAR-Sph, STAR-Non, H-Sph, and H-Non classes.  TOP Right: SVM prediction confidence map.  Bottom Left: SVM support vectors.  Bottom Right: SVM and linear, binary decision boundaries overlaid with \texttt{seed}-42 binary boundaries displayed as thick dashed lines and the linear boundaries for other seeds rendered as faint, dashed lines.}
    \label{fig:svm_decision}
\end{figure}

\begin{figure}[t]
\centering
\begin{subfigure}{0.495\textwidth}
    \includegraphics[width=\textwidth]{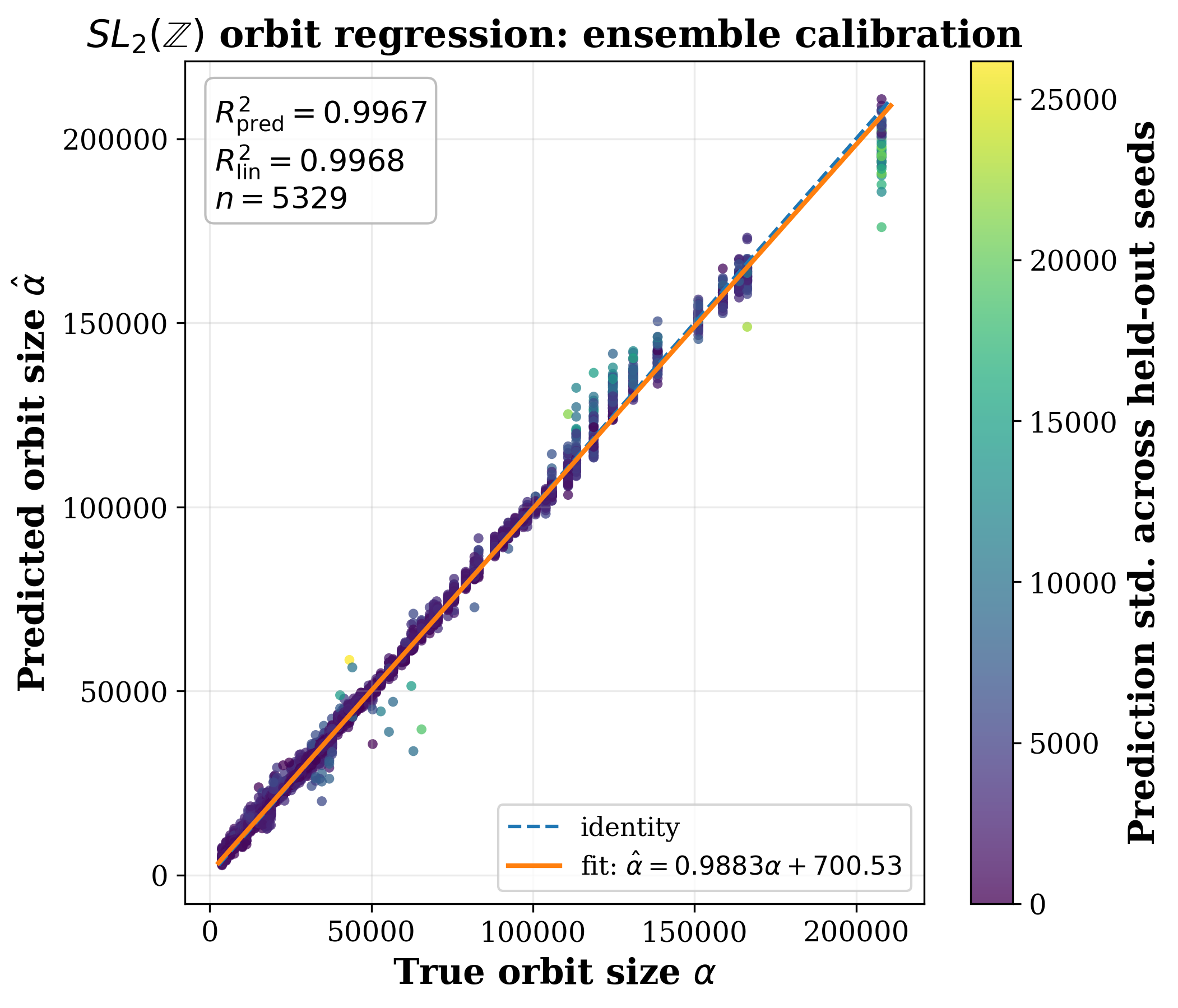}
    \caption{}
    \label{fig:sl2-indistribution-linear-fit}
\end{subfigure}
\hfill
\begin{subfigure}{0.495\textwidth}
    \includegraphics[width=\textwidth]{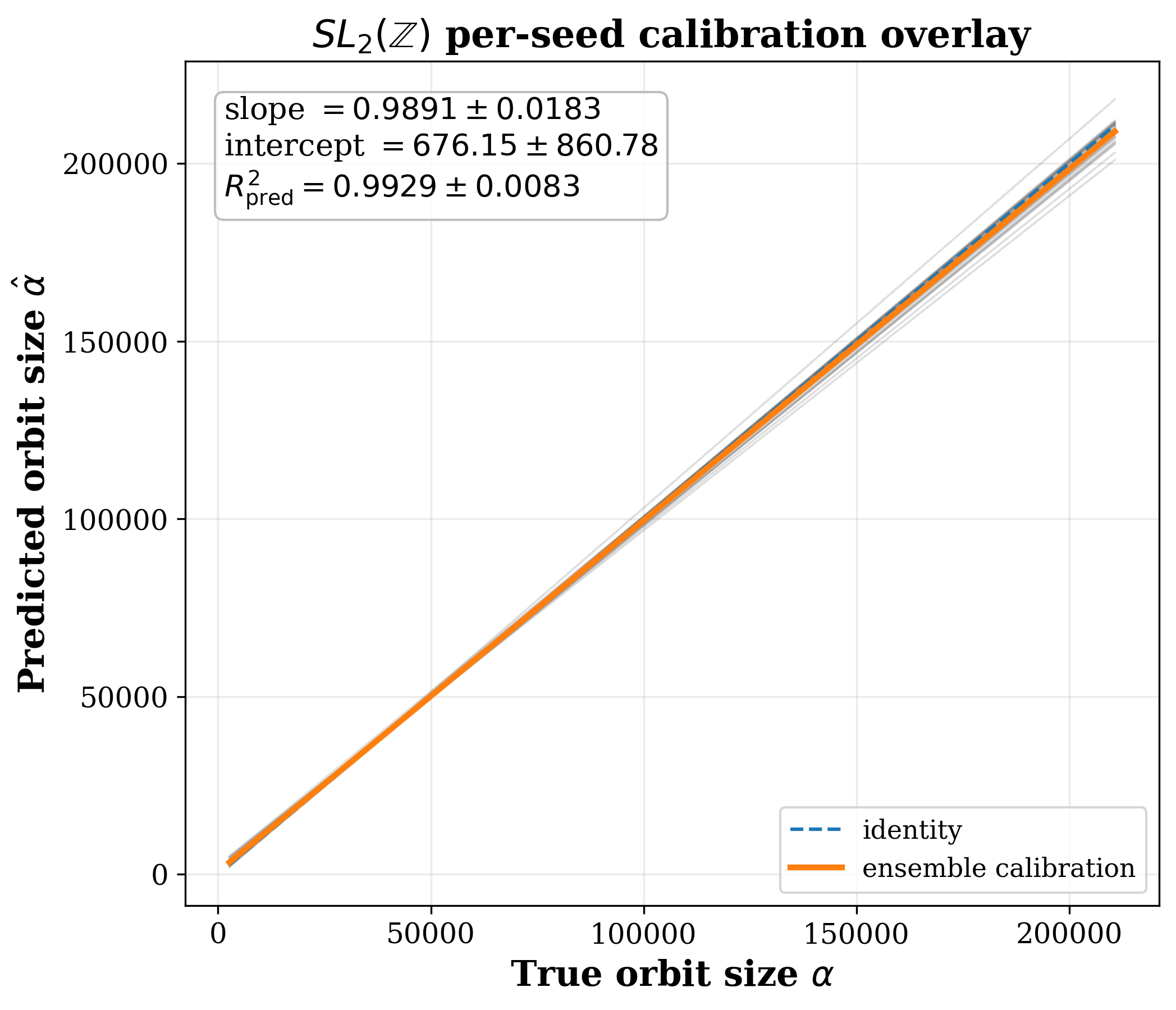}
    \caption{}
    \label{fig:sl2-seed-overlay}
\end{subfigure}
\caption{ \textbf{Learning $SL_2(\mathbb{Z})$ orbit.} \textbf{(a)} In-distribution \(SL_2(\mathbb{Z})\)-orbit regression.  Points show the ensemble-mean transformer prediction \(\widehat{\alpha}\) against the true orbit size \(\alpha\) over the unique held-out samples appearing across the \(30\)-seed robustness run.  Color indicates the standard deviation of the prediction across seeds.  The dashed line is the ideal relation \(\widehat{\alpha}=\alpha\), while the solid line is the fitted calibration.  The ensemble prediction shows strong in-distribution interpolation but visible seed sensitivity and degradation for some large-orbit examples. \textbf{(b)} Seed overlay where each point is a held-out prediction from one of the \(30\) transformer seeds.  The dashed line is the ideal relation \(\widehat{\alpha}=\alpha\).  The overlay shows that most seeds learn the same global linear trend, but the spread increases for large orbit sizes and for a small set of lower-\(\alpha\) outliers.}
\end{figure}

\begin{figure}
    \centering
    \includegraphics[width=0.85\linewidth]{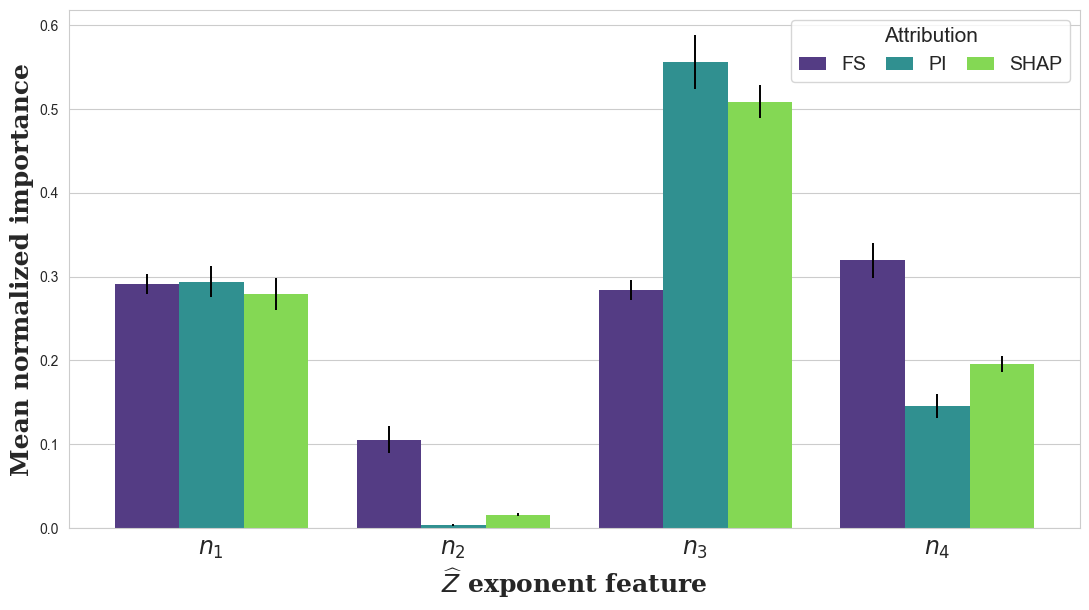}
    \caption{\textbf{Learning $d$ from $\widehat{Z}$.} Normalized mean exponent feature importance.}
    \label{fig:learning-d-from-zhat}
\end{figure}

\begin{figure}
    \centering
    \includegraphics[width=\linewidth]{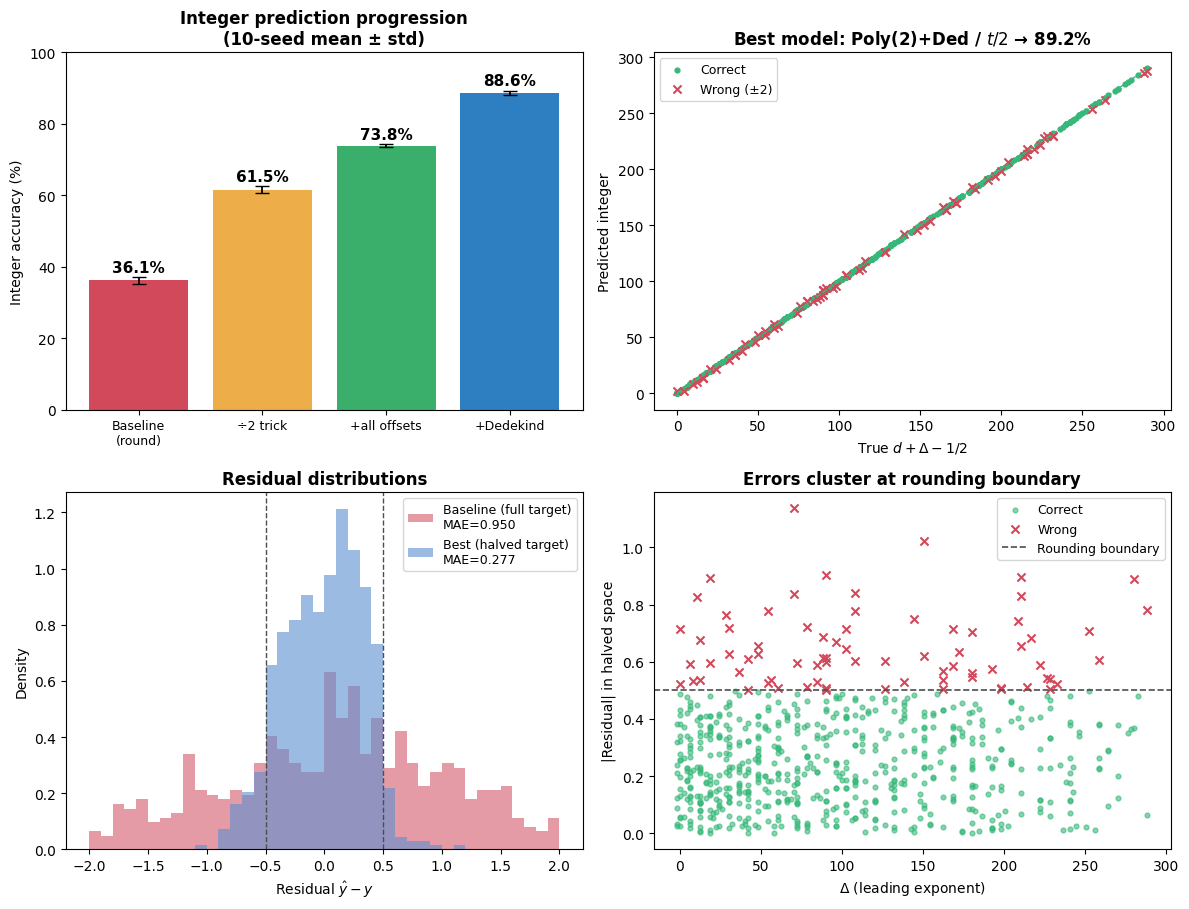}
    \caption{Error profile for learning $d$ exactly, on the $n=685$
three-fiber dataset $\mathcal{D}_3^{20}$. Top left:
exact-integer accuracy at each stage (from left: direct rounding with
$\Delta$ appended, the halved target $t/2$, all nineteen offsets, the
symmetrized Dedekind features), mean $\pm$ one standard deviation over
the ten seeds; the offsets-only starting point ($17.9\%$) is discussed
in the text. The remaining panels show out-of-fold predictions under
the seed-42 fold assignment, for which the best model is exact on
$89.2\%$ ($611/685$), every miss off by exactly $\pm 2$ in $t$. Top
right: predicted vs.\ true $t = d+\Delta-\tfrac12$. Bottom left:
residual histograms of the best model (halved target, MAE $0.277$) and
of the first bar's rounding model (full target, MAE $0.950$) under the
same folds; each is plotted in its own rounding variable, so
$|\varepsilon|<\tfrac12$ marks a correctly rounded prediction in both,
and the axis is truncated to $[-2,2]$. Bottom right: best-model
$|\varepsilon|$ against $\Delta$, with the misses marked.}
    \label{fig:exact_error_profile}
\end{figure}

\end{document}